\documentclass[twocolumn]{aastex631}
\usepackage{rotating}
\usepackage{graphicx}
\usepackage{amssymb}
\usepackage{amsmath}
\usepackage{hyperref}
\usepackage{longtable}
\usepackage{comment}


\def\gtrsim{\mathrel{\hbox{\rlap{\hbox{\lower4pt\hbox{$\sim$}}}\hbox{$>$}}}}
\def\lesssim{\mathrel{\hbox{\rlap{\hbox{\lower4pt\hbox{$\sim$}}}\hbox{$<$}}}}

\begin{document}

\title{A Multiwavelength Machine-learning Approach to Classifying X-ray Sources in the Fields of Unidentified 4FGL-DR4 sources}
\author{Hui Yang}
\affiliation{Department of Physics, The George Washington University, 725 21st St, NW, Washington, DC 20052, USA}
\author{Jeremy Hare}
\affiliation{Astrophysics Science Division, NASA Goddard Space Flight Center, 8800 Greenbelt Rd, Greenbelt, MD 20771, USA}
\affiliation{Center for Research and Exploration in Space Science and Technology, NASA/GSFC, Greenbelt, MD 20771, USA}
\affiliation{The Catholic University of America, 620 Michigan Ave., N.E. Washington, DC 20064, USA}
\author{Oleg Kargaltsev}
\affiliation{Department of Physics, The George Washington University, 725 21st St, NW, Washington, DC 20052, USA}

\email{huiyang@gwmail.gwu.edu}

\begin{abstract}

A large fraction of Fermi-Large Area Telescope (LAT) sources in the fourth Fermi-LAT 14\,yr catalog (4FGL) still remain unidentified (unIDed).  
We continued to improve our machine-learning pipeline and used it to classify 1206 X-ray sources with signal-to-noise ratios $>3$ located within the extent of 73 unIDed 4FGL sources with Chandra X-ray Observatory observations included in the Chandra Source Catalog 2.0. 
Recent improvements to our pipeline include astrometric corrections, probabilistic cross-matching to lower-frequency counterparts, and a more realistic oversampling method. 
X-ray sources are classified into eight broad predetermined astrophysical classes defined in the updated training data set, which we also release. 
We present details of the machine-learning classification, describe the pipeline improvements, and perform an additional spectral and variability analysis for brighter sources. 
The classifications give 103 plausible X-ray counterparts to 42 GeV sources. 
We identify 2 GeV sources as isolated neutron star candidates, 16 as active galactic nucleus candidates, seven as sources associated with star-forming regions, and eight as ambiguous cases. 
For the remaining 40 unIDed 4FGL sources, we could not identify any plausible counterpart in X-rays, or they are too close to the Galactic Center. Finally, we outline the observational strategies and further improvements in the pipeline that can lead to more accurate classifications. 
\end{abstract}


\section{Introduction}

Over the last 15\,yr, the Large Area Telescope (LAT) on board the Fermi $\gamma$-ray Space Telescope has revealed a large number of $\gamma$-ray sources \citep{2009ApJ...697.1071A,2022ApJS..260...53A}. However, only a small fraction, comprising 14\% for Galactic sources with $|b|<5^{\circ}$ or 6\% for all sources, are confidently identified in the fourth Fermi-LAT 14\,yr point-source catalog (4FGL-DR4, hereafter 4FGL; \citealt{2023arXiv230712546B}), among a total of 7,195 sources. 
The 4FGL catalog also includes the associated source category ($\approx$60\% of sources) where the $\gamma$-ray sources have probable counterparts at other wavelengths based purely on positional coincidence. The matched counterpart may or may not belong to a known source class.

Among the identified Galactic sources, pulsars stand out as one of the most prevalent types. The number of $\gamma$-ray pulsars has surged from six during the \emph{EGRET} era to 297 in the third Fermi-LAT pulsar catalog \citep{2023ApJ...958..191S}. 
While many pulsars are identified through their $\gamma$-ray and radio pulsations, imaging X-ray observations can also unveil young pulsar (PSR) candidates. This identification can be achieved either through their distinct spectral properties and the absence of multiwavelength (MW) counterparts or by resolving extended emission from pulsar wind nebulae (PWNe, see, e.g., \citealt{2012ApJS..201...37K,2021ApJ...914...85H}). Recent research has leveraged machine learning (ML) for classifying unidentified (unIDed) $\gamma$-ray sources based solely on their $\gamma$-ray properties \citep{2023MNRAS.521.6195M,2024MNRAS.527.1794Z}. Other studies have utilized Swift observations, emphasizing the importance of MW data for improved classification accuracy \citep{2021ApJ...923...75K,2022ApJ...940..139J}. 
Recently, a systematic search for X-ray counterparts of unassociated Fermi-LAT sources was conducted using the first four scans comprising the eROSITA all-sky survey in the western Galactic hemisphere, using their own $\gamma$-ray-based classifications and the X-ray and MW properties of potential counterparts \citep{2024arXiv240117295M}. 
We recently conducted a pilot study of X-ray sources coincident with 13 Fermi-LAT sources using the multiwavelength machine-learning classification pipeline \citep[MUWCLASS;][]{2022ApJ...941..104Y} to explore their nature \citep{2024ApJ...961...26R}.  

The Chandra X-ray Observatory has observed many unIDed GeV sources discovered since the Fermi observatory was launched. 
The Chandra Source Catalog version 2.0 \citep[hereafter CSCv2.0;][]{2010ApJS..189...37E,2020AAS...23515405E} analyzed the Chandra X-ray Observatory data obtained before the end of 2014, while the more recent Chandra Source Catalog version 2.1 (hereafter CSCv2.1) added more data (up to the end of 2021) but was not formally released at the time of writing this paper and is still undergoing improvements. 
The CSCv2.0, covering $\sim$550 deg$^2$ of the sky, provides detailed information about positional, spectral, and temporal properties for $\sim$317,000 unique sources. 
The subarcsecond Chandra X-ray Observatory source localizations greatly reduce the chance of confusion when cross-matching to counterparts at lower frequencies, which is particularly important in the crowded regions of the Galactic plane.

In this paper, we present a census and classifications of X-ray sources from the CSCv2.0 within the 95\% confidence level positional uncertainties (PUs) of unIDed 4FGL sources. We use an updated version of MUWCLASS to shed light on the nature of these 4FGL sources. Section \ref{sec:obs_and_dat} details the selection of 4FGL unIDed sources, and the X-ray and MW data analysis. It is followed by the description of the updated automated MW classification pipeline in Section \ref{sec:muwclass}. Section \ref{sec:results} presents the results of the classifications of X-ray sources. 
We discuss the relationships between X-ray, radio, and $\gamma$-ray sources, and compare our results to other independent classification studies in Section \ref{sec:disucssion}. 
The conclusions are given in Section \ref{sec:conclusion}.

\section{Observations and Data Reduction}
\label{sec:obs_and_dat}

\subsection{Selection of unIDed 4FGL sources}

We selected the unIDed 4FGL sources, which include unassociated sources, as well as associated sources that are marked as unknown class sources \footnote{Unknown class sources are sources with $|b| < 10^\circ$ that have a sole counterpart (of unknown type) from large radio and X-ray surveys according to the 4FGL catalog.} from the 4FGL catalog \citep{2023arXiv230712546B}. 
Extended GeV $\gamma$-ray sources were excluded due to the challenges posed by their large sizes, which may depend on the accuracy of the Galactic diffuse $\gamma$-ray background model. 
To select 4FGL sources covered by Chandra X-ray Observatory ACIS observations, we cross-matched the CSCv2.0 with the selected 4FGL sources, resulting in 109 4FGL sources such that at least one CSCv2.0 source, with X-ray detection significance$>3$, is found within PUs, where 90 of them are unassociated sources, and 19 are associated sources matched to the unknown type sources. 
In principle, there may be 4FGL sources that have Chandra X-ray Observatory coverage, but no X-ray sources were detected. 
Among the 109 unIDed 4FGL sources, 36 are identified as sources on top of an interstellar clump (hereafter C-type sources) potentially linked to imperfections in the model for Galactic diffuse emission, with a character ``c" appended to their names, according to the 4FGL catalog. 
While the classification results for all 109 sources are summarized in Table \ref{tab:classification}, our focus in the subsequent sections will be on the 73 non-C-type sources. 

\begin{figure*}
\begin{center}
\includegraphics[scale=0.25,trim=0 0 0 0]{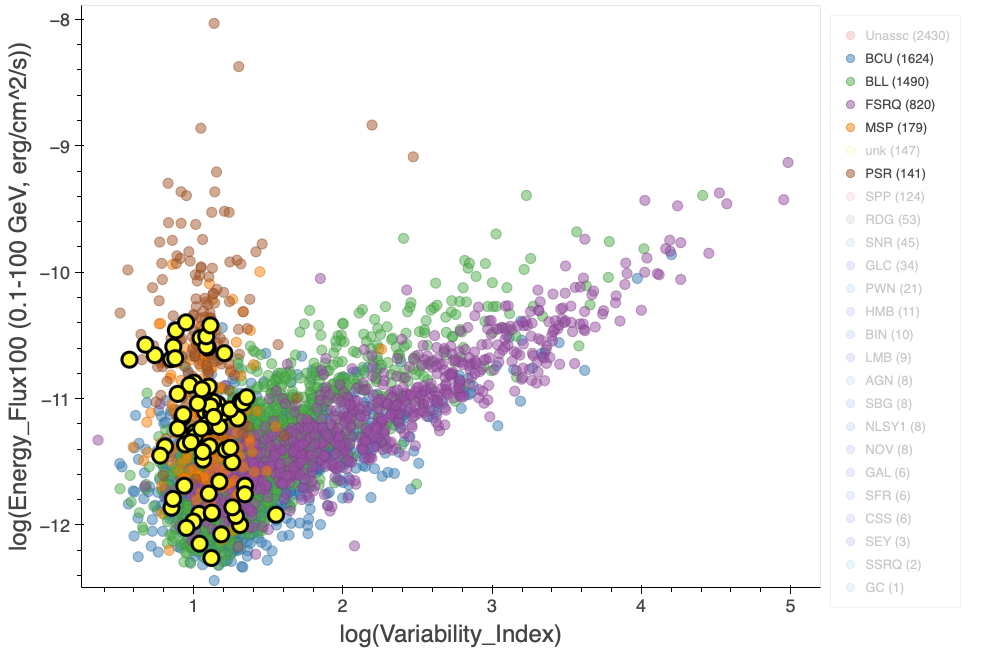}
\includegraphics[scale=0.25,trim=0 0 0 0]{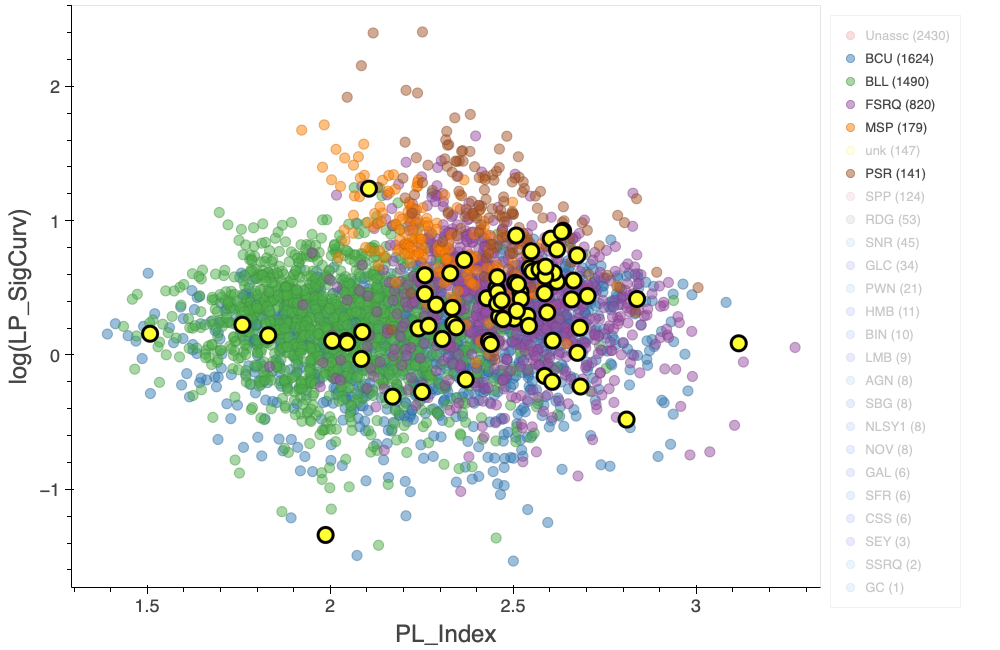}
\caption{Examples of 2D slices of the multidimensional feature space of pulsars (MSPs in orange and PSRs in brown) and blazars (BLLs in green, FSRQs in purple, and BCUs in blue) from the 4FGL catalog \citep{2023arXiv230712546B},  with 73 unIDed GeV sources studied in this paper overplotted in yellow. The visualization tool is available at \url{https://muwclass.github.io/GCLASS/}} 
\label{fig:GCLASS}
\end{center}
\end{figure*}

We have developed an interactive web-based plotting tool to visualize the $\gamma$-ray properties of the 4FGL catalog through various 2D slices of the multidimensional feature space. The tool is available online \footnote{\url{https://muwclass.github.io/GCLASS/}}. 
Two example plots, shown in Figure \ref{fig:GCLASS}, demonstrate a good degree of separation between pulsars, including millisecond pulsars (MSP) and PSR, and blazars, including BL Lac objects (BLL), flat-spectrum radio quasars (FSRQ), and blazar candidates of uncertain type (BCU) from 4FGL, overplotted with the 73 selected unIDed sources. 
The left panel shows the plot of the energy flux in the 0.1--100 GeV (Energy\_Flux100) vs.\ the $\gamma$-ray variability index (Variability\_Index), where most of the variable GeV sources are blazars. The right panel shows the plot of the significance of the GeV spectral
curvature (LP\_SigCurv) vs.\ the photon index obtained from fitting with a power-law (PL) model (PL\_Index), where BLL sources exhibit harder photon indices. All of the 73 unIDed sources are not variable in GeV with Variability\_Index$<40$, and some of them (with higher Energy\_Flux100 values) are more aligned with pulsars. A few unIDed 4FGL sources exhibit smaller PL\_Index values, making them similar to BLL sources.

\subsection{Chandra Source Analysis}

We used the methods outlined in \cite{2022ApJ...941..104Y}, 
with some updates, to extract the X-ray properties from the CSCv2.0 for both the training data set (TD) and the X-ray sources to be classified. 
One significant update involves applying an absolute astrometric correction to the X-ray source positions, a step not performed in the CSCv2.0 but to be implemented in the not-yet-released CSCv2.1. 
This was achieved by aligning the X-ray source positions to reference sources from the Gaia DR3 catalog \citep{2023A&A...674A...1G} using a method akin to that provided on the CSCv2.1 website,\footnote{\url{https://cxc.cfa.harvard.edu/csc/memos/files/Martinez-Galarza_stack_astrometry_translation_spec_doc.pdf}} and explained in the Appendix \ref{apd:astrometry}.
Additionally, we modified the calculation of the interobservation variability feature ($P_{\rm inter}$ used in \citealt{2022ApJ...941..104Y}), since we found the calculations in \cite{2022ApJ...941..104Y} were too conservative, resulting in underestimated variability probabilities in certain cases. Previously, \cite{2022ApJ...941..104Y} used the sum of per-observation soft, medium, and hard band fluxes to calculate the per-observation broadband flux. However, if one of the bands was missing for a given observation, the broadband flux was not calculated. Therefore, we now use the per-observation broadband fluxes from the CSCv2.0 to compute the variability.

\subsection{MW Data} 
\label{sec:MW-data}

We cross-matched the astrometrically corrected CSCv2.0 source positions with various MW catalogs, including Gaia DR3 \citep{2023A&A...674A...1G},  Two Micron All-Sky Survey  (2MASS; \citealt{2006AJ....131.1163S}), AllWISE \citep{2014yCat.2328....0C}, and CatWISE2020 \citep{2021ApJS..253....8M}. 
We decided to use CatWISE2020 instead of unWISE \citep{2019ApJS..240...30S} due to the additional 3\,yr of data and similar analysis methods to AllWISE, and the availability of proper motions from CatWISE2020. 

Before cross-matching with X-ray sources, we combined these MW catalogs into one merged catalog using their PUs and proper-motion measurements. We used a similar method as the one outlined for Gaia cross-matching in \cite{2019A&A...621A.144M} with a 5$\sigma$ threshold. 
We searched a large radius of 30$\arcsec$, centered on the X-ray source, for each MW catalog and used an iterative method to merge the four MW catalogs. The 2MASS, AllWISE, and CatWISE2020 sources were matched to Gaia sources, with Gaia source coordinates adjusted to each catalog's epoch using Gaia proper motions. If the separation between Gaia and other catalog sources is smaller than the 5$\sigma$ combined (in quadrature) PUs of the two catalogs, the two sources are considered a match. For MW sources not matched to Gaia, we attempted to match them with CatWISE2020 sources using proper motions from CatWISE2020. If no match with Gaia or CatWISE2020 is established, the remaining sources from 2MASS and AllWISE catalogs are matched with each other without any proper-motion corrections. For proper-motion measurements from CatWISE2020, which are not as accurate as those from Gaia, we did not use them if the ratios of their total proper motions to their uncertainties $<5$ or the reduced chi-squared of the astrometry fitting $>1.5$. We did not use proper motions from AllWISE since they did not account for parallax. 
In cases where a source is matched to multiple catalogs, we prioritize its positions and the PUs in the following order: Gaia DR3, CatWISE2020, 2MASS, and AllWISE. 

We then employed the probabilistic cross-matching algorithm NWAY \citep{2018MNRAS.473.4937S} to match the merged catalog to X-ray sources. 
The NWAY algorithm provides the probability of having a true counterpart among all considered associations ($p\_{\rm any}$), and the probability of each $i$th counterpart ($p\_{\rm i}$), including the case of having no association at all. 
We also calculated the chance coincidence probability, which is the probability of having one or more MW catalog sources within a randomly placed circle of a chosen radius,  $P_{\rm c} = 1 - \exp(- \rho \pi r_{\rm PU}^2)$, where $\rho$ ($\rm arcsec^{-2}$) is the source density from the merged MW catalog calculated for a radius of 30\arcsec, and $r_{\rm PU}$  is the 2$\sigma$ combined PU radius of the X-ray and MW sources. 
While NWAY allows for the inclusion of each source's spectral energy distribution (SED) as additional priors, we opt not to do this due to the lack of a priori knowledge regarding the classifications of the sources under consideration. The cross-matching results between the X-ray sources and the merged MW catalog are available for viewing online.\footnote{\url{https://muwclass.github.io/MUWCLASS_4FGL-DR4/}}

\section{Automated Multiwavelength Classification}
\label{sec:muwclass}

MUWCLASS pipeline developed by \cite{2022ApJ...941..104Y} is used to perform automated classification of CSCv2.0 X-ray sources within the PUs of the 109 unIDed 4FGL sources. 
The pipeline employs a random forest (RF) ML model with a TD of $\sim3,000$ CSCv2.0 sources with literature-verified classes. We have made minor updates to the TD presented in \cite{2022ApJ...941..104Y}, which are outlined in Appendix \ref{apd:TD-updates}. The TD includes active galactic nuclei (AGNs), low-mass stars (LM-STARs), high-mass stars (HM-STARs), young stellar objects (YSOs), cataclysmic variables (CVs), high-mass X-ray binaries (HMXBs), low-mass X-ray binaries (LMXBs),\footnote{This class also includes nonaccreting X-ray binaries, such as wide-orbit binaries with MSPs, as well as red-back and black widow systems.} and isolated neutron stars (NSs). In addition to X-ray properties (fluxes, hardness ratios, X-ray variability measures), each source is characterized by its MW properties in the optical/near-infrared (NIR)/mid-infrared (MIR) bands, as detailed in Section \ref{sec:MW-data}.  
The updated TD is available and visualized online.\footnote{\url{https://muwclass.github.io/XCLASS_CSCv2/}}

Certain limitations and caveats of the MUWCLASS pipeline have been discussed in \cite{2022ApJ...941..104Y}, wherein we introduced methods to mitigate them. One limitation is the selection bias in the TD, where sources are systematically brighter than those we classify because they have to be bright enough for classification through traditional methods. 
Furthermore, there is a high risk of misclassifying sources without MW counterparts as NSs, as all NSs in the TD lack counterparts, while fainter X-ray sources are more likely to be distant and thus may not be detected in the optical/NIR/MIR surveys that we used. 

To address these challenges, we have introduced a different oversampling method than the one employed in \citealt{2022ApJ...941..104Y} to produce synthetic sources (see Appendix \ref{apd:physical-oversampling}). The new method samples the (optical) extinction and (X-ray) absorption values from the distributions observed in the TD sources and applies them to the lesser-populated (excluding AGN) classes. This oversampling is more realistic (i.e., ``physical") compared to generic algorithms (e.g., SMOTE; \citealt{2011arXiv1106.1813C} used in \citealt{2022ApJ...941..104Y}). It also provides a better representation (due to the extra extinction and absorption being applied) of the fainter population of sources,  which have fluxes more consistent with those of the unIDed X-ray sources that we classify.

Another bias arises from direction-dependent intervening extinction (absorption). 
The majority of AGNs in our TD have low extinction due to their locations off the Galactic plane, making them look markedly different from AGNs viewed through the Galactic plane. 
To partially address this bias, we incorporated direction-specific extinction and absorption for AGNs in the TD as described in \cite{2022ApJ...941..104Y}. 
Specifically, the X-ray fluxes and optical/NIR/MIR magnitudes of TD AGNs are artificially reddened using the Galactic absorption/extinction values from the direction of the sources to be classified. This correction will make the TD AGNs fainter, harder, and redder when we classify sources near/within the Galactic plane with high absorption/extinction values, but will not affect the TD AGNs much when we classify sources at higher Galactic latitudes with small absorption/extinction values.
However, a more comprehensive solution is required, involving the utilization of improved 3D extinction maps together with distance measurements, to account for this bias across all source types, rather than exclusively for AGNs. We plan to incorporate this into future releases of the MUWCLASS pipeline.

The performance evaluation of the updated MUWCLASS pipeline is detailed in Appendix  \ref{apd:evaluation}, after the integration of the latest TD and other developments. 
The updated MUWCLASS pipeline is made publicly available at the  GitHub repository.\footnote{\url{https://github.com/MUWCLASS/MUWCLASS}}

\section{Results} 
\label{sec:results}

\begin{figure*}
\begin{center}
\includegraphics[width=0.495\textwidth]{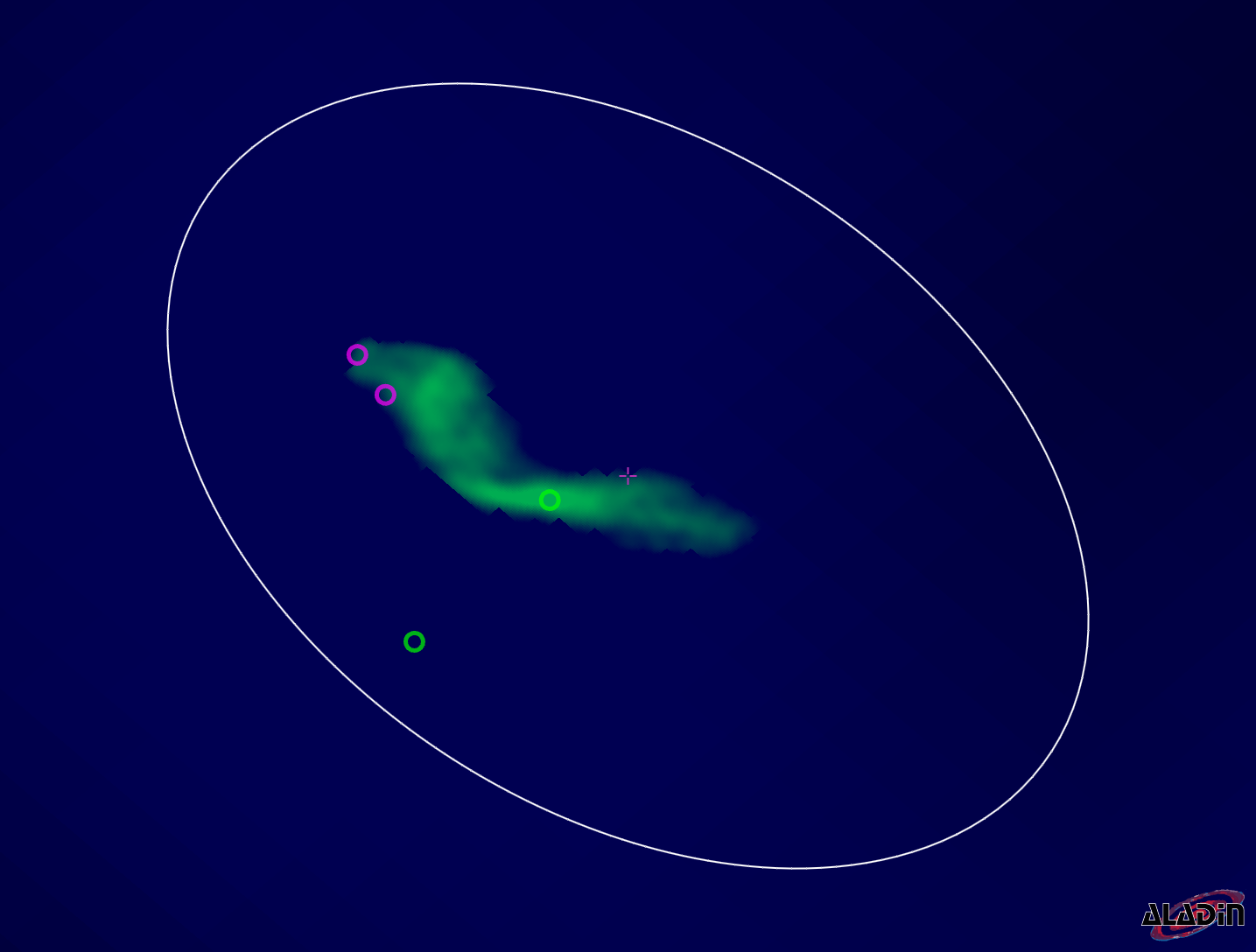}
\includegraphics[width=0.495\textwidth]{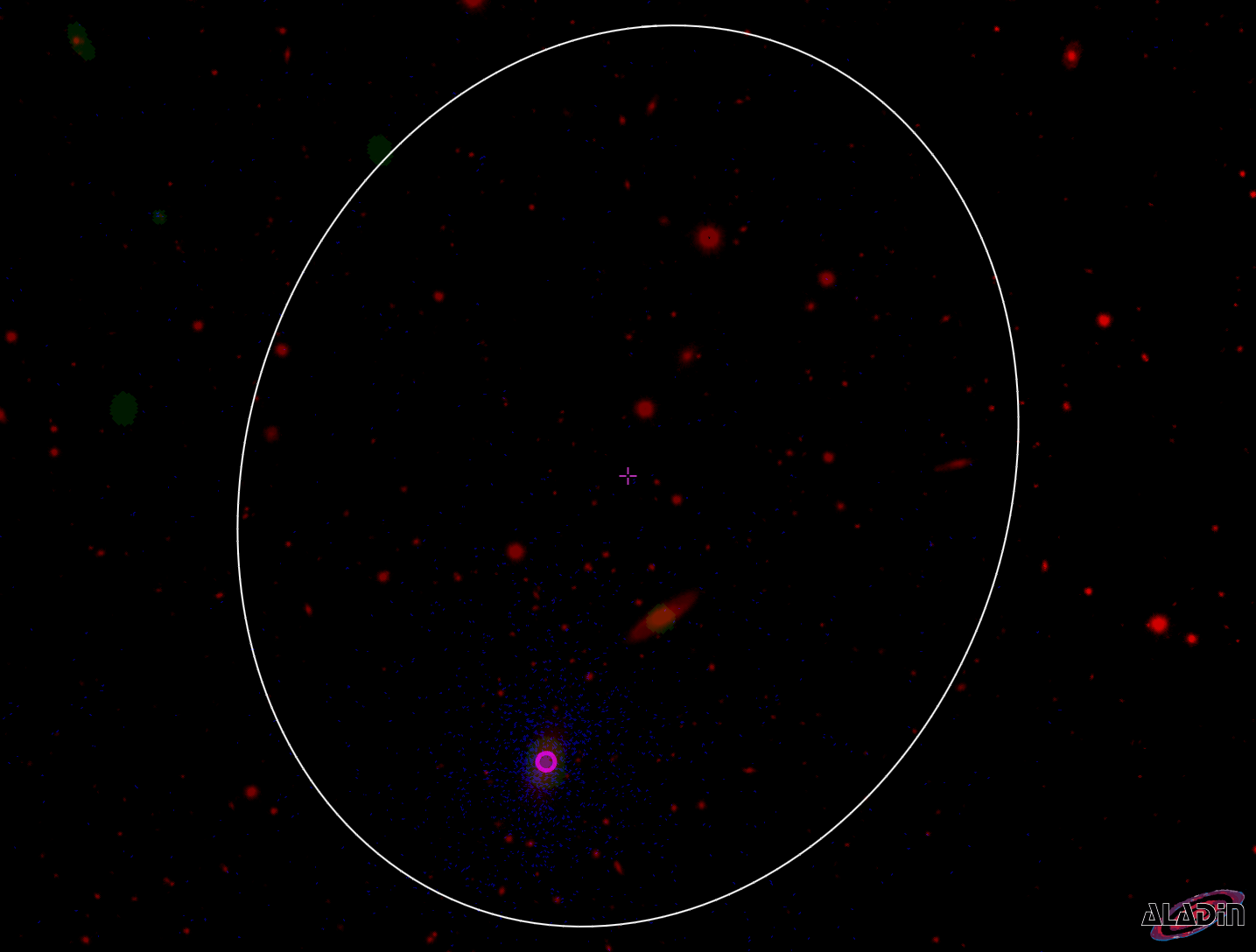}
\includegraphics[width=0.495\textwidth]{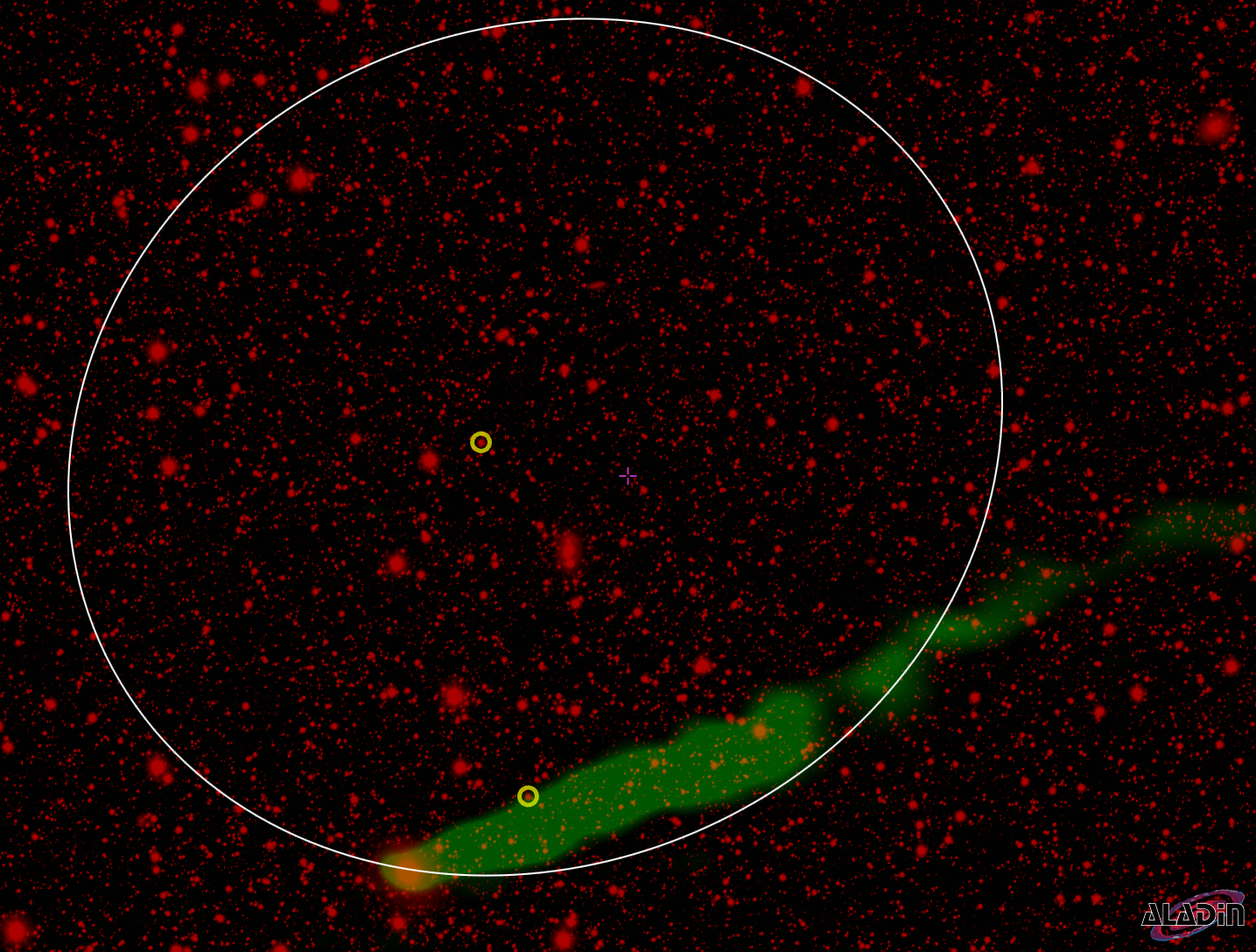}
\includegraphics[width=0.495\textwidth]{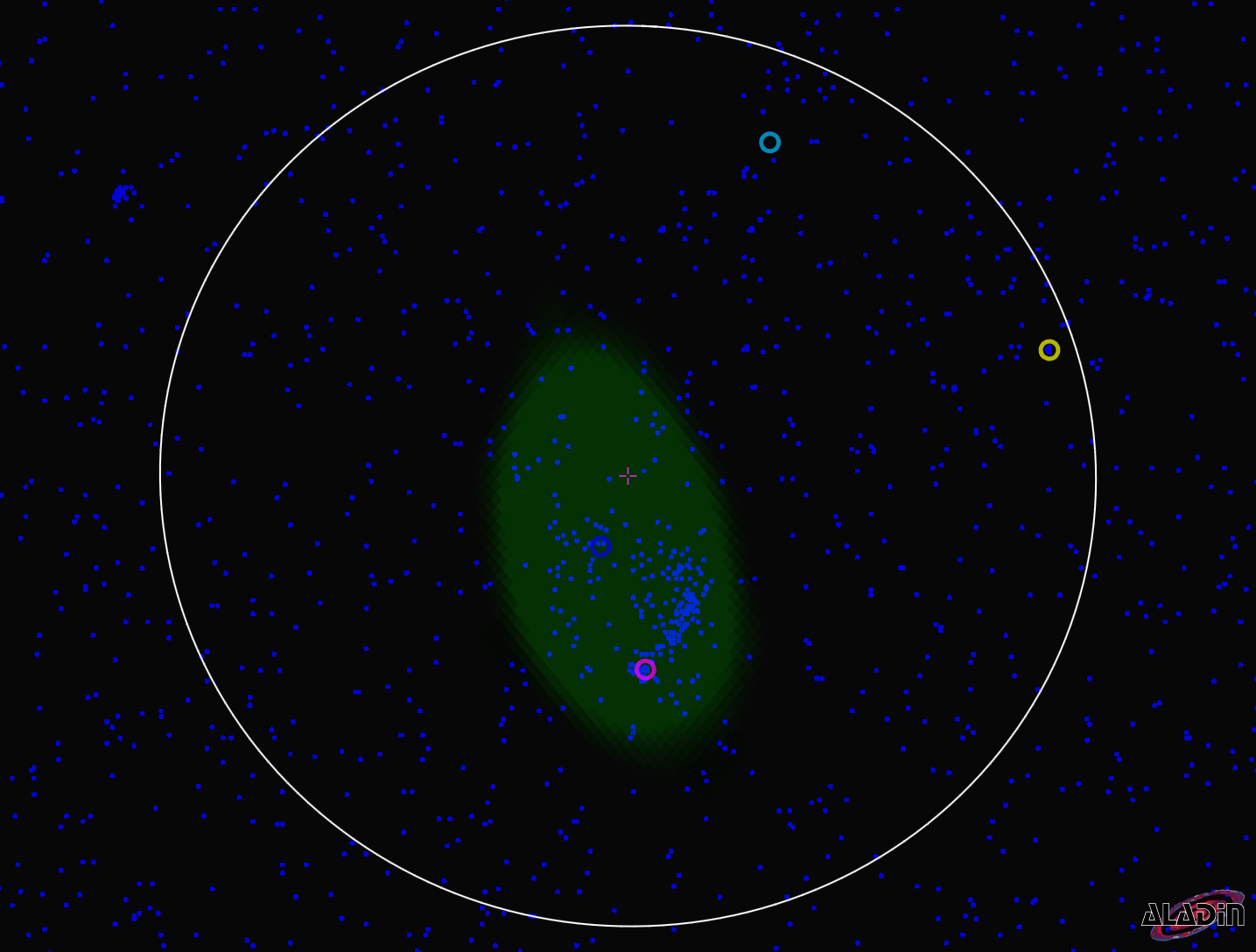}
\caption{Examples of visualization plots of selected 4FGL fields from \url{https://muwclass.github.io/MUWCLASS_4FGL-DR4/}. The top left panel shows the combined radio (RACS-low, in green) and TeV (HESS, in blue) image of 4FGL J1844.4--0306. The top right panel shows the combined radio (RACS-mid, in green), optical (PanSTARRS DR1, in red), and X-ray (Chandra X-ray Observatory, in blue) image of 4FGL J1025.9+1244. The lower left panel shows the combined radio (RACS-low, in green) and optical (DECaPS2, in red) image of 4FGL J1615.3--6034. 
The lower right panel shows the combined radio (RACS-low, in green) and X-ray (Chandra X-ray Observatory, in blue) image of 4FGL J2004.3+3339. The white ellipses represents the $\gamma$-ray error ellipses at 95\% confidence levels. The classified AGNs, YSOs, CVs, LM-STARs, and NSs are marked by cyan, lime, blue, yellow, and magenta circles, respectively.} 
\label{fig:fields} 
\end{center}
\end{figure*}

In this section, we present the classification results for X-ray sources within the 95\% PUs of the selected 4FGL sources. Comprehensive visualizations and interactive tables with all classification results can be accessed online,\footnote{\url{https://muwclass.github.io/MUWCLASS_4FGL-DR4/}} with a few examples shown in Figure \ref{fig:fields}. 
We group the discussion of the GeV sources based on the most likely classes of the X-ray sources that might be associated with the GeV sources. 
Sections \ref{sec:NS-Cand} through \ref{sec:Ambiguous} focus on NS candidates, AGN candidates, 4FGL sources associated with star-forming regions (SFRs), and ambiguous cases, respectively. 
19 out of the 73\,GeV sources are within $5^\circ$ from the complex Galactic Center environment, where the source density is exceptionally high, and therefore are not discussed below due to the challenges in cross-matching X-ray sources to potential low-frequency counterparts, which strongly affects the classification accuracy. The results of classifications of X-ray sources in these 19 GeV source fields are, however, included in Table \ref{tab:classification}.

For simplicity, in cases where multiple counterparts match an X-ray source (within its PU at the 95\% confidence level), we present only the classification results for the most likely association identified using NWAY. 
For the 1206 X-ray sources with signal-to-noise ratios (S/N)$>3$ that we classified within the 73 non-C-type GeV sources, 460 of them do not have any MW counterparts, 653 have only one MW counterpart, while the remaining 93 X-ray sources have more than one MW counterpart. 
For the rest of the paper, we will explicitly indicate whether a source has any MW counterparts. Otherwise, the counterpart will not be discussed if no counterpart is matched. 
However, detailed classification results for all associations are available.\footnote{\url{https://muwclass.github.io/MUWCLASS_4FGL-DR4/}}  
Table \ref{tab:classification} provides a summary of classifications for all X-ray sources with S/N$>3$, represented by the first numbers in columns (2)--(9) in the first row in Table \ref{tab:classification}. These numbers sum up to 1206 for the 73 non-C-type sources. Additionally, sources with S/N$>5$ and classified with classification confidence threshold (CT)$>2$ are represented by the second numbers in columns (2)--(9), amounting to 194 for the 73 non-C-type sources. 
We also summarized the number of sources that may be potential contributors to the $\gamma$-ray sources ($\gamma$-candidates) for each field  in column (10) in Table \ref{tab:classification}. The $\gamma$-candidates comprise significant sources with S/N$>5$ classified with CT$>2$, before the plus sign or in cases with no plus sign, and relatively bright sources with X-ray fluxes in the 0.5--7 \,keV $F_{\rm b}>5\times10^{-14}$\,erg\,s$^{-1}$\,cm$^{-2}$ (but not necessarily classified with CT$>2$), after the plus sign. Additionally, a few manually added sources, identified or discussed by other studies as potential candidates to $\gamma$-ray sources or matched to radio emission, are included after the plus sign as well. The flux threshold of $5\times10^{-14}$\,erg\,s$^{-1}$\,cm$^{-2}$ is chosen to include sources that have a sufficient number of photons for spectral fitting, which can be used to check the ML classifications. These are generally the brightest X-ray sources within the PUs of GeV sources, which makes the associations likely regardless of the classification confidence. 
Source types for $\gamma$-candidates that can produce $\gamma$-rays include isolated NSs, AGNs, and X-ray binaries (XRBs). For the XRBs, we consider the LMXB and CV classes along with HMXBs, as potential counterparts to the GeV sources. 
Our LMXB class includes spider binaries, which are known GeV sources, while CVs (which are not known to produce GeV emission outside novae) are simply too confused with other $\gamma$-ray emitting classes (see the confusion matrices, hereafter CMs, in Figure \ref{fig:LOOCV}). Thus, some CVs may be misclassified as LMXBs, NSs, HMXBs, or AGNs. 
The classification results and MW properties of these 103 $\gamma$-candidate sources are shown in Table \ref{tab:X-ray-classification}.

\begin{figure*}
\begin{center}
\includegraphics[width=168pt,trim=0 0 0 0]{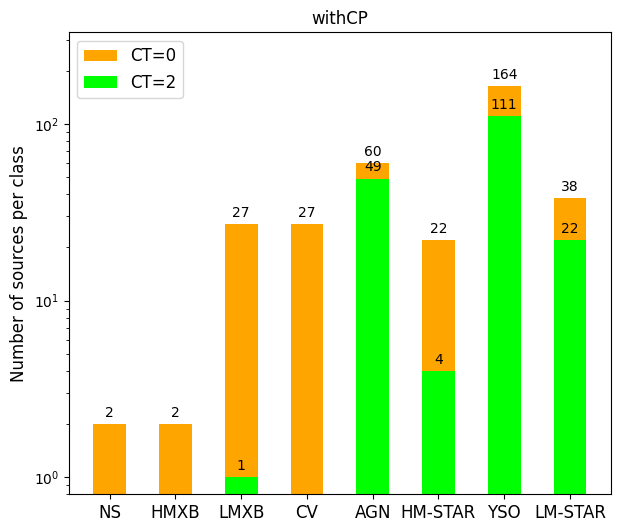}
\includegraphics[width=168pt,trim=0 0 0 0]{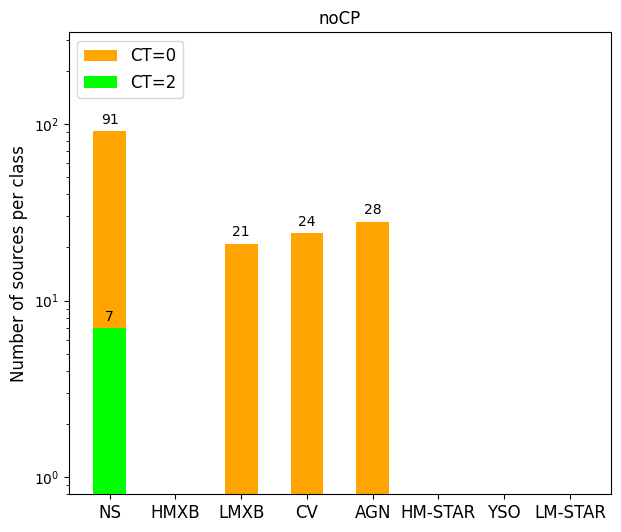}
\includegraphics[width=168pt,trim=0 0 0 0]{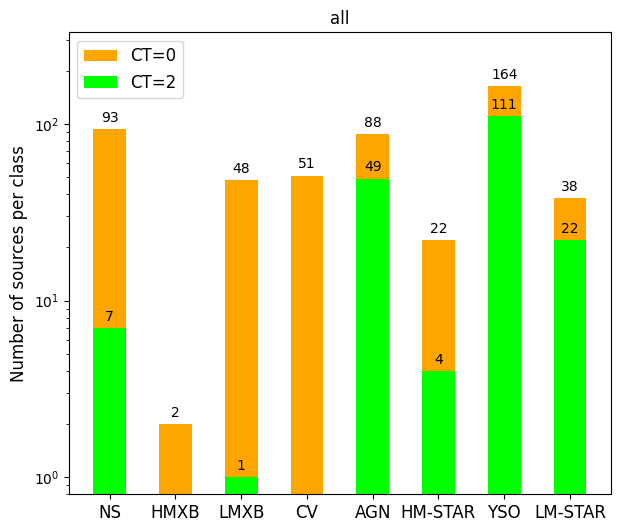}
\caption{Classification breakdown for sources detected with S/N$>=5$ with counterparts (in the left panel), without any counterparts (in the middle panel), and combined (in the right panel) within the 73 GeV sources. The different confidence threshold (CT) cuts are shown in different colors, as specified in the legend.} 
\label{fig:breakdown}
\end{center}
\end{figure*}

Figure \ref{fig:breakdown} shows the breakdown of classifications for sources detected with S/N$>=$5 at different CT levels.\footnote{See equation (7) in \cite{2022ApJ...941..104Y} for the mathematical definition of the CT values. Typically, we select a threshold of CT$=2$, where a higher CT value indicates a more confident classification.}  
Notably, seven NSs are classified with CT$>=2$, all lacking MW counterparts. Additionally, 111 YSOs, 22 LM-STARs, four HM-STARs, 49 AGNs, and one LMXB are classified with CT$>=2$, all of which have MW counterparts. 
We also checked the NS, XRB, and AGN catalogs used to create the TD to make sure that the sources we classify do not appear there.

We performed a more detailed spectral analysis for $\gamma$-candidates, requiring $\ge50$ net counts in each observation for it to be included in the simultaneous spectral fitting. 
Below, we follow the approach from \cite{2023MNRAS.526.2736Z} and perform the spectral analysis with the spectral analysis software Bayesian X-ray Analysis \citep{2014A&A...564A.125B} and Sherpa \citep{2001SPIE.4477...76F}, to generate Bayesian parameter estimation and compare models. 
The interstellar X-ray absorption is modeled using {\tt xstbabs}, with the {\tt wilm} elemental abundance \citep{2000ApJ...542..914W} and {\tt vern} photoelectric absorption cross section \citep{1996ApJ...465..487V}. 
The data are grouped to at least 5 photons per bin for fitting and to a minimum of 15 photons per bin for display purposes. The spectrum is fit with WSTAT statistics \citep{1979ApJ...230..274W}. Absorbed models, including PL (xspowerlaw), thermal plasma emission (Mekal; xsmekal), and blackbody (xsbbodyrad), are considered. 
The model comparison involves evaluating Bayesian evidence with Jeffreys' scale \citep{1939thpr.book.....J}, 
ruling out models if $\Delta \log_{10} Z = \log_{10} Z_i - \log_{10} Z_0 < -1$, where $Z_0$ represents the highest evidence among all models for a source, and $Z_i$ is the evidence of each model. 
However, for sources with $<100$ net counts, we only report the fitting results from PL model since scarce statistics makes it difficult to differentiate between the models. 
Unless explicitly stated otherwise, all uncertainties in this paper are reported at the 1$\sigma$ level. 

\startlongtable
\renewcommand{\tabcolsep}{0.1cm}
\begin{deluxetable*}{lcccccccccc}
\tablecaption{Classification Results Summary\label{tab:classification}}
\tablewidth{0pt}
\tablehead{
\colhead{4FGL Name} &    \colhead{NS} &   \colhead{HMXB} &  \colhead{LMXB} &  \colhead{AGN} &  \colhead{CV} & \colhead{YSO} & \colhead{HM-STAR} & \colhead{LM-STAR}  & \colhead{$\gamma$-candidate} & \colhead{coverage (\%)}
}
\startdata
non-C-type & 341/7 &  2/0 & 112/1 & 174/49 & 108/0 & 341/111 &    62/4 &   66/22 & 7N1L49A  &                \nodata \\
  &   &    &   &   &   &  &      &   & +14N1H6L13A12C &  \\
J0058.3--4603 &   0/0 &  0/0 &   0/0 &    1/0 &   0/0 &     0/0 &     0/0 &     0/0 &                  +1A &                  83 \\
J0335.6--0727 &   0/0 &  0/0 &   0/0 &    1/1 &   0/0 &     0/0 &     0/0 &     0/0 &                   1A &                  44 \\
J0442.8+3609 &   7/0 &  0/0 &   1/0 &    4/2 &   2/0 &     0/0 &     0/0 &     0/0 &                   2A &                  86 \\
J0615.9+1416 &   2/0 &  0/0 &   1/0 &    2/0 &   1/0 &     1/0 &     0/0 &     2/1 &                      &                  77 \\
J0639.1--8009 &   7/0 &  0/0 &   5/0 &   19/9 &   0/0 &     0/0 &     0/0 &     0/0 &              9A+1L1A &                  72 \\
J0737.4+6535 &  15/0 &  0/0 &   3/1 &    9/1 &   0/0 &     0/0 &     0/0 &     0/0 &                 1L1A &                 100 \\
J0829.3+3729 &   0/0 &  0/0 &   0/0 &    1/0 &   0/0 &     0/0 &     0/0 &     0/0 &                      &                  11 \\
J0859.2--4729$^\dag$ &  67/0 &  1/0 &  13/0 &   33/0 &  25/0 &  118/40 &     8/0 &     3/0 &              +3N2A1C &                 100 \\
J0859.3--4342$^\dag$ &  15/0 &  1/0 &   6/0 &    6/1 &   7/0 &  105/51 &     4/2 &     5/1 &            1A+1H1A2C &                  96 \\
J0900.2--4608 &   0/0 &  0/0 &   0/0 &    0/0 &   0/0 &     1/0 &     0/0 &     0/0 &                      &                   9 \\
J0933.8--6232 &   1/0 &  0/0 &   0/0 &    0/0 &   0/0 &     0/0 &     0/0 &     1/1 &                      &                 100 \\
J0938.8+5155 &   0/0 &  0/0 &   0/0 &    1/0 &   0/0 &     0/0 &     0/0 &     0/0 &                      &                  19 \\
J1018.2+0145 &   0/0 &  0/0 &   1/0 &    0/0 &   0/0 &     0/0 &     0/0 &     0/0 &                      &                  21 \\
J1025.9+1244 &   1/0 &  0/0 &   0/0 &    1/0 &   0/0 &     1/0 &     0/0 &     0/0 &                  +1N &                  98 \\
J1032.0+5725 &   4/0 &  0/0 &   0/0 &  16/10 &   0/0 &     0/0 &     0/0 &     1/1 &                  10A &                  45 \\
J1046.7--6010$^\dag$ &   4/0 &  0/0 &  12/0 &    0/0 &   6/0 &    25/2 &     1/0 &    14/4 &                      &                 100 \\
J1048.5--5923 &   0/0 &  0/0 &   0/0 &    0/0 &   1/0 &     0/0 &     0/0 &     0/0 &                      &                  53 \\
J1106.4+0859 &   0/0 &  0/0 &   0/0 &    4/1 &   0/0 &     0/0 &     0/0 &     0/0 &                   1A+1A &                   3 \\
J1110.3--6501 &   0/0 &  0/0 &   0/0 &    1/0 &   0/0 &     0/0 &     0/0 &     0/0 &                      &                  82 \\
J1115.1--6118$^\dag$ &  80/2 &  0/0 &  32/0 &    3/0 &  25/0 &    14/2 &    25/2 &     2/0 &                   2N &                 100 \\
J1116.3+1818 &   0/0 &  0/0 &   0/0 &    2/1 &   0/0 &     0/0 &     0/0 &     0/0 &                   1A &                  30 \\
J1242.6+3236 &   2/0 &  0/0 &   0/0 &    2/1 &   0/0 &     0/0 &     0/0 &     0/0 &                   1A &                 100 \\
J1256.9+2736 &   6/0 &  0/0 &   1/0 &   20/8 &   0/0 &     0/0 &     0/0 &     0/0 &                   8A+1A &                  52 \\
J1317.5--0153 &   0/0 &  0/0 &   0/0 &    1/0 &   0/0 &     0/0 &     0/0 &     0/0 &                      &                  65 \\
J1355.8--6155 &   0/0 &  0/0 &   0/0 &    0/0 &   0/0 &     0/0 &     0/0 &     1/1 &                      &                  17 \\
J1401.2--6116 &   0/0 &  0/0 &   1/0 &    0/0 &   0/0 &     0/0 &     0/0 &     0/0 &                      &                  18 \\
J1417.7--6057 &   0/0 &  0/0 &   0/0 &    0/0 &   1/0 &     2/0 &     0/0 &     0/0 &                      &                 100 \\
J1435.4+3338 &   0/0 &  0/0 &   0/0 &   16/7 &   0/0 &     0/0 &     0/0 &     1/0 &                7A+1A &                 100 \\
J1502.6+0207 &   1/0 &  0/0 &   0/0 &    5/2 &   0/0 &     0/0 &     0/0 &     0/0 &                   2A &                  29 \\
J1510.9+0551 &   1/0 &  0/0 &   0/0 &    5/4 &   0/0 &     1/0 &     0/0 &     0/0 &                   4A &                 100 \\
J1615.3--6034 &   1/0 &  0/0 &   1/0 &    0/0 &   0/0 &     0/0 &     0/0 &     3/2 &         +1A             &                  74 \\
J1616.6--5009 &   0/0 &  0/0 &   1/0 &    0/0 &   0/0 &     0/0 &     0/0 &     0/0 &                  +1L &                  44 \\
J1619.3--5047 &   2/1 &  0/0 &   4/0 &    0/0 &   3/0 &     4/0 &     0/0 &     1/0 &                   1N &                  29 \\
J1700.3--4557 &   0/0 &  0/0 &   1/0 &    0/0 &   0/0 &     0/0 &     0/0 &     0/0 &                      &                  85 \\
J1701.9--4625 &   2/0 &  0/0 &   0/0 &    0/0 &   0/0 &     0/0 &     0/0 &     0/0 &                      &                  53 \\
J1719.0--4038 &   0/0 &  0/0 &   1/0 &    0/0 &   0/0 &     0/0 &     0/0 &     0/0 &                      &                 100 \\
J1720.6--3706$^\star$ &   0/0 &  0/0 &   0/0 &    0/0 &   1/0 &     0/0 &     0/0 &     0/0 &                  +1C &                  80 \\
J1725.1--3408$^\star$$^\dag$ &   4/0 &  0/0 &   3/0 &    0/0 &   1/0 &     6/0 &     2/0 &     1/0 &                +1N1L &                  89 \\
J1732.8--3725$^\star$ &   0/0 &  0/0 &   1/0 &    0/0 &   0/0 &     0/0 &     0/0 &     0/0 &                  +1L &                  49 \\
J1734.0--2933$^\diamond$ &   1/0 &  0/0 &   0/0 &    0/0 &   0/0 &     0/0 &     0/0 &     0/0 &                  +1N &                 100 \\
J1735.4--2944$^\diamond$ &   0/0 &  0/0 &   0/0 &    0/0 &   0/0 &     0/0 &     0/0 &     1/0 &                      &                 100 \\
J1737.1--2901$^\diamond$ &   0/0 &  0/0 &   0/0 &    0/0 &   1/0 &     0/0 &     1/0 &     0/0 &                  +1C &                 100 \\
J1739.2--2717$^\diamond$ &   0/0 &  0/0 &   0/0 &    0/0 &   1/0 &     0/0 &     0/0 &     0/0 &                  +1C &                  51 \\
J1739.7--2836$^\diamond$ &   0/0 &  0/0 &   0/0 &    0/0 &   1/0 &     0/0 &     0/0 &     0/0 &                      &                 100 \\
J1740.6--2808$^\diamond$ &  10/0 &  0/0 &   1/0 &    0/0 &   4/0 &     2/0 &     2/0 &     0/0 &                +1N1C &                 100 \\
J1742.5--2833$^\diamond$ &   8/1 &  0/0 &   1/0 &    0/0 &   3/0 &     1/0 &     0/0 &     1/0 &              1N+1N1C &                 100 \\
J1743.8--3143$^\diamond$ &   0/0 &  0/0 &   0/0 &    0/0 &   0/0 &     0/0 &     0/0 &     1/1 &                      &                 100 \\
J1744.5--2612$^\star$$^\diamond$ &   0/0 &  0/0 &   1/0 &    0/0 &   0/0 &     0/0 &     0/0 &     0/0 &                      &                 100 \\
J1744.9--2905$^\star$$^\diamond$ &  24/2 &  0/0 &   0/0 &    0/0 &   5/0 &     3/1 &     1/0 &     1/0 &                   2N &                 100 \\
J1745.6--3145$^\diamond$ &   0/0 &  0/0 &   0/0 &    0/0 &   1/0 &     0/0 &     0/0 &     0/0 &                  +1C &                 100 \\
J1747.8--3006$^\diamond$ &   1/0 &  0/0 &   0/0 &    0/0 &   0/0 &     0/0 &     0/0 &     0/0 &                      &                 100 \\
J1748.3--2906$^\diamond$ &   1/0 &  0/0 &   1/0 &    0/0 &   0/0 &     0/0 &     0/0 &     0/0 &                      &                 100 \\
J1750.4--3023$^\star$$^\diamond$ &   1/0 &  0/0 &   0/0 &    0/0 &   0/0 &     0/0 &     0/0 &     1/1 &                  +1N &                 100 \\
J1751.1--3455 &   3/0 &  0/0 &   1/0 &    0/0 &   0/0 &     0/0 &     0/0 &     4/2 &                      &                  19 \\
J1751.6--3002$^\diamond$ &   0/0 &  0/0 &   0/0 &    1/0 &   0/0 &     0/0 &     0/0 &     0/0 &                  +1A &                 100 \\
J1752.3--3139$^\diamond$ &   0/0 &  0/0 &   1/0 &    0/0 &   0/0 &     0/0 &     0/0 &     0/0 &                      &                  97 \\
J1757.6--2731$^\diamond$ &   0/0 &  0/0 &   0/0 &    0/0 &   1/0 &     0/0 &     0/0 &     1/0 &                  +1C &                 100 \\
J1800.7--2355$^\star$ &   0/0 &  0/0 &   0/0 &    0/0 &   0/0 &     2/0 &     0/0 &     0/0 &                      &                  98 \\
J1800.9--2407 &   0/0 &  0/0 &   0/0 &    0/0 &   0/0 &     1/0 &     0/0 &     0/0 &                      &                  87 \\
J1804.9--3001$^\diamond$ &  23/0 &  0/0 &   4/0 &    3/0 &   9/0 &     4/0 &     9/0 &     5/2 &                +1L1C &                  84 \\
J1805.3--2734$^\diamond$ &   3/0 &  0/0 &   1/0 &    0/0 &   0/0 &     0/0 &     1/0 &     1/0 &                      &                  80 \\
J1812.8--3144 &   0/0 &  0/0 &   1/0 &    0/0 &   0/0 &     0/0 &     0/0 &     0/0 &                      &                 100 \\
J1817.7--2517 &   0/0 &  0/0 &   1/0 &    0/0 &   0/0 &     0/0 &     0/0 &     0/0 &                      &                  40 \\
J1818.5--2036 &  15/0 &  0/0 &   6/0 &    3/0 &   3/0 &    15/1 &     2/0 &     5/1 &                  +1N &                  42 \\
J1836.8--2354 &   3/0 &  0/0 &   1/0 &    1/0 &   0/0 &     0/0 &     1/0 &     2/1 &       +1L               &                  75 \\
J1843.7--3227 &   7/0 &  0/0 &   1/0 &    7/0 &   3/0 &     0/0 &     2/0 &     2/2 &                  +1N1A &                  54 \\
J1844.4--0306$^\star$ &   6/1 &  0/0 &   2/0 &    0/0 &   1/0 &     5/1 &     1/0 &     2/0 &                1N+1N &                 100 \\
J2004.3+3339$^\star$ &   1/0 &  0/0 &   0/0 &    0/0 &   1/0 &     0/0 &     1/0 &     1/0 &        +1N1C              &                 100 \\
J2038.4+4212$^\dag$ &   5/0 &  0/0 &   0/0 &    0/0 &   0/0 &    14/5 &     0/0 &     1/0 &                      &                  71 \\
J2059.1+4403$^\star$$^\dag$ &   6/0 &  0/0 &   0/0 &    1/0 &   1/0 &    16/8 &     1/0 &     2/1 &                  +1N &                  36 \\
J2144.7--5640 &   0/0 &  0/0 &   0/0 &    1/0 &   0/0 &     0/0 &     0/0 &     0/0 &                      &                   4 \\
J2236.9+1839 &   0/0 &  0/0 &   0/0 &    1/0 &   0/0 &     0/0 &     0/0 &     0/0 &                  +2A &                 100 \\
J2254.4+0108 &   1/0 &  0/0 &   1/0 &    3/1 &   0/0 &     0/0 &     0/0 &     0/0 &                   1A &                  39 \\
\hline
C-type & 146/4 &  3/0 & 62/1 & 45/7 & 34/1 & 342/98 &    20/1 &   50/11 & 4N1L7A1C &  \\
&   &    &   &   &   &  &      &   & +12N2H4L2A5C &  \\
J0552.0+0256c &   1/0 &  0/0 &  0/0 &  3/0 &  0/0 &    0/0 &     0/0 &     1/1 &                      &                  33 \\
J0613.1+1749c &   2/0 &  0/0 &  1/0 &  2/1 &  1/0 &    6/0 &     0/0 &     2/1 &                   1A &                  36 \\
J0859.8--4530c &   3/0 &  0/0 &  0/0 &  1/1 &  0/0 &    0/0 &     0/0 &     0/0 &                   1A &                   6 \\
J1038.8--5848c &   0/0 &  0/0 &  2/0 &  0/0 &  1/0 &   10/0 &     0/0 &     4/2 &                  +1L &                  17 \\
J1312.6--6231c &   9/0 &  0/0 &  0/0 &  0/0 &  2/0 &    5/1 &     0/0 &     1/0 &                  +1N &                  44 \\
J1401.9--6130c &   0/0 &  0/0 &  1/0 &  0/0 &  0/0 &    0/0 &     0/0 &     0/0 &                  +1L &                  58 \\
J1619.4--5106c &   1/0 &  0/0 &  0/0 &  0/0 &  1/0 &    1/0 &     0/0 &     0/0 &                  +1C &                  63 \\
J1620.8--5035c &  12/1 &  1/0 &  3/0 &  0/0 &  1/0 &    3/0 &     2/0 &     3/0 &                1N+1N &                  94 \\
J1634.0--4742c &   2/1 &  0/0 &  0/0 &  0/0 &  1/1 &    1/0 &     0/0 &     0/0 &                 1N1C &                 100 \\
J1636.9--4710c &   1/0 &  0/0 &  0/0 &  0/0 &  1/0 &    2/0 &     1/0 &     0/0 &                      &                 100 \\
J1638.1--4641c &   8/0 &  0/0 &  3/0 &  0/0 &  0/0 &    0/0 &     1/0 &     3/0 &                  +1N &                 100 \\
J1638.4--4715c &   0/0 &  0/0 &  1/0 &  0/0 &  0/0 &    0/0 &     0/0 &     0/0 &                      &                 100 \\
J1639.8--4642c &   0/0 &  0/0 &  0/0 &  0/0 &  0/0 &    1/0 &     0/0 &     1/0 &                      &                 100 \\
J1700.1--4013c$^\star$ &   0/0 &  0/0 &  0/0 &  1/0 &  0/0 &    0/0 &     0/0 &     0/0 &                  +1A &                  31 \\
J1741.4--3046c$^\star$ &   1/0 &  0/0 &  1/0 &  0/0 &  0/0 &    1/0 &     0/0 &     0/0 &                  +1N &                  67 \\
J1750.9--2655c &   1/0 &  0/0 &  1/0 &  0/0 &  0/0 &    0/0 &     0/0 &     0/0 &                +1N1L &                  44 \\
J1800.2--2403c &   8/1 &  0/0 &  1/0 &  0/0 &  1/0 &    4/0 &     3/0 &     1/0 &                   1N &                 100 \\
J1806.9--2455c$^\star$ &   1/0 &  0/0 &  1/0 &  0/0 &  0/0 &    0/0 &     0/0 &     0/0 &                  +1N &                  66 \\
J1809.8--2408c &   9/0 &  0/0 &  3/0 &  2/0 &  1/0 &    3/0 &     0/0 &     2/0 &                  +1N &                  80 \\
J1816.2--1654c$^\star$ &   3/0 &  0/0 &  1/0 &  0/0 &  1/0 &    3/0 &     0/0 &     1/0 &                  +1N &                  36 \\
J1818.0--1339c &  10/1 &  0/0 & 24/1 &  0/0 & 10/0 &  79/20 &     6/0 &     8/1 &                 1N1L &                  25 \\
J1826.5--1202c &   1/0 &  1/0 &  0/0 &  0/0 &  0/0 &    1/0 &     0/0 &     0/0 &                +1N1H &                  66 \\
J1830.1--0212c &   1/0 &  0/0 &  0/0 &  1/0 &  0/0 &    7/4 &     0/0 &     0/0 &                      &                  39 \\
J1831.3--0203c$^\star$ &   4/0 &  0/0 &  5/0 &  2/0 &  2/0 &  45/20 &     3/1 &     0/0 &                +1L2C &                 100 \\
J1834.2--0827c &   1/0 &  0/0 &  0/0 &  0/0 &  0/0 &    0/0 &     0/0 &     0/0 &                      &                 100 \\
J1842.5--0359c$^\star$ &   2/0 &  0/0 &  3/0 &  0/0 &  1/0 &    1/0 &     0/0 &     1/0 &                      &                  75 \\
J1845.8--0236c &   0/0 &  0/0 &  0/0 &  0/0 &  0/0 &    1/0 &     0/0 &     0/0 &                      &                 100 \\
J1846.9--0247c &   0/0 &  0/0 &  0/0 &  0/0 &  0/0 &    1/0 &     0/0 &     0/0 &                      &                  76 \\
J1848.6--0202c$^\star$ &   1/0 &  1/0 &  1/0 &  0/0 &  0/0 &    2/0 &     1/0 &     0/0 &                  +1H &                  90 \\
J1910.2+0904c &   4/0 &  0/0 &  0/0 &  0/0 &  0/0 &    0/0 &     0/0 &     0/0 &                      &                 100 \\
J2022.6+3716c &   2/0 &  0/0 &  1/0 &  1/0 &  0/0 &    2/0 &     0/0 &     2/1 &                      &                  37 \\
J2026.5+3718c$^\star$ &  24/0 &  0/0 &  5/0 &  9/0 &  5/0 &   42/3 &     2/0 &    10/1 &                +1N1C &                  27 \\
J2056.4+4351c$^\star$ &  12/0 &  0/0 &  1/0 & 11/4 &  0/0 &  39/12 &     0/0 &     5/3 &                4A+2N &                  51 \\
J2143.4+6608c &   1/0 &  0/0 &  1/0 &  1/1 &  0/0 &   10/6 &     0/0 &     2/0 &                   1A &                 100 \\
J2152.0+4718c &   0/0 &  0/0 &  0/0 &  1/0 &  0/0 &    1/0 &     0/0 &     0/0 &                      &                  32 \\
J2256.5+6212c &  21/0 &  0/0 &  2/0 & 10/0 &  5/0 &  71/32 &     1/0 &     3/1 &                +1A1C &                  56 \\
\hline
All combined & 487/11 &  5/0 & 174/2 & 219/56 & 142/1 & 683/209 &    82/5 &  116/33 & 11N2L56A1C &                \nodata \\
  &   &    &   &   &   &  &      &   & +26N3H10L15A17C &         \\
\enddata
\tablecomments{The first number in each of the second--ninth columns shows the total number of sources (S/N$>3$) per 4FGL field that are classified as the class in the column name. The second number shows the number of sources with S/N$>5$ that are classified with CT$>=2$. 
The 10th column shows a summary of X-ray sources that may be potential contributors to $\gamma$-ray sources ($\gamma$-candidates) for each field, classified as NS (N), HMXB (H), LMXB (L), CV (C), or AGN (A) types. The $\gamma$-candidates include the significant (S/N$>5$) sources that are classified with CT$>=2$ before the plus sign or in cases of no plus sign, and the sources with $F_{\rm b}>5\times10^{-14}$\,erg\,s$^{-1}$\,cm$^{-2}$ (but not necessarily classified with CT$>$2) or a few sources that are manually added, after the plus sign. The last column shows the X-ray coverage fraction of each 4FGL source. The sources are ordered by their names. The first part of the table presents the non-C-type sources whereas the second part presents the C-type sources. Sources with $^\star$, $^\dag$, or $^\diamond$ in their names are the 4FGL sources associated with unknown type sources, non-C-type 4FGL sources in SFRs discussed in Section \ref{sec:SFR-Cand}, or non-C-type 4FGL sources within 5$^\circ$ from the Galactic Center.}
\end{deluxetable*}

\subsection{NS Candidates}
\label{sec:NS-Cand}

We classify 21 NSs from the non-C-type 4FGL fields.  
To further investigate the NS classifications, we cross-match them with deeper surveys, including the DECam Plane Survey 2 \citep[DECaPS2;][]{2023ApJS..264...28S}, the VPHAS+ survey \citep{2014MNRAS.440.2036D}, UKIRT Infrared Deep Sky Survey-DR6 Galactic Plane Survey \citep{2008MNRAS.391..136L}, and the Pan-STARRS1 Surveys \citep{2016arXiv161205560C}. Four of these NS candidates are matched to optical and infrared (IR) sources, with three of these NS counterparts being within the PUs of 4FGL J1750.4--3023, and 4FGL J1744.9--2905 located close to the Galactic Center. We do not discuss these candidates because the MW matching is unreliable due to the extreme crowding in this region. 
The remaining NS candidate with a MW match is 2CXO J111513.6--611657 within the PU of 4FGL J1115.1--6118, which we discuss in Section \ref{sec:J1115.1-6118}.

\subsubsection{4FGL J1619.3--5047} 
\label{sec:J1619.3-5047}

Only 29\% of the PU of 4FGL J1619.3--5047 is covered by Chandra X-ray Observatory observations from the CSCv2.0. 2CXO J161945.8--504222 is the only significantly detected X-ray source (S/N$=6.3$), classified as an NS with a CT=2.8. 
This source exhibits a relatively faint X-ray flux with a value of $F_{\rm b}=2.65\times10^{-14}$\,erg\,s$^{-1}$\,cm$^{-2}$, and is located at an off-axis angle (OAA) of 9.2\arcmin. It has 67 net counts, and the spectral fitting with an absorbed PL model yields a photon index $\Gamma=2.6^{+2.2}_{-1.4}$ which is typical for a pulsar, albeit very uncertain. 
The source was observed with the VPHAS+ DR3 survey but undetected with a 5$\sigma$ sensitivity of $g=22.79$ of AB magnitude, which provides a lower limit on the X-ray to optical flux ratio $f_{\rm X}/f_{\rm O}$\footnote{The $g$-band optical energy flux (or its upper limit) for VPHAS+ and PanSTARRS surveys is calculated from the $g$-band AB magnitude (or the limiting magnitude) using $F = 10^{(g+48.6)/-2.5} \nu_{\rm ref}$ in units of \,erg\,s$^{-1}$\,cm$^{-2}$, with $\nu_{\rm ref}=6.38\times10^{14}$\,Hz for VPHAS+, and $\nu_{\rm ref}=6.19\times10^{14}$\,Hz for PanSTARRS.} of 1.5.  
The complex environment surrounding 4FGL J1619.3–-5047 includes two nearby unIDed 4FGL sources, several supernova remnants (SNRs), and one PSR (PSR J1617--5055 with an age of 8 kyr and a spin down energy loss rate $\dot{E}=1.6\times10^{37}$\,erg\,s$^{-1}$; see \citealt{2021ApJ...923..249H}).

Although the NS candidate 2CXO J161945.8--504222 might be a potential X-ray counterpart to 4FGL J1619.3--5047, the limited coverage in the CSCv2.0 does not allow for a more conclusive assessment.

\subsubsection{4FGL J1844.4--0306}
\label{sec:J1844.4-0306}

\cite{2023ApJ...952..158Z} proposed an association between 4FGL J1844.4--0306, coincident with the TeV source HESS J1844--030, and the SNR candidate G29.37+0.1. They suggested that G29.37+0.1 is a composite SNR, hosting a PWN powered by a PSR. \cite{2017A&A...602A..31C} previously identified 2CXO J184443.3--030518 as the pulsar candidate powering the PWN. Our classification results confidently support this association by classifying 2CXO J184443.3--030518 as an NS with a CT=2.8. 

The brightest ($F_{\rm b}=10^{-13}$\,erg\,s$^{-1}$\,cm$^{-2}$) X-ray source  within the field, 2CXO J184441.8--030551,  is classified as an NS with a lower CT=1.2. Its X-ray spectrum is best fit by a PL model with $\Gamma=0.9\pm0.1$. It is worth noting that such a small photon index is atypical for an NS (although not unseen). 
Based on their nondetection in the VPHAS+ survey, which has a 5$\sigma$ sensitivity of $g=22.97$ in AB magnitude, lower limits on $f_{\rm X}/f_{\rm O}$ are 2.2 and 6.6 for 2CXO J184443.3--030518 and 2CXO J184441.8--030551, respectively. 
Both 2CXO J184443.3--030518 and 2CXO J184441.8--030551 overlap with a bright extended radio source (see the top left panel in Figure \ref{fig:fields}), which might be associated with the GeV and TeV emission in a scenario with a PWN powered by a PSR.   
Given our classification results, we conclude that 2CXO J184443.3--030518 is a potential NS X-ray counterpart to 4FGL J1844.4--0306. We also note that a young pulsar PSR J1844--0310 with an age of 813 kyr and an $\dot{E}=2.8\times10^{34}$\,erg\,s$^{-1}$ \citep{2005AJ....129.1993M} is located a few arcminutes outside the GeV PU, which is not detected in X-rays.

\subsection{AGN Candidates}
\label{sec:AGN-Cand}

A total of 62 AGN candidates are classified from 20 4FGL fields. The majority (57) of these AGN candidates are located off the Galactic plane, with $|b|>5^{\circ}$. Only five of them are located in the plane, specifically within the PUs of 4FGL J0859.2--4729, 4FGL J0859.3--4342, and 4FGL J1751.6--3002. 
We found that most of the AGN candidates have MIR counterparts from Wide-field Infrared Survey Explorer (WISE) catalogs. 
The detection of a coincident radio source provides strong support to the AGN classification \citep[e.g., ][]{2020ApJ...892..105A}, and we checked the 77 firmly identified AGNs (including blazars of different types and radio galaxies) in the 4FGL and found that they do match to radio sources from the Combined Radio Multi-Survey Catalog \citep[MSC;][]{2023ApJ...943...51B} and the Rapid ASKAP Continuum Surveys \citep[RACS;][]{2020PASA...37...48M}, with the exception of one extended 4FGL source, 4FGL J1324.0--4330e, which is a complicated case of the Centaurus A lobes \citep{2016A&A...595A..29S}. Hence, we also searched for radio detection for these AGN candidates and found that nine of them matched to radio sources. 
The relatively low fraction of matches may be due to the candidate AGN having weaker or no jets, thus being fainter at radio (and other) wavelengths, compared to AGNs firmly identified in the 4FGL catalog, allowing the former to evade the detection in the radio surveys.

\begin{figure}
\begin{center}
\includegraphics[width=0.45\textwidth,trim=0 0 0 0]{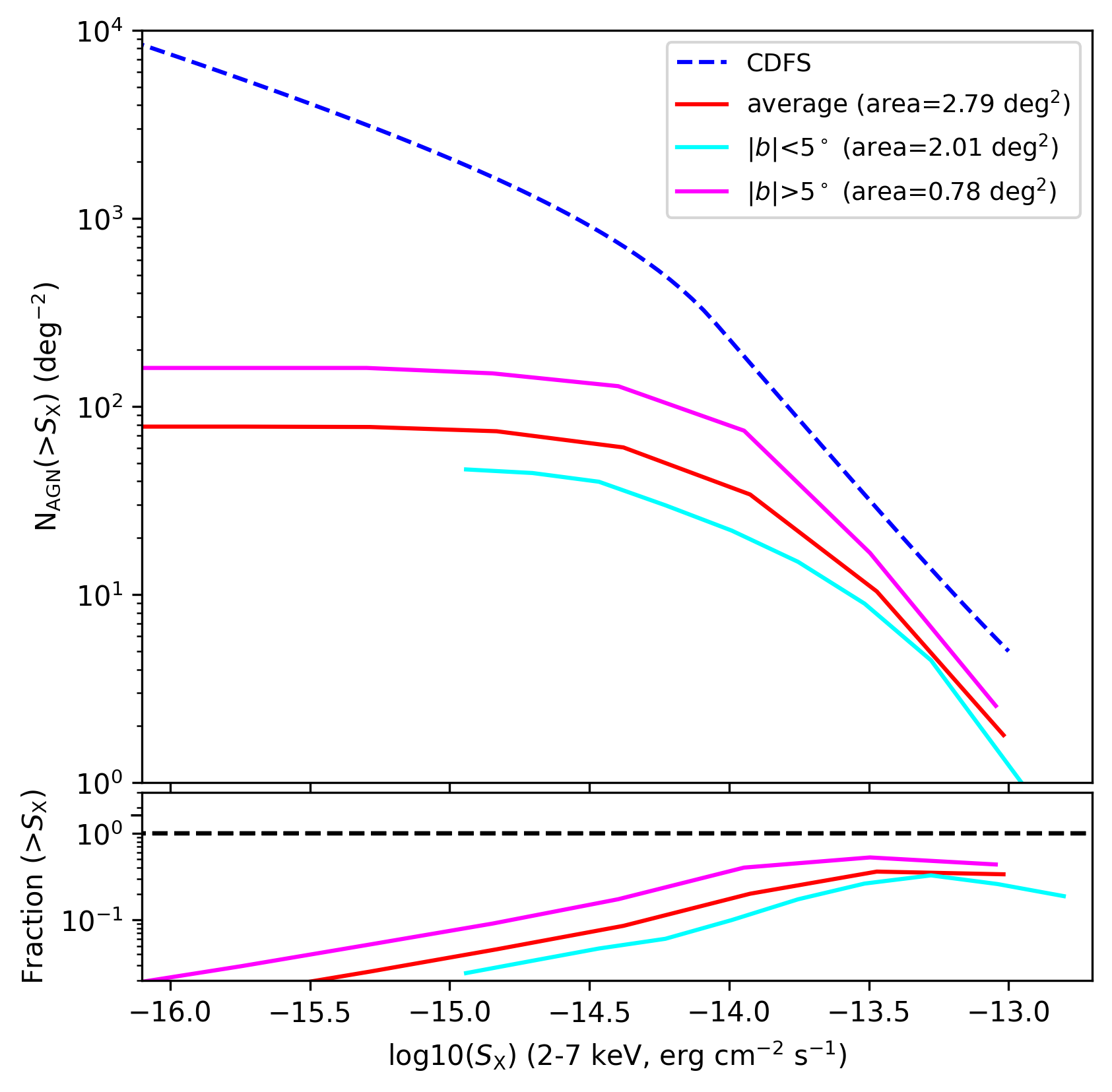}
\caption{Cumulative number count density (number density of sources brighter than a given flux) in the hard band (2--7\,keV) for the classified AGNs at all Galactic latitudes (solid red line), $|b|<5^{\circ}$ (solid cyan line), $|b|>5^{\circ}$ (solid magenta line), compared to the CDFS AGN catalog from \cite{2017ApJS..228....2L} (dashed blue line). The bottom panel shows the fraction of each cumulative number count density to that of the CDFS.} 
\label{fig:AGN_loglogS}
\end{center}
\end{figure}

We compared the number of classified AGNs brighter than a given flux in the hard band  (2--7\,kev) without the significance and CT cuts to that from the AGN catalog in the same band from the Chandra Deep Field South \citep[CDFS;][]{2017ApJS..228....2L} in Figure \ref{fig:AGN_loglogS}. 
We found that the pipeline's chance of classifying an AGN becomes lower at low Galactic latitudes compared to higher latitudes. This difference is likely attributed to higher extinction and a more complex environment within the Galactic plane. 
Furthermore, we noticed that the slope of the cumulative counts of classified AGNs as a function of their fluxes above  $2\times10^{-14}$\,erg\,s$^{-1}$\,cm$^{-2}$ is consistent with that from the CDFS, as shown in Figure \ref{fig:AGN_loglogS}, albeit being a factor of $\sim2-5$ lower in absolute value. 
The decrease can be explained by a much deeper Chandra X-ray Observatory exposure for the CDFS area, and by possible differences in the detection and analysis  methods used for the CSCv2.0 and CDFS.

\subsubsection{4FGL J0058.3--4603}
\label{sec:J0058.3-4603}

Within the PU of 4FGL J0058.3--4603, only one X-ray source, 2CXO J005806.2--460419, is significantly detected (S/N$=7.5$).  
It is classified as an AGN (CT=0.5) with the optical--MIR counterpart having Gmag=20.5 and W1mag=15.3. 
Its 0.5--7\,keV X-ray flux is $\approx2.3\times10^{-13}$\,erg\,s$^{-1}$\,cm$^{-2}$. 
The X-ray spectrum (with 77 net counts) can be fit by a PL model with $\Gamma=2.9\pm0.6$. 
Additionally, this source is associated with a radio source from the RACS-mid survey \citep{2023PASA...40...34D}. 

Based on the rather uncertain AGN classification and its association with a radio source, 2CXO J005806.2--460419 might be a counterpart to 4FGL J0058.3--4603.

\subsubsection{4FGL J0335.6--0727}
\label{sec:J0335.6-0727}

Within the PU of 4FGL J0335.6--0727, only a singly X-ray source, 2CXO J033532.7--072741, is detected (S/N$=8.2$).
The source is classified as an AGN with a CT=2.5. It has an optical--MIR counterpart with Gmag=20.3, Jmag=17.1, and W1mag=14.7. 
Its X-ray flux is $\approx1.1\times10^{-13}$\,erg\,s$^{-1}$\,cm$^{-2}$. 
The X-ray spectrum is best fit by a PL model with $\Gamma=2.2\pm0.2$. Additionally, it has a radio counterpart 
with a flux of $3.1\pm0.4$\,mJy at 2.049\,GHz from the MSC. The source has a spectroscopic redshift of 0.36147 from the ninth data release of the Sloan Digital Sky Survey \citep[SDSS;][]{2012ApJS..203...21A}, supporting its AGN classification. 

Therefore, 2CXO J033532.7--072741 can be a counterpart to 4FGL J0335.6--0727.  

\subsubsection{4FGL J0442.8+3609}
\label{sec:J0442.8+3609}

Within the PU of 4FGL J0442.8+3609, two significantly detected (S/N$>5$) X-ray sources are classified with CT$>2$. 
Both 2CXO J044308.1+361310 and 2CXO J044309.2+360856 are matched to MIR counterparts with W1mag\,=\,17.01 and W1mag\,=\,16.99, and classified as AGNs with CT=38 and CT=34, respectively. The X-ray spectrum of 2CXO J044308.1+361310 can be fit by a PL model with $\Gamma=1.9\pm0.6$. 
It is noteworthy that there are several radio sources detected in the field, but none of them are matched to any X-ray source. 

Given our classifications, we suggest that either 2CXO J044308.1+361310 or 2CXO J044309.2+360856 can be a potential X-ray counterpart to 4FGL J0442.8+3609 as an AGN. 

\subsubsection{4FGL J0639.1--8009}
\label{sec:J0639.1-8009}

The brightest ($F_{\rm b}=2.8\times10^{-13}$\,erg\,s$^{-1}$\,cm$^{-2}$) X-ray source within the PU of 4FGL J0639.1--8009 is 2CXO J063805.1--801854, classified as an AGN with a very low CT=0.1 and matched to an optical--MIR counterpart. 
Neither of the three X-ray spectral models provides a good fit to its spectrum, where we see residuals around 5--7 keV as well as a soft excess at energies lower than 2 keV (view its spectrum fitting with a PL model at \url{https://muwclass.github.io/MUWCLASS_4FGL-DR4/images/2CXOJ063805.1-801854_pl_spectrum.png}). 
The source exhibits X-ray variability with the intraobservation variability probability $P_{\rm intra}=1.0$, and its X-ray spectrum is notably hard, with the majority of photons in the 2.0--7\,keV. Furthermore, it has a Gaia counterpart at a distance\footnote{In cases where an X-ray source is matched to a Gaia counterpart with the distance listed in the Gaia distances catalog, we only consider those with parallax to parallax error (RPlx) $\ge$3 as reliable  measurements.} of $d=850^{+203}_{-138}$\,pc, corresponding to $L_{\rm X}=2.4\times10^{31}$\,erg\,s$^{-1}$ in the 0.5--7\,keV, making the AGN classification less convincing. 
Moreover, a period of $13,151\pm330$\,s has been detected for this source from the Chandra X-ray Observatory data, suggesting its classification as a CV \citep{2016MNRAS.462.4371I}. 
Additionally, there are nine AGN candidates classified with CT$>2$ within the $\gamma$-ray PU  where all of these AGN candidates have an MIR counterpart. 

Given 
that a CV is unlikely to be a GeV emitter, we suggest that 4FGL J0639.1--8009 is a GeV AGN that might be associated with one or several of the X-ray AGNs classified in this field.

\subsubsection{4FGL J1025.9+1244}
\label{sec:J1025.9+1244}

The brightest X-ray sources within the PU of 4FGL J1025.9+1244 are 2CXO J102557.9+124109 and 2CXO J102557.9+124108X, with a separation of 1.3$\arcsec$, and appear to be extended out to $2\arcmin-3\arcmin$. The X-ray flux of 2CXO J102557.9+124108X is $F_{\rm b}=5.2\times10^{-12}$\,erg\,s$^{-1}$\,cm$^{-2}$, derived from the source region defined in the CSCv2.0, while the point-spread function (PSF) 90\% aperture region fails to measure the flux. 
2CXO J102557.9+124109 is classified as an NS with a low CT=0.7. 
Its X-ray spectrum of 2CXO J102557.9+124109 is best fit by a PL model with $\Gamma=2.6^{+0.5}_{-0.3}$. 
2CXO J102557.9+124109 also coincides with the galaxy 2MASX J10255796+1241086 having a spectroscopic red-shift of 0.14199 \citep{2012ApJS..203...21A} within a galaxy cluster \citep{2008ApJS..176...39Q,2016MNRAS.460.3669Y} and a bright radio source in NVSS and RACS surveys with an extent of $1-2\arcmin$ in diameter (see the top right panel in Figure \ref{fig:fields}). Another galaxy 2MASX J10255304+1242396 is observed coinciding with a radio source, close to 2CXO J102557.9+124109 with a separation of $\sim2\arcmin$ within the $\gamma$-ray PU. 
The extended X-ray emission makes the PU of 2CXO J102557.9+124109 underestimated; thus, no MW counterpart is matched leading to an NS classification by MUWCLASS. The X-ray extended emission is of similar sizes from the radio, optical, and X-ray images, suggesting the emission is all produced by the radio galaxy.

\subsubsection{4FGL J1032.0+5725}
\label{sec:J1032.0+5725}

In the region of 4FGL J1032.0+5725, located at the edge of the Lockman Hole Northwest, 10 significant X-ray sources are confidently classified as AGNs. 
Among them, eight are genuine AGNs with measured redshifts according to \cite{2010AA...518A..10V}. They were excluded from our TD as we filtered out X-ray sources with PUs$>1\arcsec$ when we cross-matched AGNs to literature-verified catalogs, to maintain the reliability of the populous classes in the TD. 
The remaining two sources, 
2CXO J103220.2+573211 and 2CXO J103121.9+573134, are matched to MIR counterparts with W1mag=17.22 and W1mag=17.37, and classified as AGNs with CT=24 and CT=9. 
Spectra of all eight AGNs, having $>50$ counts, can be fit with an absorbed  PL model. The photon indices for these sources range from $\Gamma=1.1^{+0.3}_{-0.2}$ for 2CXO J103145.7+573401  to $2.5^{+0.7}_{-0.6}$ for 2CXO J103225.1+572814. However, 2CXO J103225.1+572814 has a Gaia counterpart at a distance of $490^{+167}_{-103}$\,pc with RPlx=4.1, which corresponds to an X-ray luminosity of $8\times10^{29}$\,erg\,s$^{-1}$ in the 0.5--7 keV band. 
The spectroscopic redshift measurement from \cite{2010AA...518A..10V} indicates that the reliability of the Gaia distance measurement for 2CXO J103225.1+572814 may be questionable. 
A few radio sources are detected in the field from the MSC catalog, but many of them are located in the area not covered by Chandra X-ray Observatory observations, and none of them are matched to any X-ray source. 

Based on our X-ray classifications, we suggest that 4FGL J1032.0+5725 is a GeV AGN that might be associated with one or several of the X-ray AGNs classified in this field.

\subsubsection{4FGL J1106.4+0859}
\label{sec:J1106.4+0859}

For 4FGL J1106.4+0859, the X-ray coverage of the $\gamma$-ray PU is limited to only 3\%. Within the covered part, a single X-ray source is detected above S/N$=5$, which is 2CXO J110823.6+091239 with S/N=6.9. It is classified as an AGN with a CT=57, matching to an optical--MIR counterpart with Gmag=19.09 and W1mag=16.35. Its X-ray flux is $3.0\times10^{-14}$\,erg\,s$^{-1}$\,cm$^{-2}$. 
The X-ray spectrum is satisfactorily fit by a PL model with $\Gamma=2.0\pm0.3$. 
The optical counterpart has a spectroscopic redshift of 2.676 \citep{2018AA...613A..51P}, supporting its AGN classification. 
Another source, 2CXO J110807.9+091722, is detected at $\approx15'$ off-axis, which was excluded from our automated classification for that reason. 
It has a fairly high flux of $1.7\times10^{-13}$\,erg\,s$^{-1}$\,cm$^{-2}$ in 0.5--7\,keV\footnote{For this source (2CXO J110807.9+091722),  2CXO J205855.7+44033, and 2CXO J223704.7+184056 located at large OAAs, we used the fluxes measured within the source region (as defined in CSC) unlike the PSF 90\% aperture for the point sources.}. 
It displays variability between the observations with $P_{\rm inter}=1.0$. Its spectrum is well fitted by a PL model with $\Gamma=2.5\pm0.3$. Within its PU of 2.57$\arcsec$, it is matched with an optical--MIR source with Gmag=19.44 and W1mag=14.73, with a separation of 1.5$\arcsec$ and $P_{\rm c}=0.7\%$, due to the fairly low density of optical--MIR sources at the high Galactic latitude of the source ($b=59.8^{\circ}$). The optical--MIR source is a quasar with a spectroscopic redshift of 0.638 \citep{2018AA...613A..51P}. 
We also note that there are tens of radio sources detected from the MSC within the GeV PU, mostly located in the area not covered by Chandra X-ray Observatory observations.

Due to the very limited coverage, it is premature to say that 2CXO J110823.6+091239 or 2CXO J110807.9+091722 is the counterpart of 4FGL J1106.4+0859.

\subsubsection{4FGL J1116.3+1818}
\label{sec:J1116.3+1818}

The X-ray coverage of the PU of 4FGL J1116.3+1818 is $\approx$30\%. There is only one X-ray source, 2CXO J111633.3+181420, detected significantly with S/N$=8$ within the covered part. 
It is matched to an MIR counterpart with W1mag=16.82, and classified as an AGN with a CT=24. The X-ray flux for this source is 3.5$\times10^{-14}$\,erg\,s$^{-1}$\,cm$^{-2}$. 
Its X-ray spectrum (with 102 net counts) can be fit by an absorbed PL model with $\Gamma=1.6^{+0.4}_{-0.3}$. 
It has a spectroscopic redshift of 0.6954 \citep{2014AA...563A..54P}, confirming its AGN classification. 
There are a few radio sources detected from the MSC that are located in the area not covered by Chandra X-ray Observatory observations. 

Given the classification results, we suggest that 2CXO J111633.3+181420, classified as an AGN, may be a potential X-ray counterpart to 4FGL J1116.3+1818. However, the limited Chandra X-ray Observatory coverage makes this association less certain.

\subsubsection{4FGL J1242.6+3236}
\label{sec:J1242.6+3236}

\cite{2023RAA....23b5007M} reported that the spatial offset and the hard $\gamma$-ray spectrum make 4FGL J1242.6+3236 an unlikely counterpart of a nearby star-forming galaxy NGC 4631. Instead, the authors suggested that an unIDed background blazar may be responsible for the GeV emission. 

We classify 2CXO J124235.1+323340, the only significant X-ray source  (S/N$=6.1$) within the $\gamma$-ray PU, as an AGN with a CT=15. It has an MIR counterpart with W1mag\,=\,16.91 and its spectrum can be fit by a PL model with $\Gamma=1.8^{+0.7}_{-0.4}$. However, \cite{2012MNRAS.419.2095M} identified 2CXO J124235.1+323340 as an HMXB associated with NGC 4631, while also flagging the source as a potential AGN due to its significant offset from the galactic center.

Therefore, 2CXO J124235.1+323340 is a plausible X-ray counterpart to 4FGL J1242.6+3236.

\subsubsection{4FGL J1256.9+2736}
\label{sec:J1256.9+2736}

We classified eight X-ray sources with S/N\,$>5$  within the PU of 4FGL J1256.9+2736 as AGNs with CT\,$>2$. Among them, 2CXO J125751.0+273231, 2CXO J125745.0+273210, and 2CXO J125740.2+273118 are identified as AGNs with measured redshifts by \cite{2010AA...518A..10V,2018AJ....155..189D}, which were removed from our TD due to their relatively large PUs ($>1\arcsec$). 
All eight AGN candidates have MIR counterparts while some of them have optical counterparts. 
The X-ray spectra of six classified AGNs with $>50$ net counts are well fitted with an absorbed PL model. The photon indices for these sources range from $\Gamma=1.5^{+0.2}_{-0.1}$ for 2CXO J125734.0+272730 to $2.1^{+0.4}_{-0.3}$ for 2CXO J125751.0+273231.  
Additionally, some of these sources exhibit variability, including 2CXO J125740.2+273118 ($P_{\rm intra}=0.96$), 2CXO J125730.0+272612 ($P_{\rm inter}=1.0$), and 2CXO J125734.0+272730 ($P_{\rm inter}=1.0$). 
2CXO J125724.4+272952, which is excluded from our automated classification due to its extended X-ray emission, coincides with the  radio galaxy NGC 4839 \citep{2012MNRAS.421.1569B} and is associated with the radio source RACS-DR1 J125724.2+272954 from the RACS survey. 

Based on our classifications, we suggest that 4FGL J1256.9+2736 is a GeV-emitting AGN associated with one (or more) X-ray AGNs that we classify in this field, particularly the radio galaxy 2CXO J125724.4+272952.

\subsubsection{4FGL J1435.4+3338}
\label{sec:J1435.4+3338}

Within the PU of 4FGL J1435.4+3338, there are eight AGN candidates classified.  
All of them have MIR counterparts while some of them also have optical counterparts. 
Among these, three AGN candidates with sufficient ($>50$ per observation) counts can be fit by a PL model, with $\Gamma$ ranging from $2.1\pm0.2$ for 2CXO J143542.6+333404 to $2.6^{+0.5}_{-0.4}$ for 2CXO J143513.2+333118. 2CXO J143513.2+333118 exhibits long-term variability across multiple X-ray observations with $P_{\rm inter}=1.0$. 
Additionally, it coincides with a radio source in the RACS-mid and Faint Images of the Radio Sky at Twenty cm surveys with an integrated flux density of 2.15\,mJy at 1.4 GHz \citep{2015ApJ...801...26H}. It is identified as an AGN with a spectroscopic redshift of 0.851 \citep{2012ApJS..200....8K}. 

We suggest that 2CXO J143513.2+333118 is the most likely X-ray counterpart to 4FGL J1435.4+3338, based on our classification as an AGN and its radio detection.

\subsubsection{4FGL J1502.6+0207}
\label{sec:J1502.6+0207}

Within the PU of 4FGL J1502.6+0207, the only two sources with S/N$>5$ are classified as AGNs with CT$>2$. 
The brightest X-ray source, 2CXO J150237.4+015813, exhibits an X-ray spectrum that can be fit with a PL model with $\Gamma=1.6^{+0.3}_{-0.2}$. 
The other source, 2CXO J150234.5+015205 shows variability with $P_{\rm intra}=0.97$, and its X-ray spectrum is well described by a PL model with $\Gamma=1.4^{+0.3}_{-0.2}$. 
Both of them have MIR counterparts from WISE while 2CXO J150234.5+015205 is also matched to a Gaia counterpart. 
Several radio sources are detected from the MSC within the GeV PU where they are not covered by the limited (30\% coverage) Chandra X-ray Observatory observations. 

Based on our X-ray classifications, we suggest that either 2CXO J150237.4+015813 or 2CXO J150234.5+015205 may be the potential X-ray AGN counterpart to 4FGL J1502.6+0207.

\subsubsection{4FGL J1510.9+0551}
\label{sec:J1510.9+0551}

Within the PU of 4FGL J1510.9+0551, the only four sources with S/N$>5$ are all classified as AGNs with CT$>2$. 
Spectra of all four AGN candidates can be well fitted with a PL model. Their photon indices vary from $\Gamma=1.6^{+0.3}_{-0.2}$ for 2CXO J151100.5+054912 to $2.7^{+0.2}_{-0.1}$ for 2CXO J151100.4+054921. 
Additionally, 2CXO J151100.4+054921 is matched to a radio source and classified as a $\gamma$-ray emitting blazar candidate based on WISE colors and radio images by \cite{2019ApJS..242....4D}. 

Therefore, 2CXO J151100.4+054921 is the most likely potential X-ray counterpart to 4FGL J1510.9+0551, based on our classification as an AGN and its radio detection.

\subsubsection{4FGL J1615.3--6034}
\label{sec:J1615.3-6034}

Two X-ray sources, 2CXO J161519.3--603852 and 2CXO J161524.6--603401, are classified as LM-STARs with CT$>2$, both with an optical--MIR counterpart. 2CXO J161519.3--603852 exhibits intraobservation variability with $P_{\rm intra}=1.0$. Its X-ray spectrum can be fit by an absorbed Mekal model with a  plasma temperature kT$=2.4^{+0.5}_{-0.4}$\,keV. 
There is another X-ray source, 2CXO J161532.9--603954, just outside the PU of 4FGL J1615.3--6034, which is located at one end of the elongated jet-like radio structure in the RACS survey extending up to 30$\arcmin$ (see the lower left panel in Figure \ref{fig:fields}). 
The jet-like structure belongs to the galaxy ESO 137-007, which is part of Norma galaxy cluster at $d=$68\,Mpc \citep{2022ApJ...939L...4D}. 

Therefore, 2CXO J161532.9--603954 is the most likely X-ray counterpart to 4FGL J1615.3--6034 as an AGN with a one-sided jet pointing toward us.

\subsubsection{4FGL J2236.9+1839}
\label{sec:J2236.9+1839}

The only X-ray source within the PU of 4FGL J2236.9+1839 is 2CXO J223708.5+183939 with S/N$=6.4$. It is classified as an AGN with a modest CT=1.1. It has an MIR counterpart, and its X-ray spectrum can be fit by a PL model with $\Gamma=1.31^{+0.60}_{-0.42}$. 
There is another X-ray source 2CXO J223704.7+184056 with a 0.5--7\,keV flux of $4.7\times10^{-14}$\,erg\,s$^{-1}$\,cm$^{-2}$, which was not classified due to its placement at a large OAA (11.1$\arcmin$). The source exhibits intraobservation variability with $P_{\rm intra}=0.97$ and is matched to an optical--MIR counterpart. It is also matched to a radio source detected from the RACS-mid survey with a total flux density of $5\pm0.5$\,mJy at 750--1500\,MHz. 
\cite{2021ApJS..257...30M} identified the source as a BLL type blazar with a spectroscopic redshift of 0.7222, and suggested its association with 4FGL J2236.9+1839. 

So we suggest that 2CXO J223704.7+184056 is the most likely X-ray AGN counterpart to 4FGL J2236.9+1839.

\subsubsection{4FGL J2254.4+0108}
\label{sec:J2254.4+0108}

Three X-ray sources are detected with S/N$>5$ within the PU of 4FGL J2254.4+0108. 
2CXO J225435.8+010158 is the only one classified with CT$>2$. It is classified as an AGN with a CT=6.1, matching to an MIR counterpart. 
There are a few radio sources detected from the MSC within the GeV PU where they are not covered by the limited (39\% coverage) Chandra X-ray Observatory observations. 

Based on our classification, we suggest that 2CXO J225435.8+010158, classified as an AGN, is a possible X-ray counterpart to 4FGL J2254.4+0108.

\subsection{UnIDed sources in Star-Forming Regions}
\label{sec:SFR-Cand}

SFRs have been proposed to be associated with $\gamma$-ray sources \citep{2008AIPC.1085...97R}. For instance, \cite{2018A&A...611A..77Y} suggests that the stellar winds of luminous stars in Westerlund 2 are likely sites of acceleration of particles responsible for the diffuse $\gamma$-ray emission of the surrounding interstellar medium. However, SFRs that are more than a few Myr old can also host SNRs and compact objects (isolated or in binaries with a normal star), which can also contribute to the $\gamma$-ray emission.

\subsubsection{4FGL J0859.2--4729}
\label{sec:J0859.2-4729}

4FGL J0859.2--4729 is spatially coincident with the massive SFR RCW 38, situated at a distance of 1.7 kpc \citep{2006AJ....132.1100W}. It is a very young ($<$ 1\,Myr) and strongly obscured ($A_V\sim10$) stellar cluster. 
Among the 40 X-ray sources classified with CT$>2$, all are identified as YSOs.  
Proper-motion measurements derived from Gaia-matched counterparts reveal a consistent motion indicative of their cluster memberships. This shows that the pipeline can reliably identify YSOs although it has no knowledge about the age of this environment.  
Assuming a distance of 1.7\,kpc, the X-ray luminosity of these 40 YSOs spans a range from $3.6\times10^{29}$ to $2.5\times10^{32}$\,erg\,s$^{-1}$.

Additionally, six bright sources ($F_{\rm b}>5\times10^{-14}$\,erg\,s$^{-1}$\,cm$^{-2}$) are classified with low CT $<2$ as one CV (2CXO J085903.6--473040), two AGNs (2CXO J085854.2--472841 and 2CXO J085902.0--473209), and three NSs (2CXO J085905.7--473009, 2CXO J085906.9--473102, and 2CXO J085906.6--473022). 
The absence of Gaia counterparts hampers the determination of their cluster membership through proper-motion measurements. 
Notably, 2CXO J085906.6--473022 exhibits a very hard X-ray spectrum  with a PL model fit yielding  $\Gamma=0.4^{+0.9}_{-0.8}$ and shows intraobservation variability with $P_{\rm intra}=1.0$. This makes it less likely to be an NS since NSs usually do not have such a hard spectrum. 
2CXO J085903.6--473040 is best fit by a PL model with $\Gamma=2.0\pm0.3$ and shows intraobservation variability with $P_{\rm intra}=0.99$. 
Both sources classified as AGNs can be fit with an absorbed PL model, with 2CXO J085854.2--472841 having $\Gamma=1.9\pm0.4$ and $N_{\rm H}=(7\pm1)\times10^{22}$\,cm$^{-2}$, and 2CXO J085902.0--473209 having $\Gamma=1.8\pm0.3$ and $N_{\rm H}=(6\pm1)\times10^{22}$\,cm$^{-2}$, where the absorption values are higher than those fit for the YSOs in the field. 
2CXO J085902.0--473209 also exhibits X-ray variability with $P_{\rm intra}=1.0$ and is matched to an NIR counterpart. 
If either AGN or NS classifications are correct, one or more of these sources could be responsible for the GeV emission from this region. 
However, we also note that the MIR image from WISE in the central part of the cluster is dense and saturated, leading to the lack of WISE matches despite the likely presence of multiple MIR sources, 
which can result in less accurate classification results for sources near the central cluster region. 
Therefore, we are more inclined to attribute the GeV emission to the winds driven by young stars in RCW 38.

\subsubsection{4FGL J0859.3--4342}
\label{sec:J0859.3-4342}

4FGL J0859.3--4342 is situated near the HII region RCW 36, which is ionized by a large and young ($\sim$2--3\,Myr) star cluster, [BDB2003] G265.14+01.45\footnote{See details in \url{http://galaxymap.org/cat/view/rcw/36}.}.  
Among the 55 sources (with S/N$>5$) classified with CT$>2$, 51 are YSOs, and two are HM-STARs. 
Proper-motion measurements from Gaia indicate that most of these sources are cluster members with motions consistent with that of the cluster. 
Assuming a distance of 700\,pc \citep{2004ApJ...614..818B}, the X-ray luminosity of the 51 YSOs ranges from $1.0\times10^{29}$ to $1.3\times10^{31}$\,erg\,s$^{-1}$.

Additionally, one source is classified as an AGN (2CXO J085926.8--434933) with CT$>2$, and four more sources with fluxes $>5\times10^{-14}$\,erg\,s$^{-1}$\,cm$^{-2}$ are classified as two CVs (2CXO J085927.0--434528 and 2CXO J085932.2--434602), an AGN (2CXO J085928.0--434510), and an HMXB (2CXO J085910.9--434343) with lower CT values.  
Both 2CXO J085927.0--434528 and 2CXO J085932.2--434602 show intraobservation variability with $P_{\rm intra}=1.0$ and are matched to optical counterparts, while 2CXO J085932.2--434602 is also matched to an NIR counterpart.  Their X-ray spectra are best fit by the Mekal model, with kT$=7.82^{+0.81}_{-0.66}$\,keV for 2CXO J085927.0--434528 showing prominent 6.7 keV iron line complex, and kT$=11.9^{+4.8}_{-2.3}$\,keV for 2CXO J085932.2--434602. 
The Gaia counterparts for the two classified CVs show proper motions consistent with the cluster stars' proper motion, raising doubt about their CV classifications. 
However, the distance to 2CXO J085927.0--434528, as measured by \cite{2021AJ....161..147B}, is $1580^{+821}_{-438}\,$pc, which is larger than that of the cluster, albeit with large uncertainty. 
Classified as an HMXB with a CT=1.5, 2CXO J085910.9--434343 exhibits a proper motion and a distance consistent with that of the cluster stars, based on Gaia DR3. Besides the optical counterpart, it also has an NIR--MIR counterpart from 2MASS and WISE. Its X-ray spectrum is best fit by a Mekal model, with kT$=13^{+8}_{-4}$\,keV, and it shows intraobservation variability with $P_{\rm intra}=1.0$. 
However, based on the Gaia distance of $755^{+250}_{-140}$\,pc and $E(B-V)=1.12$ measured from 3D  extinction maps of \cite{2016ApJ...818..130B}, the dereddened Gaia absolute magnitude, $M_{\rm G}=8.2$  indicates that the source is too faint to be an HMXB, and may be a YSO residing in the cluster. 
2CXO J085928.0--434510 is classified as an AGN with a low CT=0.7, but its X-ray spectrum is best fit with a Mekal model, with kT$=8^{+4}_{-2}$\,keV, because it accounts for excess around 6.7 keV. However, we note that the limited set of models we are fitting does not include a PL model with a broad iron feature, which could be more appropriate for an AGN. 
2CXO J085926.8--434933 is classified as an AGN with a CT=2.6, with its X-ray spectrum fit by a PL model with $\Gamma=1.9\pm0.3$. 
It is also matched to an NIR--MIR counterpart. Both AGN candidates show intraobservation variability with $P_{\rm intra}=1.0$. 
We also note that the central part of the cluster is saturated in IR images from WISE, making cross-matching and classifications less reliable for those sources near the cluster center.

\subsubsection{4FGL J1046.7--6010}
\label{sec:J1046.7-6010}

4FGL J1046.7--6010 is situated at the southeastern outskirts of the Eta Carinae Nebula (NGC 3372), a massive SFR that harbors some of the most luminous stars and coincides with Bochum 11, a massive 6\,Myr old open cluster that might be associated with NGC 3372\footnote{See the star clusters near NGC 3372 in \url{http://www.atlasoftheuniverse.com/nebulae/ngc3372.html}} \citep{2002A&A...389..871D}. 
Two YSOs are classified with CT$>2$, one of which is identified as a YSO based on their MIR excess emission \citep{2011ApJS..194...14P}, and may be members of NGC 3372.  Additionally, we classified four LM-STARs, which could be foreground or background stars.

We note that, 4FGL J1046.7--6010 and two other 4FGL sources (4FGL J1045--5940 and 4FGL J1048.5--5923) overlap with NGC 3372. The GeV $\gamma$-ray emission around NGC 3372 has been previously modeled, with a central point source, 4FGL J1045.1--5940, originating from the massive binary $\eta$ Carinae and multiple extended components likely arising from the interaction of accelerated protons  with the cluster's ambient gas \citep{2022MNRAS.517.5121G}.

\subsubsection{4FGL J1115.1--6118}
\label{sec:J1115.1-6118}

4FGL J1115.1--6118 has been studied by \cite{2020ApJ...897..131S}, who proposed an association with the young massive stellar cluster NGC 3603. 
NGC 3603 is an SFR located in the Carina spiral arm of the Milky Way at a distance of approximately $7\pm1$\,kpc, with an average age between 2 and 3\,Myr. The Fermi-LAT analysis conducted by \cite{2020ApJ...897..131S} with 10\,yr data provides the localization of 4FGL J1115.1--6118 more consistent with NGC 3603, compared to the $\gamma$-ray position from the 4FGL using 14\,yr data, which is shifted to the south, and only the northern part of 4FGL J1115.1--6118 overlaps with NGC 3603. 

We classify 14 YSOs and 25 HM-STARs within the $\gamma$-ray PU for sources with S/N$>3$ classified without any CT cut. 
These sources are spatially concentrated toward NGC 3603. Furthermore, among sources with S/N$>5$, we classified two NSs (2CXO J111513.6--611657 and 2CXO J111459.1--611707) with CT$>2$. 
Using deeper optical--IR surveys, 2CXO J111513.6--611657 is matched to an NIR counterpart ($i=20.6$) from the VPHAS+ DR2 with a separation of 0.02$\arcsec$ and $P_{\rm c}=9.3\%$, which makes the NS classification unlikely to be correct. 

\subsubsection{4FGL J1725.1--3408} 
\label{sec:J1725.1-3408}

4FGL J1725.1--3408 overlaps with the SFR NGC 6357, also known as the Lobster Nebula, hosting many young stars with ages of a few Myr. The brightest ($F_{\rm b}=1.2\times10^{-13}$\,erg\,s$^{-1}$\,cm$^{-2}$) X-ray source within the $\gamma$-ray PU is 2CXO J172508.8--341112, which is a Wolf-Rayet star \citep{1988A&A...199..217V} but classified as a YSO with a CT\,=\,0.6, exhibiting extended X-ray emission with tens of arcseconds in size. It has an optical--NIR counterpart, and its X-ray spectrum is best fit by a Mekal model with kT$=2.14^{+0.08}_{-0.07}$\,keV. The Gaia counterpart provides a distance measurement of $1.9\pm0.1$\,kpc, consistent with the distance of NGC 6357. The second brightest X-ray source, 2CXO J172510.9--340843, is classified as an NS with a low CT\,=\,0.4. 
It has no counterpart at lower frequencies. Based on its nondetection in the VPHAS+ survey, which has a 5$\sigma$ sensitivity of $g=22.46$ in AB magnitude, the lower limit on $f_{\rm X}/f_{\rm O}$ is 3.2. 
Its X-ray spectrum can be fit by a PL model with $\Gamma=1.5\pm0.6$. The third brightest source, 2CXO J172516.8--341211, is identified as a YSO \citep{2021ApJ...916...32G} but classified as an LMXB with a CT\,=\,0.2. It has an optical--NIR counterpart, and its X-ray spectrum is fit with a PL model ($\Gamma=2.1^{+0.4}_{-0.3}$). Both 2CXO J172510.9--340843 and 2CXO J172516.8--341211 show variability with $P_{\rm intra}=1.0$. 
A complex extended radio structure, tracing the gas and interactive shocks of NGC 6357, also overlaps with 4FGL J1725.1--3408.

\subsubsection{4FGL J2038.4+4212}
\label{sec:J2038.4+4212}

4FGL J2038.4+4212 is located southwest to the SFR DR 21 lying at a distance of $1.50^{+0.08}_{-0.07}$\,kpc \citep{2012A&A...539A..79R}, exhibiting a bright complex radio emission from both RACS-mid and NVSS surveys.
Five X-ray sources are classified as YSOs with CT$>2$ within the PU of 4FGL J2038.4+4212, which were also classified as YSOs by \cite{2013ApJS..209...32B}.

\subsubsection{4FGL J2059.1+4403}
\label{sec:J2059.1+4403}

4FGL J2059.1+4403 is located near the Cygnus Wall, a turbulent region within the SFR NGC 7000 at a distance of $795\pm25$ pc \citep{2020ApJ...899..128K}. 
Eight YSOs and one LM-STAR are classified with CT$>2$. 
One X-ray source, 2CXO J205743.9+435858, lacks any counterpart and is classified as an NS with a CT=1.0. Its spectrum can be described by a PL model with $\Gamma=2.3^{+1.3}_{-1.1}$. If it is located at the distance of NGC 7000, its flux of $F_{\rm b}=5.2\times10^{-14}$\,erg\,s$^{-1}$\,cm$^{-2}$ would correspond to a luminosity of $1.6\times10^{30}$\,erg\,s$^{-1}$. 
Based on its nondetection in the PanSTARRS DR1 survey \citep{2016arXiv161205560C}, which has a mean 5$\sigma$ sensitivity of $g=23.3$ in AB magnitude, the lower limit on  $f_{\rm X}/f_{\rm O}$ is 4.9 for 2CXO J205743.9+435858. 
Another source, 2CXO J205855.7+440337, has a 0.5--7 keV flux of $3.8\times10^{-13}$\,erg\,s$^{-1}$\,cm$^{-2}$, and was not classified due to its placement at a large OAA (12.3$\arcmin$). 
It appears that the source is pointlike since the extent of the source ($\sim7\arcsec$) is consistent with the large PSF size at an OAA of 12.3$\arcmin$.  Its X-ray spectrum is very soft, and cannot be described by a PL model. The Mekal model provides a reasonable fit with kT$=0.52\pm0.04$\,keV. 
It is matched to a very bight optical--MIR source with Gmag=6.355, Jmag=5.095, and W1mag=4.09, with a separation of 0.9$\arcsec$ and $P_{\rm c}=2.1\%$. 
The optical source is a K0.5III type star at a distance of 223\,pc \citep{2018A&A...612A..96F}. $f_{\rm X}/f_{\rm O}$ of $9\times10^{-6}$  is typical for a low-mass star. However, the X-ray source shows no signs of variability; hence, X-rays are probably not attributable to a flare.  
This source could be an active binary, but it is unlikely to produce GeV emission.

\subsection{Ambiguous Cases}
\label{sec:Ambiguous}

Below, we discuss the unIDed sources 
where we could not identify the  most plausible  X-ray counterpart mostly due to the low confidence of classifications, although these classifications are consistent with the   $\gamma$-ray producing classes.

\subsubsection{4FGL J0737.4+6535}
\label{sec:J0737.4+6535}

4FGL J0737.4+6535 spatially coincides with the intermediate spiral star-forming galaxy NGC 2403, located approximately 3.5\,Mpc away.  \cite{2020ApJ...896L..33X} proposed an association between 4FGL J0737.4+6535 and supernova (SN) 2004dj,  one of the nearest and brightest SNs in NGC 2403. They interpreted the fading $\gamma$-ray emission arising from the SN ejecta interacting with a surrounding high-density shell, decelerating the ejecta and converting $\sim$1\% of the SN kinetic energy to relativistic protons. 2CXO J073717.0+653557, which coincides with SN 2004dj, is classified as an LMXB with a CT=2.0 and matched with an optical--NIR counterpart. Its X-ray spectrum is best fit with the PL model with $\Gamma=2.5\pm0.1$, and it exhibits long-term variability across multiple X-ray observations. The LMXB classification is not surprising because we do not have an SN class in our TD.

\cite{2021ApJ...923...75K} reported the association of a Swift-X-Ray Telescope (XRT) counterpart, SwXF4 J073711.3+653348, with 4FGL J0737.4+6535. The XRT source was classified as a blazar by \cite{2021ApJ...923...75K} and is located just outside the edge of the $\gamma$-ray PU, so we did not classify it in an automated way. We classify one more AGN (with CT$>2$) in this field -- 2CXO J073726.0+653656, matched to an MIR counterpart. 
However, all the three X-ray sources mentioned above were classified as HMXBs associated with NGC 2403 by \citep{2012MNRAS.419.2095M}, while the flag was raised for 2CXO J073726.0+653656 as an AGN due to its large offset from the galactic center.

\subsubsection{4FGL J1616.6--5009}
\label{sec:J1616.6-5009}

Only one X-ray source, 2CXO J161701.5--501628, is detected with S/N=5.1 in the CSCv2.0 within the 44\% Chandra X-ray Observatory coverage of the PU of 4FGL J1616.6--5009. It is matched to an optical--NIR counterpart and classified as an LMXB with a low 
CT=0.7. With an X-ray flux of $1.6\times10^{-13}$\,erg\,s$^{-1}$\,cm$^{-2}$ in the 0.5--7\,keV, the corresponding $L_{\rm X}=1.2\times10^{32}$\,erg\,s$^{-1}$, at a Gaia distance of $2.5^{+0.7}_{-0.6}$\,kpc. 
Additionally, it shows variability, with $P_{\rm intra}=1.0$. 
In addition, PSR J1616--5017, with an age of 167 kyr and an $\dot{E}=1.6\times10^{34}$\,erg\,s$^{-1}$, is located just outside the GeV PU, which is not detected in the CSCv2.0. 

\subsubsection{4FGL J1720.6--3706}
\label{sec:J1720.6-3706}

There is only one X-ray source, 2CXO J172026.9--370322, detected with S/N$=3.6$ within the $\gamma$-ray PU of 4FGL J1720.6--3706, with a flux of $F_{\rm b}=9.4\times10^{-14}$\,erg\,s$^{-1}$\,cm$^{-2}$. The X-ray source has a relatively large PU of 4.56$\arcsec$ due to its off-axis placement in two observations (6.8 and 7.7\,ks long), with OAAs of 8.7$\arcmin$ and 12.9$\arcmin$, respectively. 
The cross-matching to optical--MIR catalogs results in a large $P_{\rm c}=74$\% with two potential counterparts identified. With the first counterpart, the X-ray source is classified as a CV with a low CT=0.2, while with the second counterpart the X-ray source is classified as an LMXB with a CT=0.1. If no counterpart is considered, the X-ray source is classified as an NS with a CT=0.2. 
We also note that there is another bright ($F_{\rm b}=2\times10^{-13}$\,erg\,s$^{-1}$\,cm$^{-2}$) X-ray source 2CXO J172051.8--371037, located just outside the $\gamma$-ray PU, which was not included in the automated classification for that reason. The source is very soft with a PL model photon index $\Gamma=4.7\pm0.7$. It is matched to the bright optical--IR counterpart TYC 7374-9-1. The luminosity of $1.5\times10^{30}$\,erg\,s$^{-1}$ at Gaia distance of 253\,pc and $f_{\rm X}/f_{\rm O}$\footnote{The Gaia $G$-band optical energy flux is calculated from the Gaia $G$-band magnitude using $F = f_{\rm zp} 10^{-{\rm G}/2.5} \lambda_{\rm ref}$ with $f_{\rm zp}=2.5\times10^{-9}$ erg\,s$^{-1}$cm$^{-2}$\,\AA$^{-1}$ and $\lambda_{\rm ref}=6218$\,\AA.} of $4\times10^{-4}$ suggest that the X-rays are likely due to the coronal activity of the star, making this source an unlikely $\gamma$-ray emitter. 
Additionally, 4FGL J1720.6--3706 has several bright radio sources from the MSC in its proximity.

\subsubsection{4FGL J1732.8--3725}
\label{sec:J1732.8-3725}

There is only one  X-ray source, 2CXO J173249.6--372111 with a flux of $F_{\rm b}=1.3\times10^{-13}$\,erg\,s$^{-1}$\,cm$^{-2}$, within the PU of 4FGL J1732.8--3725. 
Due to the large OAA (9.5\arcmin) and only 2.0\,ks exposure, the X-ray source has a large PU of 4.92\arcsec. The cross-matching to optical--IR catalogs, with a significant likelihood of confusion ($P_{\rm c}=85\%$), reveals three potential counterparts. The primary counterpart (p\_i$=37\%$) results in the classification of the X-ray source as an LMXB with a low CT=0.5. This classification is maintained even when considering the source without any counterpart. 
We also note that a 356\,kyr old pulsar PSR B1730--37 with an $\dot{E}=1.5\times10^{34}$\,erg\,s$^{-1}$ is located a few arcminutes outside the GeV PU.

\subsubsection{4FGL J1818.5--2036}
\label{sec:J1818.5-2036}

About 42\% of the PU of 4FGL J1818.5--2036 is covered by Chandra X-ray Observatory observations from the CSCv2.0. 
2CXO J181920.3--203952, the brightest ($F_{\rm b}=10^{-13}$\,erg\,s$^{-1}$\,cm$^{-2}$) X-ray source detected within the PU of 4FGL J1818.5--2036, is classified as an NS with a low CT=0.1. 
Spectral fitting with an absorbed PL model yields $\Gamma=1.5\pm0.3$ for 2CXO J181920.3--203952. 
Assuming the X-ray source is not detected with a mean 5$\sigma$ sensitivity of $g=23.3$ in AB magnitude from PanSTARRS DR1, it provides a lower limit of 9.2 on $f_{\rm X}/f_{\rm O}$ for 2CXO J181920.3--203952. 
Moreover, a number of radio sources from the MSC have been detected within the $\gamma$-ray PU, but many of them are not covered by the Chandra X-ray Observatory observations, and none of them match any of the detected X-ray sources.

\subsubsection{4FGL J1836.8--2354}
\label{sec:J1836.8-2354}

2CXO J183659.8--235129, the brightest ($F_{\rm b}=4.5\times10^{-14}$\,erg\,s$^{-1}$\,cm$^{-2}$) X-ray source within the PU of 4FGL J1836.8--2354, is classified as an LMXB with a low CT=0.9, with an optical--NIR counterpart matched (Gmag=17.5 from Gaia and Jmag=15.5 from 2MASS with $P_{\rm c}=59.8\%$, but no accurate distance was measured). It is worth noting that 4FGL J1836.8--2354 is suggested to be associated with the binary millisecond PSR J1836--2354A in the globular cluster M22 \citep{2019MNRAS.486.3992A}, located half an arcminute outside of the 95\% $\gamma$-ray PU. Additionally, a bright radio source, PMN J1836--2406, is located to the south outside the $\gamma$-ray PU of 4FGL J1836.8--2354.

\subsubsection{4FGL J1843.7--3227}
\label{sec:J1843.7-3227}

The globular cluster NGC 6681 (M 70) is located at the northwest corner within the PU of 4FGL J1843.7--3227. NGC 6681 has an angular size of 8$\arcmin$ at a distance of 9\,kpc, which is significantly smaller than the GeV PU.  
2CXO J184316.0--322414, the brightest ($F_{\rm b}=8\times10^{-14}$\,erg\,s$^{-1}$\,cm$^{-2}$) X-ray source detected within the PU of 4FGL J1843.7--3227, is classified as an NS with a CT=1.8. 
A potential MIR counterpart from CatWISE2020 is located just outside the X-ray PU (0.95\arcsec), with a separation of 0.976$\arcsec$ and $P_{\rm c}=8\%$, hence not matched by our pipeline.  
The MIR source is associated with a Gaia source with Gmag=21.0, BPmag=21.4, RPmag=19.9, and an NIR source detected from The VISTA Hemisphere Survey catalog DR5 \citep{2021yCat.2367....0M} with Jpmag=17.9, Kspmag=17.0. 
Despite relatively large OAAs (6.5\arcmin) for both Chandra X-ray Observatory observations, the X-ray source may be extended, as indicated by the corresponding flag raised by the CSCv2.0. 
The X-ray spectra can be fit by a PL model with $\Gamma=1.9\pm0.2$. 
Additionally, 2CXO J184316.0--322414 is coincident with a radio source J184315.84--322419.9 from the MSC catalog, with a flux of $12.5\pm1.9$\,mJy at 1.086\,GHz. 
Another source, 2CXO J184310.4-322043, is classified as an AGN with a low CT=0.2 and coincides with two radio sources from the MSC, resembling a two-sided jet seen in the RACS survey.  

\subsubsection{4FGL J2004.3+3339}
\label{sec:J2004.3+3339}

4FGL J2004.3+3339 is overlapped with a molecular cloud, G70.7+1.2, which is suggested to be associated with two unrelated objects (a B-star XRB and a Be star) interacting with a dense molecular cloud \citep{1992Natur.360..139K,2007ApJ...665L.135C}. The B-star XRB is coincident with a hard X-ray source (2CXO J200423.4+333906), with a photon index $\Gamma=1.8$, an X-ray luminosity of $\sim4\times10^{31}$\,erg\,s$^{-1}$ (assuming a fiducial distance of 4.5\,kpc), and an NIR counterpart with Jmag=15.56, Hmag=14.51, and K$'$mag=13.97 observed from the Keck II telescope \citep{2007ApJ...665L.135C}. The NIR counterpart of the B-star XRB is not resolved by the 2MASS survey due to its faintness and vicinity to the bright Be star. However, a faint Gaia counterpart with Gmag=20.1 is matched. 
Our pipeline classified 2CXO J200423.4+333906 as a CV with a low CT=0.6. 
Another X-ray source, 2CXO J200422.7+333844 (S/N=3.2), which overlaps G70.7+1.2 and arcminute scale extended X-ray emission to the north of the point source (see the lower right panel in Figure \ref{fig:fields}), is classified as an NS with a CT=2.8. 
Based on its nondetection in the PanSTARRS survey, which has a mean 5$\sigma$ sensitivity of $g=23.3$ in AB magnitude, we calculate the lower limit on $f_{\rm X}/f_{\rm O}$ as  0.06 for 2CXO J200422.7+333844. 
However, both X-ray sources have detection S/N $<$5. Since their significance might be affected by the surrounding extended emission, we decided to discuss these two sources here. We also note that the classifications of CVs have some confusion with HMXBs, which indicates that the classified CV can be an HMXB as suggested by \cite{2007ApJ...665L.135C}. 

\begin{longrotatetable}
\begin{deluxetable*}{lccccccccccccr}
\tablecaption{Classifications and Multiwavelength Properties of Potential X-ray Counterparts for UnIDed 4FGL Sources \label{tab:X-ray-classification}}
\tablewidth{700pt}
\tabletypesize{\scriptsize}
\tablehead{
\colhead{4FGL name$^{\rm a}$} & \colhead{2CXO name$^{\rm b}$} & 
\colhead{Class$^{\rm c}$} & \colhead{Class Prob$^{\rm d}$} & \colhead{CT} & \colhead{$F_\gamma$} & \colhead{$F_{\rm X}$} & 
\colhead{$HR_{\rm h(ms)}^{\rm e}$} &  \colhead{$P_{\rm intra}$} & 
\colhead{Gmag} & \colhead{Jmag} & 
\colhead{W1mag} & \colhead{$F_{\rm R}$} & \colhead{$f_{\rm X}/f_{\rm O}$} \\ 
\colhead{} & \colhead{} & \colhead{} &  \colhead{} & \colhead{} & \colhead{($10^{-13}$\,cgs)} & \colhead{($10^{-13}$\,cgs)} & \colhead{} & \colhead{} & \colhead{(mag)} & \colhead{(mag)} & \colhead{(mag)} & \colhead{(mJy)} &  \colhead{}
} 
\startdata
J0058.3--4603 & J005806.2--460419 &             AGN &       $46\pm9$ &    0.5 &     10 &    2.3 &       -0.1 &               0.47 &     20.5 &     \nodata &      15.3 &      6.7 &     2.3 \\
J0335.6--0727 & J033532.7--072741 &   AGN (AGN [1]) &       $60\pm9$ &    2.5 &     12 &    1.1 &       -0.1 &               0.91 &     20.3 &     17.1 &      14.7 &      4.4 &     1.0 \\
J0442.8+3609 & J044308.1+361310 &             AGN &       $99\pm2$ &     38 &     60 &    0.2 &        0.5 &               0.01 &     \nodata &     \nodata &      17.0 &     \nodata &    \nodata \\
J0442.8+3609 & J044309.2+360856 &             AGN &       $99\pm2$ &     34 &     60 &    0.1 &        0.4 &               0.76 &     \nodata &     \nodata &      17.0 &     \nodata &    \nodata \\
J0639.1--8009 & J063554.8--800814 &             AGN &       $98\pm3$ &     17 &     20 &    0.7 &        0.6 &               0.52 &     \nodata &     \nodata &      15.7 &     \nodata &    \nodata \\
J0639.1--8009 & J063623.7--801259 &             AGN &       $97\pm4$ &     11 &     20 &    0.3 &       -0.0 &               0.51 &     \nodata &     \nodata &      15.9 &     \nodata &    \nodata \\
J0639.1--8009 & J063640.9--801126 &             AGN &       $98\pm3$ &     23 &     20 &    0.2 &        0.4 &               0.25 &     \nodata &     \nodata &      15.6 &     \nodata &    \nodata \\
J0639.1--8009 & J063704.1--801401 &             AGN &       $99\pm2$ &     24 &     20 &    0.1 &        0.1 &               0.94 &     \nodata &     \nodata &      17.5 &     \nodata &    \nodata \\
J0639.1--8009 & J063719.7--801230 &             AGN &       $93\pm7$ &    6.6 &     20 &    0.5 &        0.8 &               0.96 &     \nodata &     \nodata &      14.2 &     \nodata &    \nodata \\
J0639.1--8009 & J063805.1--801854 &    AGN (CV [2]) &      $40\pm10$ &    0.1 &     20 &    2.8 &        0.8 &               1.00 &     20.1 &     16.2 &      16.3 &     \nodata &     2.0 \\
J0639.1--8009 & J063850.7--801522 &             AGN &      $74\pm11$ &    2.8 &     20 &    0.2 &        0.9 &               0.93 &     \nodata &     \nodata &      15.6 &     \nodata &    \nodata \\
J0639.1--8009 & J063905.7--801957 &             AGN &       $99\pm3$ &     23 &     20 &    0.2 &        0.4 &               0.40 &     \nodata &     \nodata &      17.6 &     \nodata &    \nodata \\
J0639.1--8009 & J063944.2--801454 &             AGN &       $99\pm3$ &     20 &     20 &    0.1 &        0.3 &               0.76 &     \nodata &     \nodata &      16.8 &     \nodata &    \nodata \\
J0639.1--8009 & J064013.0--801724 &             AGN &      $82\pm13$ &    3.2 &     20 &    0.3 &        0.5 &               0.98 &     \nodata &     \nodata &      17.7 &     \nodata &    \nodata \\
J0639.1--8009 & J064126.4--801048 &            LMXB &      $36\pm20$ &    0.0 &     20 &    0.6 &        0.2 &               0.99 &     \nodata &     \nodata &      \nodata &     \nodata &    \nodata \\
J0737.4+6535 & J073717.0+653557 &   LMXB (SN [3]) &       $53\pm9$ &    2.0 &     12 &    0.3 &       -0.3 &               0.77 &     19.4 &     16.2 &      \nodata &     \nodata &     0.1 \\
J0737.4+6535 & J073726.0+653656 &             AGN &       $98\pm3$ &     18 &     12 &    0.1 &        0.4 &               0.68 &     \nodata &     \nodata &      16.4 &     \nodata &    \nodata \\
J0859.2--4729 & J085854.2--472841 &             AGN &      $49\pm18$ &    0.3 &    205 &    0.7 &        1.0 &               0.95 &     \nodata &     \nodata &      \nodata &     \nodata &    \nodata \\
J0859.2--4729 & J085902.0--473209 &             AGN &      $31\pm12$ &    0.5 &    205 &    0.7 &        0.9 &               1.00 &     \nodata &     17.3 &      \nodata &     \nodata &    \nodata \\
J0859.2--4729 & J085903.6--473040 &              CV &      $31\pm21$ &    0.0 &    205 &    1.0 &        0.9 &               0.99 &     \nodata &     \nodata &      \nodata &     \nodata &    \nodata \\
J0859.2--4729 & J085905.7--473009 &              NS &      $50\pm24$ &    0.6 &    205 &    0.7 &        0.8 &               0.83 &     \nodata &     \nodata &      \nodata &     \nodata &    \nodata \\
J0859.2--4729 & J085906.6--473022 &              NS &      $29\pm14$ &    0.0 &    205 &    0.9 &        1.0 &               1.00 &     \nodata &     \nodata &      \nodata &     \nodata &    \nodata \\
J0859.2--4729 & J085906.9--473102 &              NS &      $44\pm17$ &    0.7 &    205 &    0.5 &        1.0 &               1.00 &     \nodata &     \nodata &      \nodata &     \nodata &    \nodata \\
J0859.3--4342 & J085910.9--434343 &            HMXB &      $57\pm14$ &    1.5 &     90 &    1.2 &        1.0 &               1.00 &     19.4 &     14.5 &      11.1 &     \nodata &     0.4 \\
J0859.3--4342 & J085926.8--434933 &             AGN &      $65\pm13$ &    2.6 &     90 &    0.7 &        0.9 &               1.00 &     \nodata &     16.7 &      13.4 &     \nodata &    \nodata \\
J0859.3--4342 & J085927.0--434528 &              CV &      $61\pm12$ &    1.8 &     90 &     16 &        0.8 &               1.00 &     18.1 &     \nodata &      \nodata &     \nodata &     1.8 \\
J0859.3--4342 & J085928.0--434510 &             AGN &      $48\pm17$ &    0.7 &     90 &    2.8 &        1.0 &               1.00 &     \nodata &     \nodata &      \nodata &     \nodata &    \nodata \\
J0859.3--4342 & J085932.2--434602 &              CV &      $30\pm11$ &    0.2 &     90 &    4.1 &        0.8 &               1.00 &     20.2 &     14.4 &      \nodata &     \nodata &     3.2 \\
J1025.9+1244 & J102557.9+124109 &    NS (AGN [4]) &      $59\pm26$ &    0.7 &     18 &    0.6 &       -0.1 &               0.42 &     \nodata &     \nodata &      \nodata &      155 &    \nodata \\
J1032.0+5725 & J103121.9+573134 &             AGN &       $96\pm7$ &    9.0 &    7.1 &    0.1 &        0.6 &               0.44 &     \nodata &     \nodata &      17.4 &     \nodata &    \nodata \\
J1032.0+5725 & J103142.0+573015 &   AGN (AGN [5]) &       $99\pm2$ &     26 &    7.1 &    0.3 &        0.1 &               0.29 &     \nodata &     \nodata &      15.4 &     \nodata &    \nodata \\
J1032.0+5725 & J103143.2+573252 &   AGN (AGN [5]) &       $98\pm2$ &     24 &    7.1 &    1.1 &        0.3 &               0.15 &     \nodata &     \nodata &      16.1 &     \nodata &    \nodata \\
J1032.0+5725 & J103143.3+573157 &   AGN (AGN [5]) &      $100\pm0$ &    145 &    7.1 &    0.2 &       -0.2 &               0.06 &     19.1 &     \nodata &      16.1 &     \nodata &     0.1 \\
J1032.0+5725 & J103145.7+573401 &   AGN (AGN [5]) &      $100\pm1$ &     79 &    7.1 &    0.2 &        0.6 &               0.42 &     20.0 &     \nodata &      16.3 &     \nodata &     0.2 \\
J1032.0+5725 & J103202.9+573208 &   AGN (AGN [5]) &       $88\pm7$ &    6.4 &    7.1 &    0.7 &        0.3 &               0.36 &     \nodata &     \nodata &      17.3 &     \nodata &    \nodata \\
J1032.0+5725 & J103220.2+573211 &             AGN &       $99\pm2$ &     24 &    7.1 &    0.1 &        0.1 &               0.31 &     \nodata &     \nodata &      17.2 &     \nodata &    \nodata \\
J1032.0+5725 & J103224.8+573153 &   AGN (AGN [5]) &       $99\pm2$ &     36 &    7.1 &    0.3 &        0.4 &               0.46 &     \nodata &     \nodata &      15.3 &     \nodata &    \nodata \\
J1032.0+5725 & J103225.1+572814 &   AGN (AGN [5]) &       $77\pm9$ &    3.5 &    7.1 &    0.3 &        0.0 &               0.47 &     19.5 &     \nodata &      15.4 &     \nodata &     0.1 \\
J1032.0+5725 & J103302.5+572833 &   AGN (AGN [5]) &       $97\pm4$ &     13 &    7.1 &    0.3 &        0.1 &               0.99 &     \nodata &     \nodata &      17.2 &     \nodata &    \nodata \\
J1106.4+0859 & J110807.9+091722 &       (AGN [6]) &    \nodata &   \nodata &     31 &    1.7 &       -0.2 &               0.49 &     19.4 &     \nodata &      14.7 &     \nodata &     0.7 \\
J1106.4+0859 & J110823.6+091239 &   AGN (AGN [6]) &       $99\pm1$ &     57 &     31 &    0.3 &       -0.0 &               0.16 &     19.1 &     \nodata &      16.3 &     \nodata &     0.1 \\
J1115.1--6118 & J111459.1--611707 &              NS &      $76\pm15$ &    2.2 &    257 &   0.01 &        0.5 &               0.19 &     \nodata &     \nodata &      \nodata &     \nodata &    \nodata \\
J1115.1--6118 & J111513.6--611657 &              NS &      $81\pm13$ &    3.0 &    257 &   0.01 &        0.0 &               0.20 &     \nodata &     \nodata &      \nodata &     \nodata &    \nodata \\
J1116.3+1818 & J111633.3+181420 &   AGN (AGN [7]) &       $99\pm3$ &     24 &     14 &    0.4 &        0.3 &               0.64 &     \nodata &     \nodata &      16.8 &     \nodata &    \nodata \\
J1242.6+3236 & J124235.1+323340 &             AGN &       $99\pm4$ &     15 &    5.5 &    0.1 &        0.1 &               0.55 &     \nodata &     \nodata &      16.9 &     \nodata &    \nodata \\
J1256.9+2736 & J125705.5+273252 &             AGN &       $97\pm5$ &     13 &     18 &    0.2 &        0.1 &               0.89 &     \nodata &     \nodata &      17.8 &     \nodata &    \nodata \\
J1256.9+2736 & J125724.3+273606 &             AGN &       $98\pm4$ &     16 &     18 &    0.1 &        0.0 &               0.83 &     \nodata &     \nodata &      17.9 &     \nodata &    \nodata \\
J1256.9+2736 & J125724.4+272952 &   AGN (AGN [8]) &    \nodata &   \nodata &     18 &    0.2 &       -0.7 &               0.92 &     20.7 &     12.8 &      10.6 &      133 &     0.3 \\
J1256.9+2736 & J125730.0+272612 &             AGN &       $92\pm8$ &    6.7 &     18 &    0.4 &       -0.4 &               0.88 &     20.8 &     \nodata &      16.5 &     \nodata &     0.5 \\
J1256.9+2736 & J125734.0+272730 &             AGN &       $99\pm1$ &     48 &     18 &    0.3 &        0.2 &               0.73 &     \nodata &     \nodata &      16.0 &     \nodata &    \nodata \\
J1256.9+2736 & J125740.2+273118 & AGN (AGN [5,9]) &       $99\pm1$ &     50 &     18 &    0.7 &       -0.1 &               0.96 &     19.5 &     \nodata &      15.8 &     \nodata &     0.3 \\
J1256.9+2736 & J125745.0+273210 & AGN (AGN [5,9]) &       $98\pm2$ &     22 &     18 &    0.2 &       -0.1 &               0.31 &     19.3 &     \nodata &      16.0 &     \nodata &     0.1 \\
J1256.9+2736 & J125746.6+273137 &             AGN &       $99\pm2$ &     32 &     18 &    0.2 &        0.2 &               0.94 &     \nodata &     \nodata &      16.4 &     \nodata &    \nodata \\
J1256.9+2736 & J125751.0+273231 & AGN (AGN [5,9]) &       $96\pm3$ &     14 &     18 &    0.3 &       -0.1 &               0.01 &     19.7 &     \nodata &      16.9 &     \nodata &     0.1 \\
J1435.4+3338 & J143513.2+333118 &  AGN (AGN [10]) &       $90\pm9$ &    6.2 &     12 &    0.5 &        0.0 &               0.58 &     \nodata &     \nodata &      16.4 &     2.2 &    \nodata \\
J1435.4+3338 & J143522.1+333816 &             AGN &      $81\pm13$ &    3.2 &     12 &    0.6 &        0.3 &               0.46 &     \nodata &     \nodata &      17.7 &     \nodata &    \nodata \\
J1435.4+3338 & J143525.5+334605 &             AGN &       $99\pm2$ &     32 &     12 &    0.2 &       -0.0 &               0.25 &     \nodata &     \nodata &      16.9 &     \nodata &    \nodata \\
J1435.4+3338 & J143529.6+334733 &             AGN &      $100\pm1$ &     85 &     12 &    0.3 &        0.1 &               0.38 &     20.0 &     \nodata &      17.1 &     \nodata &     0.2 \\
J1435.4+3338 & J143530.0+334335 &             AGN &       $96\pm4$ &     15 &     12 &    0.2 &       -0.3 &               0.90 &     \nodata &     \nodata &      16.2 &      \nodata &    \nodata \\
J1435.4+3338 & J143542.6+333404 &             AGN &      $100\pm0$ &    172 &     12 &    0.7 &        0.0 &               0.83 &     19.6 &     \nodata &      15.3 &     \nodata &     0.3 \\
J1435.4+3338 & J143550.3+333338 &             AGN &      $100\pm1$ &     70 &     12 &    0.3 &        0.1 &               0.49 &     19.6 &     \nodata &      16.4 &     \nodata &     0.1 \\
J1435.4+3338 & J143551.5+333909 &             AGN &       $95\pm7$ &    8.8 &     12 &    0.4 &        0.3 &               0.40 &     \nodata &     \nodata &      17.2 &     \nodata &    \nodata \\
J1502.6+0207 & J150234.5+015205 &             AGN &      $100\pm1$ &     69 &     11 &    0.4 &        0.1 &               0.97 &     19.8 &     \nodata &      15.0 &     \nodata &     0.2 \\
J1502.6+0207 & J150237.4+015813 &             AGN &       $99\pm1$ &     42 &     11 &    0.7 &        0.4 &               0.20 &     \nodata &     \nodata &      15.4 &     \nodata &    \nodata \\
J1510.9+0551 & J151051.3+054937 &             AGN &       $99\pm2$ &     31 &     13 &    0.1 &        0.0 &               0.98 &     \nodata &     \nodata &      17.8 &     \nodata &    \nodata \\
J1510.9+0551 & J151100.4+054921 &  AGN (AGN [11]) &       $67\pm7$ &    5.0 &     13 &    0.1 &       -0.6 &               0.91 &     19.3 &     16.9 &      14.6 &       10 &    0.04 \\
J1510.9+0551 & J151100.5+054912 &             AGN &       $99\pm1$ &     43 &     13 &    0.2 &        0.4 &               0.77 &     20.2 &     \nodata &      15.5 &     \nodata &     0.2 \\
J1510.9+0551 & J151104.1+055136 &             AGN &       $99\pm1$ &     51 &     13 &    0.2 &        0.1 &               0.51 &     \nodata &     \nodata &      16.3 &     \nodata &    \nodata \\
J1615.3--6034 & J161532.9--603954 &      (AGN [12]) &    \nodata &   \nodata &     38 &    1.3 &        0.2 &               0.74 &     19.9 &     12.3 &      10.3 &     6739 &     0.8 \\
J1616.6--5009 & J161701.5--501628 &            LMXB &      $46\pm16$ &    0.7 &    267 &    1.6 &        0.5 &               1.00 &     17.8 &     15.0 &      \nodata &     \nodata &     0.1 \\
J1619.3--5047 & J161945.8--504222 &              NS &      $85\pm17$ &    2.8 &    209 &    0.3 &        1.0 &               0.30 &     \nodata &     \nodata &      \nodata &     \nodata &  $>1.5$ \\
J1720.6--3706 & J172026.9--370322 &              CV &      $42\pm17$ &    0.2 &    203 &    0.9 &        0.9 &               0.03 &     19.9 &     15.0 &      \nodata &     \nodata &     0.5 \\
J1725.1--3408 & J172510.9--340843 &              NS &      $46\pm19$ &    0.4 &    109 &    0.8 &        0.9 &               1.00 &     \nodata &     \nodata &      \nodata &     \nodata &   $>3.2$ \\
J1725.1--3408 & J172516.8--341211 & LMXB (YSO [13]) &      $40\pm12$ &    0.2 &    109 &    0.5 &        0.6 &               1.00 &     18.3 &     14.5 &      \nodata &     \nodata &     0.1 \\
J1732.8--3725 & J173249.6--372111 &            LMXB &      $34\pm16$ &    0.5 &     80 &    1.3 &        0.1 &               0.76 &     20.9 &     \nodata &      \nodata &     \nodata &     2.0 \\
J1734.0--2933 & J173411.3--293117 &              NS &      $46\pm20$ &    0.3 &     95 &    2.3 &        0.7 &               0.92 &     \nodata &     \nodata &      \nodata &     \nodata &    \nodata \\
J1737.1--2901 & J173657.4--285630 &              CV &      $42\pm13$ &    0.8 &    116 &    2.0 &        0.9 &               0.74 &     19.3 &     14.6 &      \nodata &     \nodata &     0.7 \\
J1739.2--2717 & J173927.6--272112 &              CV &      $62\pm22$ &    0.8 &     77 &    1.9 &        0.9 &               0.71 &     19.2 &     \nodata &      \nodata &     \nodata &     0.6 \\
J1740.6--2808 & J174036.5--280840 &              NS &      $38\pm19$ &    0.2 &     87 &    1.5 &        0.9 &               1.00 &     \nodata &     \nodata &      \nodata &     \nodata &    \nodata \\
J1740.6--2808 & J174042.0--280724 &              CV &      $46\pm16$ &    0.2 &     87 &    2.2 &        1.0 &               1.00 &     \nodata &     \nodata &      \nodata &     \nodata &    \nodata \\
J1742.5--2833 & J174216.9--283707 &              CV &      $58\pm15$ &    1.4 &    124 &    4.8 &        0.8 &               1.00 &     \nodata &     \nodata &      \nodata &     \nodata &    \nodata \\
J1742.5--2833 & J174237.6--283726 &              NS &      $84\pm13$ &    3.5 &    124 &    1.2 &        1.0 &               0.93 &     \nodata &     \nodata &      \nodata &     \nodata &    \nodata \\
J1742.5--2833 & J174249.7--283552 &              NS &      $72\pm17$ &    1.9 &    124 &    0.6 &        1.0 &               0.91 &     \nodata &     \nodata &      \nodata &     \nodata &    \nodata \\
J1744.9--2905 & J174457.8--290509 &              NS &      $77\pm14$ &    2.5 &    301 &    0.8 &        1.0 &               0.90 &     \nodata &     \nodata &      \nodata &     \nodata &    \nodata \\
J1744.9--2905 & J174507.0--290357 &              NS &      $73\pm15$ &    2.4 &    301 &    0.1 &        0.8 &               0.96 &     \nodata &     \nodata &      \nodata &     \nodata &    \nodata \\
J1745.6--3145 & J174548.1--314318 &              CV &      $42\pm17$ &    0.2 &     92 &    0.7 &       -0.5 &               0.66 &     15.7 &     13.9 &      \nodata &     \nodata &    0.01 \\
J1750.4--3023 & J175039.6--302056 &              NS &      $53\pm25$ &    0.5 &     62 &    1.8 &        0.9 &               0.50 &     \nodata &     \nodata &      \nodata &     \nodata &    \nodata \\
J1751.6--3002 & J175147.9--301046 &             AGN &       $30\pm8$ &    0.0 &     60 &    1.1 &       -0.7 &               0.94 &     \nodata &     \nodata &       8.5 &     \nodata &    \nodata \\
J1757.6--2731 & J175734.9--273654 &              CV &      $34\pm18$ &    0.2 &     72 &    0.6 &       -0.5 &               0.26 &     14.0 &     \nodata &      \nodata &     \nodata &   0.001 \\
J1804.9--3001 & J180445.6--300418 &              CV &      $49\pm16$ &    0.8 &     35 &    0.7 &        0.8 &               0.97 &     18.4 &     \nodata &      \nodata &     \nodata &     0.1 \\
J1804.9--3001 & J180510.9--300900 &            LMXB &      $42\pm18$ &    0.7 &     35 &    0.5 &        0.1 &               0.92 &     19.9 &     \nodata &      \nodata &     \nodata &     0.3 \\
J1818.5--2036 & J181920.3--203952 &              NS &      $34\pm19$ &    0.1 &    118 &    1.0 &        0.9 &               0.93 &     \nodata &     \nodata &      \nodata &     \nodata &  $>9.2$ \\
J1836.8--2354 & J183659.8--235129 &            LMXB &      $51\pm19$ &    0.9 &     58 &    0.4 &       -0.2 &               0.98 &     17.6 &     15.5 &       \nodata &      \nodata &    0.03 \\
J1843.7--3227 & J184310.4--322043 &             AGN &      $50\pm17$ &    0.2 &     20 &    0.2 &        0.4 &               0.98 &     \nodata &     \nodata &      \nodata &       74 &    \nodata \\
J1843.7--3227 & J184316.0--322414 &              NS &      $69\pm18$ &    1.8 &     20 &    0.8 &        0.7 &               0.79 &     \nodata &     \nodata &      \nodata &       16 &    \nodata \\
J1844.4--0306 & J184441.8--030551 &              NS &      $58\pm16$ &    1.2 &    401 &    1.0 &        0.7 &               0.92 &     \nodata &     \nodata &      \nodata &     \nodata &  $>6.6$ \\
J1844.4--0306 & J184443.3--030518 &              NS &      $77\pm14$ &    2.8 &    401 &    0.3 &        0.9 &               0.93 &     \nodata &     \nodata &      \nodata &     \nodata &  $>2.2$ \\
J2004.3+3339 & J200422.7+333844 &              NS &      $88\pm15$ &    2.8 &    310 &   0.01 &       -1.0 &               0.67 &     \nodata &     \nodata &      \nodata &     \nodata &    \nodata \\
J2004.3+3339 & J200423.4+333906 &              CV &      $56\pm19$ &    0.6 &    310 &    0.1 &        0.6 &               0.42 &     20.1 &     \nodata &      \nodata &     \nodata &     0.1 \\
J2059.1+4403 & J205743.9+435858 &              NS &      $58\pm22$ &    1.0 &     45 &    0.5 &        0.9 &               0.87 &     \nodata &     \nodata &      \nodata &     \nodata &    $>4.9$ \\
J2236.9+1839 & J223704.7+184056 &      (AGN [14]) &    \nodata &   \nodata &     10 &    0.5 &       -0.2 &               0.97 &     20.0 &     \nodata &      15.1 &      5.0 &     0.4 \\
J2236.9+1839 & J223708.5+183939 &             AGN &      $58\pm19$ &    1.1 &    9.5 &    0.6 &        0.6 &               0.73 &     \nodata &     \nodata &      17.3 &     \nodata &    \nodata \\
J2254.4+0108 & J225435.8+010158 &             AGN &       $91\pm9$ &    6.1 &     22 &    0.2 &        0.4 &               0.41 &     \nodata &     \nodata &      17.4 &     \nodata &    \nodata \\
\enddata
\tablecomments{$^\star$ These sources are removed from our classification; either they appear extended or outside the 95\% $\gamma$-ray PU. $^{\rm a}$ The shortened 4FGL name from the 4FGL catalog by removing the leading ``4FGL" string. $^{\rm b}$ The shortened 2CXO name from the CSCv2.0 by removing the leading ``2CXO" string. $^{\rm c}$ The predicted class by MUWCLASS and the classification from the literature (if applicable) are provided in the parentheses. [1] \cite{2012ApJS..203...21A},  [2] \cite{2016MNRAS.462.4371I}, [3] \cite{2020ApJ...896L..33X}, [4] \cite{2012ApJS..203...21A}, [5] \cite{2010AA...518A..10V}, [6] \cite{2018AA...613A..51P}, [7] \cite{2014AA...563A..54P}, [8] \cite{2012MNRAS.421.1569B}, [9] \cite{2018AJ....155..189D}, [10] \cite{2012ApJS..200....8K}, [11] \cite{2019ApJS..242....4D}, 
[12] \cite{2022ApJ...939L...4D}, [13] \cite{2021ApJ...916...32G}, and [14] \cite{2021ApJS..257...30M}. $^{\rm d}$ The classification probability of the predicted class calculated from MUWCLASS, and its 1$\sigma$ uncertainty. $ ^{\rm e}$ The hardness ratio  $HR_{\rm h(ms)}=(F_{\rm h}-F_{\rm m}-F_{\rm s})/(F_{\rm h}+F_{\rm m}+F_{\rm s})$  where $F_{\rm s}$, $F_{\rm m}$, and $F_{\rm h}$ are the fluxes in the 0.5--1.2 keV, 1.2--2\,keV, and 2--7 keV bands.}
\end{deluxetable*}
\end{longrotatetable}

\section{Discussion}
\label{sec:disucssion}

\subsection{Correlation between X-Ray, Radio, and $\gamma$-Ray Fluxes}

\begin{figure*}
\begin{center}
\includegraphics[width=0.49\textwidth,trim=0 0 0 0]{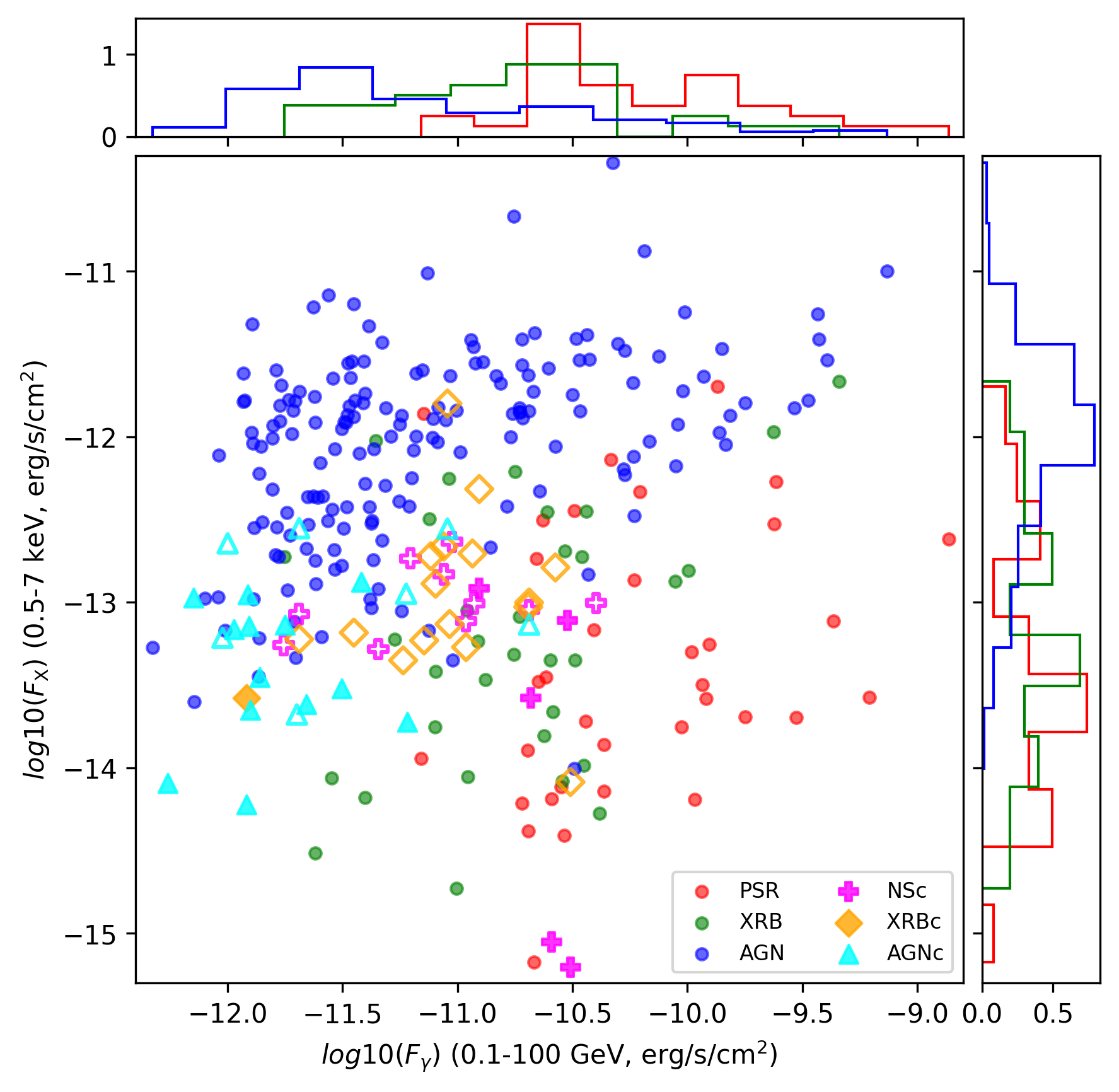}
\includegraphics[width=0.49\textwidth,trim=0 0 0 0]{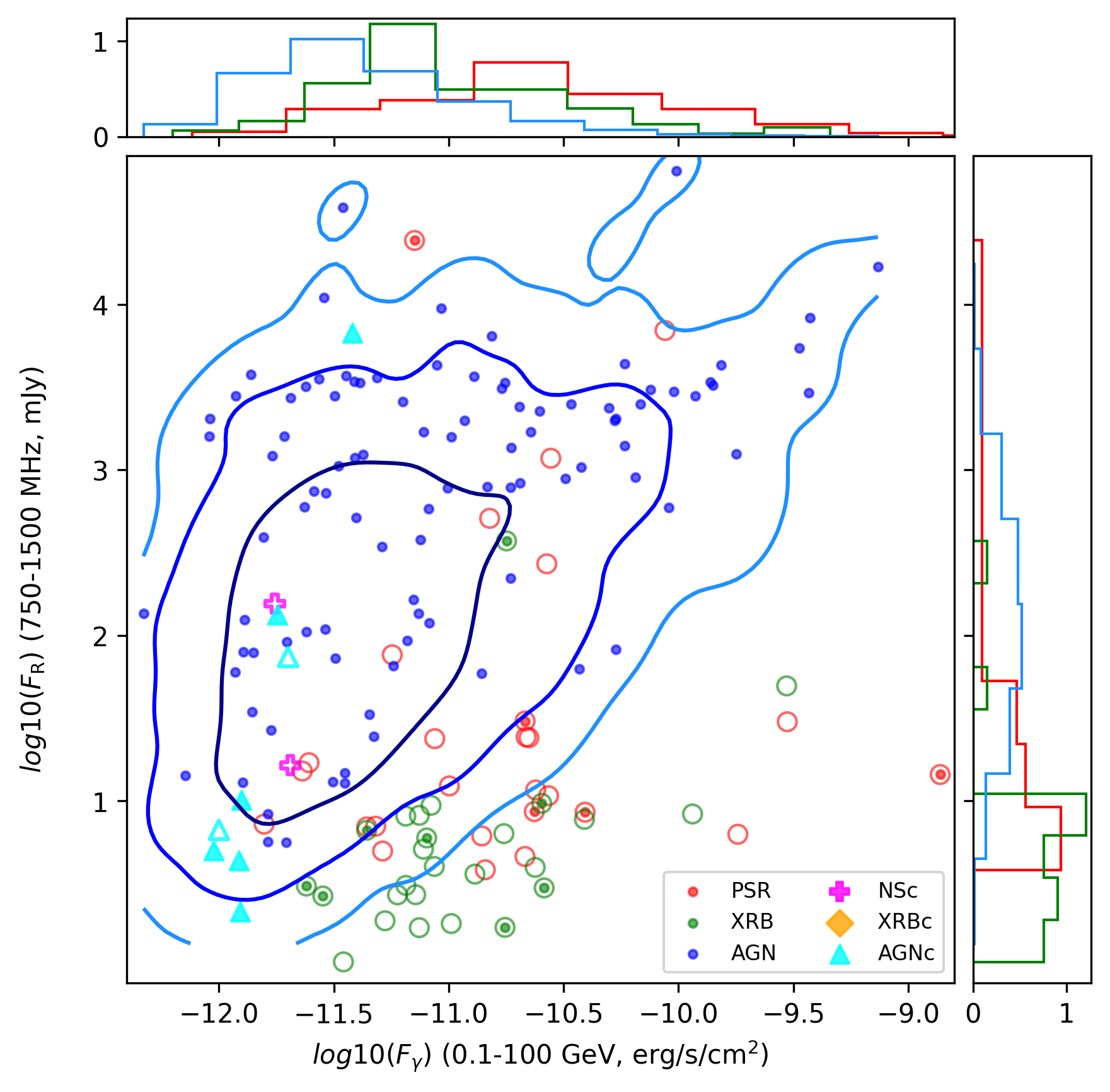}
\caption{Left panel: The distribution of 0.5--7\,keV X-ray fluxes ($F_{\rm X}$) and 0.1--100 GeV $\gamma$-ray fluxes ($F_{\gamma}$) for $\gamma$-ray-emitting PSRs, XRBs, and AGNs is shown with red, green, and blue dots, respectively. 
Additionally, the brightest sources within the PUs of each 4FGL sources classified as NS candidates (NSc), XRB candidates (XRBc, including LMXB, HMXB, and CV classes), and AGN candidates (AGNc) are marked as magenta crosses, orange diamonds, and cyan triangles. Solid markers represent sources with CT$\ge2$, while hollow markers represent sources with CT$<2$.  
The histograms show the distribution of $F_{\rm X}$ and $F_{\gamma}$ on vertical and horizontal axes for the $\gamma$-ray-emitting PSR, XRBs, and AGNs, using the same colors. Right panel: The distribution of 750--1500\,MHz radio fluxes ($F_{\rm R}$) and $F_{\gamma}$ for PSRs and XRBs identified from 4FGL is shown with open red and green circles. To maintain clarity, the distribution of AGNs is shown with contours after applying the kernel density estimation using Gaussian kernels, outlining their 1$\sigma$, 2$\sigma$, and 3$\sigma$ regions. The histograms on the sides show the distribution of $F_{\rm R}$ and $F_{\gamma}$ for these 4FGL sources detected in both radio and GeV bands. Red, green, and blue dots represent a subset of the identified PSRs, XRBs, and AGNs that are also detected in the CSCv2.0. The classified candidate sources are marked with the same symbols as in the left panel.} 
\label{fig:FG_FX}
\end{center}
\end{figure*}

Objects producing GeV $\gamma$-rays typically emit radiation across a broad range of frequencies from radio to $\gamma$-rays. Therefore, studying the correlation of emissions across different frequencies is fundamental for understanding the underlying physical processes. 
In the case of AGNs and XRBs (excluding spider pulsars and colliding wind binaries), X-ray emission mostly traces accretion onto a supermassive black hole or compact object, while radio and $\gamma$-ray radiation is typically attributed to relativistic jets. For instance, the correlation between $\gamma$-rays and radio emission has been observed for Fermi-LAT AGNs \citep[e.g.,][]{2011A&A...535A..69N}. However, it is worth noting that not all AGNs exhibit strong jets, with only $\sim10\%$ showing significant emission in radio wavelengths \citep{1989AJ.....98.1195K}. Furthermore, the correlation between X-ray, radio, and $\gamma$-ray bands can vary depending on the specific subclass of AGNs and the surrounding environment where particle acceleration occurs. 
For isolated pulsars, nonthermal X-ray emission as well as radio and $\gamma$-ray emission originate from various sites in the magnetosphere while thermal emission comes from the hot NS surface. In the case of pulsars in binaries, broadband emission can result from the intrabinary shock between the pulsar wind and the  wind of the companion star (high-mass $\gamma$-ray binaries), or from the shock near the surface of the ablated low-mass companion (black widow spider binaries). 
The distinct emission mechanisms result in different broadband SEDs for various types of $\gamma$-ray sources. For example, AGNs are known to be more X-ray efficient than rotation-powered pulsars, exhibiting higher X-ray to $\gamma$-ray flux ratios \citep[e.g.,][]{2021AJ....161..154K}.

We selected the 4FGL sources that are identified or associated (with association probabilities $>$95\%), and cross-matched their identified counterpart positions\footnote{We take precise positions of counterparts rather than 4FGL source positions.} with the CSCv2.0. We found 35 isolated pulsars (including isolated PSRs and MSPs), 33 XRBs (mostly binary MSPs like spiders, and HMXBs), and 172 AGNs (including blazars of different kinds and radio galaxies). We also cross-matched the same positions with the RACS survey, using a matching radius of 10\arcsec.  Among the 4FGL sources covered by the RACS survey, 28 (17\%) PSRs, 26 (28\%) XRBs, and 2209 (83\%) AGNs are found. There are five PSRs, eight XRBs, and 99 AGNs that are present in both the CSCv2.0 and the RACS survey. 

In the left panel of Figure \ref{fig:FG_FX} we plot the 0.5--7 keV X-ray fluxes ($F_{\rm X}$) versus 0.1--100 GeV $\gamma$-ray fluxes ($F_{\gamma}$) for the PSRs, XRBs, and AGNs matched to the CSCv2.0. 
We also show the brightest X-ray sources that were classified by us as an NS, XRB, or AGN in each 4FGL field. If a 4FGL source field has multiple X-ray sources classified as a PSR, XRB, or AGN, then only the brightest X-ray source from each of these classes is plotted. We chose to plot only the brightest X-ray source to provide a cleaner plot because X-ray sources within the same 4FGL source field will be sharing the same value of $F_{\gamma}$. 

For each of the three classes of known GeV $\gamma$-ray emitters identified from the 4FGL catalog, we fit a linear function to the logarithms of their $\gamma$-ray and X-ray fluxes. 
The flux correlation is significant for $\gamma$-ray-emitting AGNs with a Pearson's correlation coefficient of $r=0.42$ and a $p$-value of $p=10^{-8}$, where $r=0$ would indicate no correlation, and $p$ gives the probability of no correlation. XRBs exhibit an intermediate correlation ($r=0.39$ and $p=0.03$) while PSRs show no correlation ($r=0.23$ and $p=0.2$). 
It also shows that PSRs tend to display lower X-ray fluxes but higher $\gamma$-ray fluxes compared to XRBs and AGNs, while AGNs exhibit the opposite flux behavior.

We plot the 750--1500\,MHz radio fluxes ($F_{\rm R}$) against $F_{\gamma}$ for the $\gamma$-ray emitting PSRs, XRBs, and AGNs matched to the RACS survey in the right panel in Figure \ref{fig:FG_FX}. Additionally, we indicate sources with X-ray counterparts from the CSCv2.0, as well as outcomes of our classifications of X-ray sources.  
A linear regression is fit to the logarithm of $F_{\rm R}$ and $F_{\gamma}$ for each class of identified 4FGL sources. A significant correlation was found for AGNs ($r=0.47$ and $p=10^{-6}$), which has been observed before \citep[e.g.,][]{2011A&A...535A..69N}. However, no correlation was found between the fluxes of XRBs and PSRs. We also note that PSRs and XRBs exhibit systematically lower radio fluxes compared to AGNs.

The classified AGNs (cyan triangles) are comparatively fainter than previously known $\gamma$-ray AGNs in terms of their X-ray, $\gamma$-ray, and radio fluxes,\footnote{Note that the MUWCLASS pipeline does not use $F_{\gamma}$, or $F_{\rm R}$, or any other $\gamma$-ray or radio features.} with the exception of 2CXO J161532.9--603954, which exhibits
a very high radio flux associated with a one-sided jet. 
This likely represents a natural bias in the $\gamma$-ray AGN sample (based on the 4FGL catalog), where AGNs need to be bright to be reliably identified and associated with the $\gamma$-ray source. 
In the case of classified NS candidates (magenta crosses), their associated 4FGL sources generally have higher $F_\gamma$ compared to the classified AGNs, consistent with the distribution observed for identified PSRs and AGNs. The distribution of XRB candidates (orange diamonds) is consistent with identified XRBs in terms of their $F_{\rm X}$ and $F_\gamma$.

It is important to note that the distribution of identified sources from 4FGL and the classified brightest sources should not be expected to match perfectly in Figure \ref{fig:FG_FX}. 
We only plot the brightest X-ray sources within each 4FGL field, while the true counterpart could potentially be a fainter X-ray source that we do not plot. Additionally, some 4FGL source fields are only partially covered by X-ray observations; therefore, the true X-ray counterpart may not be among the classified sources.

\subsection{Comparison to other ML classification results}

We have conducted a comparative analysis of our classification outcomes with those of \cite{2021ApJ...923...75K}, \cite{2023MNRAS.521.6195M}, and \cite{2024MNRAS.527.1794Z} who used different approaches for the 4FGL source classification. \cite{2021ApJ...923...75K} employed a neural network classifier to classify potential X-ray/UV/optical counterparts to 4FGL-DR2 identified through Swift-XRT and UVOT telescopes, while both \cite{2023MNRAS.521.6195M} and \cite{2024MNRAS.527.1794Z} classified 4FGL-DR3 sources based solely on their $\gamma$-ray properties. \cite{2023MNRAS.521.6195M} employed a hierarchical approach for class determination to address imbalanced data, while \cite{2024MNRAS.527.1794Z} divided sources into high Galactic latitude and low Galactic latitude categories to account for the distinct source populations in these regions.

The summary of classification results for 36 unIDed GeV sources is presented in Table \ref{tab:comparison}. \cite{2021ApJ...923...75K} calculated Pbzr, which is the probability of the potential X-ray/UV/optical counterpart to be classified as a blazar using a binary classifier between blazars and pulsars, with Pbzr$>0.99$ indicating a likely blazar, Pbzr$<0.01$ indicating a likely pulsar, and others as ambiguous cases. We compared the classification results of the ensemble voting classifier, which combined the results of four individual classifiers from \cite{2024MNRAS.527.1794Z}, where they have a binary classifier between AGNs and PSRs for high Galactic latitude ($|b|>10^{\circ}$) sources and a ternary classifier of AGNs, PSRs, and other types for low Galactic latitude ($|b|\le10^{\circ}$) sources. We caution that the PSR class in \cite{2024MNRAS.527.1794Z} includes the MSPs where some of them belong to our LMXB class (i.e., spiders). As for \cite{2023MNRAS.521.6195M}, we compared their classification results where they determined the hierarchical groups of classes based on an RF model with a minimum of 100 sources for each subgroup, and the RF algorithm for the classification. For the class groups of Seyfert galaxies, novae, BCU, and AGNs  (hereafter ``sey,nov,bcu,agn"), FSRQ type of blazars, low-mass binaries, narrow-line Seyfert 1 galaxies, compact steep spectrum
radio sources, steep spectrum radio quasars, and binaries (hereafter ``fsrq,lmb,nlsy1,css,ssrq,bin"), and starburst galaxies, SNRs, radio galaxies, Galactic Center, BLL type of blazars, and normal galaxies (hereafter ``sbg,snr,rdg,gc,bll,gal"), they can be treated as AGN classes since most of sources are from the blazar and AGN classes. For the subgroups of MSPs, and globular clusters (hereafter ``msp,glc"), and high-mass binaries, SFRs, PWNe, and SNRs/PWNe (hereafter ``hmb,sfr,pwn,spp"), it is difficult to directly compare with our and other classifications due to their mixed nature. 

Out of the 36 sources, 13 sources (4FGL J0058.3--4603, 4FGL J0335.6--0727, 4FGL J0442.8+3609, 4FGL J0639.1--8009, 4FGL J1032.0+5725, 4FGL J1106.4+0859, 4FGL J1116.3+1818, 4FGL J1256.9+2736, 4FGL J1502.6+0207, 4FGL J1510.9+0551, 4FGL J1615.3--6034, 4FGL J2236.9+1839, and 4FGL J2254.4+0108) have been identified as promising AGN candidates in Section \ref{sec:AGN-Cand}, and exhibit consistent classifications with \cite{2021ApJ...923...75K,2023MNRAS.521.6195M,2024MNRAS.527.1794Z}.  
Two sources (4FGL J0737.4+6535 and 4FGL J1843.7--3227) are classified as AGNs by both \cite{2024MNRAS.527.1794Z,2023MNRAS.521.6195M}, while we found at least one AGN candidate within their PUs. 
For nine sources (4FGL J1616.6--5009, 4FGL J1720.6--3706, 4FGL J1725.1--3408, 4FGL J1732.8--3725,  4FGL J1737.1--2901, 4FGL J1739.2--2717, 4FGL J1742.5--2833, 4FGL J1745.6--3145, 4FGL J1757.6--2731), XRB candidates have been classified within their PUs, which are consistent with the classification of other type from \cite{2024MNRAS.527.1794Z} and ``hmb,sfr,pwn,spp" class from \cite{2023MNRAS.521.6195M}. 
For five sources (4FGL J1619.3--5047, 4FGL J1734.0--2933, 4FGL J1744.9--2905, 4FGL J1818.5--2036, 4FGL J1844.4--0306) that have NS candidates classified within their PUs, 
\cite{2024MNRAS.527.1794Z} classified them as the other type, and \cite{2023MNRAS.521.6195M} categorized them as ``hmb,sfr,pwn,spp" or ``psr" class, while \cite{2021ApJ...923...75K} supports our classification of 4FGL J1844.4--0306 as an NS. 
4FGL J1836.8--2354 was classified as a pulsar by \cite{2024MNRAS.527.1794Z} and ``msp,glc" class by \cite{2023MNRAS.521.6195M}, which is consistent with our classification of an LMXB class including spider binaries. 
Regarding 4FGL J0859.3--4342, previous classifications labeled it as a pulsar \citep{2021ApJ...923...75K,2023MNRAS.521.6195M,2024MNRAS.527.1794Z}, while we did not find any NS candidate within its PU in a crowded SFR. 
For the remaining sources, comparisons of classifications from \cite{2024MNRAS.527.1794Z,2023MNRAS.521.6195M} yielded inconsistencies, making it challenging to compare to our results.

\startlongtable
\renewcommand{\tabcolsep}{0.11cm}
\begin{deluxetable*}{lcccccr}
\tablecaption{Comparison of Classifications to Other Catalogs\label{tab:comparison}}
\tablewidth{0pt}
\tablehead{
\colhead{4FGL Name} &     \colhead{$\gamma$-candidate$^a$} &    \colhead{Pbzr$^b$} &    \colhead{class\_Z$+$24$^c$} &    \colhead{prob\_Z$+$24$^d$} &    \colhead{class\_MB23$^e$} &    \colhead{prob\_MB23$^f$}
}
\startdata
4FGL J0058.3--4603 &       +1A &      1.0 &        agn &          0.997 &             sey,nov,bcu,agn &           0.732 \\
4FGL J0335.6--0727 &        1A &  \nodata &        agn &          0.999 &             sey,nov,bcu,agn &           0.560 \\
4FGL J0442.8+3609 &        2A &  \nodata &        agn &          0.529 &             sey,nov,bcu,agn &           0.568 \\
4FGL J0639.1--8009 &   9A+1L1A &  \nodata &        agn &          0.988 &             sey,nov,bcu,agn &           0.613 \\
4FGL J0737.4+6535 &      1L1A &    0.999 &        agn &          1.000 &             sey,nov,bcu,agn &           0.473 \\
4FGL J0859.2--4729 &   +3N2A1C &    0.731 &      other &          0.666 &                         psr &           0.280 \\
4FGL J0859.3--4342 & 1A+1H1A2C &    0.003 &        psr &          0.491 &                         psr &           0.331 \\
4FGL J1032.0+5725 &       10A &  \nodata &        agn &          0.959 &      sbg,snr,rdg,gc,bll,gal &           0.528 \\
4FGL J1106.4+0859 &     1A+1A &  \nodata &        agn &          1.000 & fsrq,lmb,nlsy1,css,ssrq,bin &           0.582 \\
4FGL J1115.1--6118 &        2N &  \nodata &      other &          0.885 &      sbg,snr,rdg,gc,bll,gal &           0.314 \\
4FGL J1116.3+1818 &        1A &  \nodata &        agn &          0.999 &             sey,nov,bcu,agn &           0.389 \\
4FGL J1256.9+2736 &     8A+1A &  \nodata &        agn &          0.997 &             sey,nov,bcu,agn &           0.484 \\
4FGL J1502.6+0207 &        2A &  \nodata &        agn &          0.999 &      sbg,snr,rdg,gc,bll,gal &           0.580 \\
4FGL J1510.9+0551 &        4A &  \nodata &        agn &          0.998 &      sbg,snr,rdg,gc,bll,gal &           0.672 \\
4FGL J1615.3--6034 &       +1A &  \nodata &        agn &          0.562 &             sey,nov,bcu,agn &           0.717 \\
4FGL J1616.6--5009 &       +1L &  \nodata &      other &          0.523 &             hmb,sfr,pwn,spp &           0.400 \\
4FGL J1619.3--5047 &        1N &  \nodata &      other &          0.782 &             hmb,sfr,pwn,spp &           0.432 \\
4FGL J1720.6--3706 &       +1C &  \nodata &      other &          0.607 &             hmb,sfr,pwn,spp &           0.403 \\
4FGL J1725.1--3408 &     +1N1L &  \nodata &      other &          0.852 &             hmb,sfr,pwn,spp &           0.609 \\
4FGL J1732.8--3725 &       +1L &  \nodata &      other &          0.733 &             hmb,sfr,pwn,spp &           0.446 \\
4FGL J1734.0--2933 &       +1N &    0.245 &      other &          0.788 &             hmb,sfr,pwn,spp &           0.478 \\
4FGL J1737.1--2901 &       +1C &  \nodata &      other &          0.666 &             hmb,sfr,pwn,spp &           0.488 \\
4FGL J1739.2--2717 &       +1C &  \nodata &      other &          0.649 &             hmb,sfr,pwn,spp &           0.394 \\
4FGL J1742.5--2833 &   1N+1N1C &  \nodata &      other &          0.638 &             hmb,sfr,pwn,spp &           0.405 \\
4FGL J1744.9--2905 &        2N &  \nodata &      other &          0.636 &                         psr &           0.402 \\
4FGL J1745.6--3145 &       +1C &  \nodata &      other &          0.734 &             hmb,sfr,pwn,spp &           0.392 \\
4FGL J1750.4--3023 &       +1N &  \nodata &      other &          0.585 &             sey,nov,bcu,agn &           0.410 \\
4FGL J1757.6--2731 &       +1C &  \nodata &      other &          0.627 &             hmb,sfr,pwn,spp &           0.480 \\
4FGL J1804.9--3001 &     +1L1C &  \nodata &      other &          0.552 &             sey,nov,bcu,agn &           0.351 \\
4FGL J1818.5--2036 &       +1N &  \nodata &      other &          0.775 &             hmb,sfr,pwn,spp &           0.506 \\
4FGL J1836.8--2354 &       +1L &  \nodata &        psr &          0.490 &                     msp,glc &           0.673 \\
4FGL J1843.7--3227 &     +1N1A &  \nodata &        agn &          0.991 &             sey,nov,bcu,agn &           0.747 \\
4FGL J1844.4--0306 &     1N+1N &    0.004 &      other &          0.699 &             hmb,sfr,pwn,spp &           0.542 \\
4FGL J2004.3+3339 &     +1N1C &  \nodata &      other &          0.525 &      sbg,snr,rdg,gc,bll,gal &           0.644 \\
4FGL J2236.9+1839 &       +2A &  \nodata &        agn &          0.999 &             sey,nov,bcu,agn &           0.661 \\
4FGL J2254.4+0108 &        1A &  \nodata &        agn &          1.000 &             sey,nov,bcu,agn &           0.612 \\
\enddata
\tablecomments{$^a$ The same $\gamma$-candidate column from Table \ref{tab:classification}. $^b$ Probability of being a blazar type from \cite{2021ApJ...923...75K}. $^c$ Classification from \cite{2024MNRAS.527.1794Z}. $^d$ Corresponding classification probability. $^e$ Classification from \cite{2023MNRAS.521.6195M}. $^f$ Corresponding classification probability.}
\end{deluxetable*}

\section{Summary and Conclusions}
 \label{sec:conclusion}

We applied the updated MUWCLASS to X-ray sources from the CSCv2.0 located within the PUs of 73 unIDed 4FGL sources. 
The main scientific products and the results are summarized below: 

\begin{itemize}
    
\item We incorporated several updates into the MUWCLASS pipeline, outlined in  \cite{2022ApJ...941..104Y}. These updates include astrometric correction, modifications to X-ray interobservation variability calculations, the adoption of a probabilistic cross-matching algorithm, and the introduction of a novel oversampling method. 
\item We classified 1206 X-ray sources with S/N$>3$ within the PUs of 73 GeV sources. Detailed results and visualizations are publicly available.\footnote{\url{https://muwclass.github.io/MUWCLASS_4FGL-DR4/}}  Among these, 103 sources were identified as potential contributors to 42 GeV sources. 
However, for sources belonging to classes associated with compact objects (CVs, LMXBs, NSs), the classification reliability is low in many cases. This is largely due to the underrepresentation of these classes in the training data set and/or the faintness of many classified sources (hence, their X-ray properties are uncertain). In a number of cases, we performed additional analysis using data inaccessible to MUWCLASS. 

\item Due to the large sizes of GeV source PUs, it is not uncommon to encounter multiple X-ray sources belonging to different classes of potential $\gamma$-ray emitters within a single 4FGL source, particularly those with deeper Chandra X-ray Observatory observations. While it is likely that only one of these X-ray sources serves as the dominant contributor to the GeV emission, distinguishing among them is not feasible within the MUWCLASS pipeline framework alone. Therefore, in such cases, we considered ML classification outcomes  jointly with the GeV and radio properties and more detailed X-ray properties (such as spectra and light curves), along with data from deeper optical/NIR/MIR surveys or observations. As a result of our ML classification efforts, we associated two 4FGL sources with NS (pulsar) candidates,  16 with AGN candidates,  and seven with SFRs. For 8 GeV sources, we could not determine the preferred counterpart among multiple possible $\gamma$-ray emitters. For the remaining 40 GeV sources, we could not identify any plausible counterpart in X-rays. Of these, 19 are fully covered by Chandra X-ray Observatory observations but with varying depths. Among these,  15 sources are too close to the Galactic Center, making reliable classifications challenging due to the large degrees of confusion while cross-matching. Moreover, 40 GeV sources only have partial ($<95\%$) Chandra X-ray Observatory coverage, and hence, there can be other X-ray sources within their extent that we could not study.    
\end{itemize}

The challenges encountered in identifying compact objects in this study underscore the need for deeper optical/NIR/MIR surveys and improved X-ray photon statistics. The latter can help reduce PUs, crucial for addressing confusion, especially as surveys delve deeper. Additionally, integrating data from deep radio surveys can help break the degeneracies between AGN and various compact object classes. 
In the future, a larger data set comprising firmly classified GeV sources observed by Chandra X-ray Observatory could facilitate the development of a comprehensive MW training data set spanning from radio to X-rays. 
Training ML models with such data could improve classification performance, although multiple combinations of GeV and lower energy counterparts will need to be considered due to the large sizes of PUs of the GeV sources. Accumulating more GeV data and improving the Galactic diffuse background model can further reduce the GeV PUs, leading to more reliable classification outcomes. A Chandra legacy program targeting unexplored 4FGL sources would enable further advancements in our understanding of GeV $\gamma$-ray emitters. The exceptional angular resolution and substantial effective area of Chandra X-ray Observatory are critical for identifying optical/NIR/MIR counterparts of X-ray sources, particularly in dense regions of the Galactic plane. Among the currently studied future X-ray mission concepts, AXIS \citep{2018SPIE10699E..29M} and Lynx \citep{2017SPIE10397E..0SG} hold significant promise in this regard.

\newpage

\medskip\noindent{\bf Acknowledgments: }

We are grateful to Steven Chen, Rafael Martínez-Galarza, and Jordan Eagle for fruitful discussions that helped to advance this work. 

This work was supported by the National Aeronautics and Space Administration through Chandra Award AR0-21007X issued by the Chandra X-ray Center, operated by the Smithsonian Astrophysical Observatory for the National Aeronautics Space Administration under contract NAS803060.  Partial support for this work was also provided by the National Aeronautics and Space Administration through the Fermi Gamma-ray Observatory Guest Investigator program award 80NSSC22K1575. 
J.H. acknowledges support from NASA under award No. 80GSFC21M0002.

\bibliography{references}{}
\bibliographystyle{aasjournal}

\appendix

\section{Astrometry Correction}
\label{apd:astrometry}

The CSCv2.0 provides three levels of data products: per-observation level, stack-level, and master-level data sets. 
For the stack-level images, multiple X-ray observations are stacked into a single image after an astrometric correction is applied to align individual images to the reference image. 
However, an absolute astrometric correction has not been incorporated into the CSCv2.0. 
Therefore, we have applied a translation-only absolute astrometry correction, following the methodology outlined online.\footnote{\url{https://cxc.cfa.harvard.edu/csc/memos/files/Martinez-Galarza_stack_astrometry_translation_spec_doc.pdf}} 
This method, originally designed for CSCv2.1, involves a transformation that aligns X-ray sources from the stack-level detections with the Gaia reference catalog sources. 
Unlike the original method, we have opted for a more conservative searching radius of 1$\arcsec$ when cross-matching X-ray sources to Gaia DR3 sources.

The weighted least square method is used to calculate the transformation by minimizing the Euclidean distance between the CSCv2.0 sources and reference sources. Standard coordinates $(\xi, \eta)$ defined on the tangent plane in units of arcseconds are used:

\begin{equation}
\begin{aligned}
    \xi &= \frac{\cos\delta\sin(\alpha-\alpha_0)}{\sin\delta\sin\delta_0+\cos\delta\cos\delta_0\cos(\alpha-\alpha_0)} \\
    \eta &= \frac{\sin\delta\cos\delta_0-\cos\delta\sin\delta_0\cos(\alpha-\alpha_0)}{\sin\delta\sin\delta_0+\cos\delta\cos\delta_0\cos(\alpha-\alpha_0)}
    \end{aligned}
\end{equation}

\noindent where $(\alpha_0,\delta_0)$ refer to the coordinates of the tangent plane reference point.

The source list from the stack-level detections is cross-matched with the Gaia reference source list with a search radius of 1$\arcsec$. 
If the number of matches, $n_{\rm match}$, is $\ge$3, the astrometric correction is computed. 
The correction in standard coordinates $(\Delta\xi, \Delta\eta)$ is determined by minimizing the weighted sum of the residual square of all suitable pairings:

\begin{subequations}
\begin{align}
    S = \sum_{i=1}^{n} w_i R_i^2
\end{align}
\end{subequations} 

where $w=1/(r_0 r_1)$, $r_0$ and $r_1$ being the semi-major axis and semi-minor axis of the X-ray PU, and the residual distance $R_i$ is calculated as the Euclidean distance between the transformed coordinates of X-ray sources and the Gaia reference coordinates, 

\begin{subequations}
\begin{align}
    R_i = \sqrt{(\xi_i^{\rm Gaia}-\xi_i^{\rm trans})^2+(\eta_i^{\rm Gaia}-\eta_i^{\rm trans})^2}
\end{align}
\end{subequations}

\noindent with $\xi^{{\rm trans}}=\xi^{{\rm X}}+\Delta\xi$ and  $\eta^{{\rm trans}}=\eta^{{\rm X}}+\Delta\eta$. 

Finally, the transformed standard coordinates of X-ray sources are converted back to celestial coordinates. 
We also update the X-ray PU by adding (in quadrature) the alignment uncertainty (${\rm PU_{\rm align}}$)\footnote{${\rm PU_{\rm align}}$ is assumed at a 68\% level and is converted to a 95\% level assuming the position errors follow Rayleigh distributions before combining with the 95\% X-ray PU.} defined as

\begin{subequations}
\begin{align}
    {\rm PU_{align}} = \sqrt{\frac{\sum_{i=1}^{n} w_i R_i^2}{\sum_{i=1}^{n} w_i}}
\end{align}
\end{subequations} 

where $R_i$ is the residual distance after the transformation. 
For those stacked observations with no astrometric correction applied, we use a systematic uncertainty of $0.71\arcsec$ at a 95\% level (adopted from the CSCv2.0) for the alignment uncertainty.

After applying the absolute astrometric correction to sources in each stacked observation, the coordinates of the stack-level detections are combined to produce the statistically averaged positions and PU ellipses, following the CSC memo.\footnote{\url{https://cxc.cfa.harvard.edu/csc/memos/files/Davis_ellipse_rev2.pdf}}

\section{Training Data Updates}
\label{apd:TD-updates}

Several updates have been implemented in the TD since its last release \citep{2022ApJ...941..104Y}. These updates involve adding additional sources, particularly for less populous classes, 
refining and cleaning the classifications of ambiguous classes, specifically for YSOs and LM-STARs, and incorporating a probabilistic cross-matching method \citep{2018MNRAS.473.4937S}.  
The updated TD is now available and accessible on the visualization website,\footnote{\url{https://muwclass.github.io/XCLASS_CSCv2/}} with a total number of 3,119 sources including 1596 AGNs, 826 LM-STARs, 352 YSOs, 107 HM-STARs, 110 NSs, 52 LMXBs, 44 CVs, and 32 HMXBs.

The updated TD has been expanded with sources from additional catalogs containing literature-verified sources. 
This includes sources from the Open CV catalog \citep{2020RNAAS...4..219J}, 
confirmed CVs from \cite{2021AJ....162...94S}, the Swift Burst Alert Telescope 105 month hard X-ray survey catalog \citep{2018ApJS..235....4O}, HM-STARs (spectroscopically classified OB stars) from LAMOST-DR5 \citep{2019ApJS..241...32L}, and LM-STARs (spectroscopically classified as AFGKM stars) from LAMOST-DR8 \footnote{\url{https://www.lamost.org/dr8/}}. 
Additionally, sources showing evidence of binarity, from the APOGEE-2 data in SDSS DR16, have been retained as part of our LM-STAR class.

Various filters have been applied to enhance the quality of the TD. Specifically, sources in the central regions of SFRs were removed due to MIR image saturation, after manual inspection of AllWISE images. 
Additionally, sources marked as confused in the CSCv2.0 (with 2CXO names ending with X or A), extended sources with Gaia BP/RP flux excess factor $>20$, or AllWISE extended source flag$=5$ were removed. 
For cross-matching with the SIMBAD database, a radius of 3$\arcsec$ was applied, and sources with classifications conflicting with  those in the original publications were eliminated. This step led to the removal of several hundred YSOs used in the previous TD version, as SIMBAD classifications were deemed to be unreliable in those cases. 
To ensure the purity of less populated classes, a manual investigation was conducted for rare-type sources, and any source with suspicious or possibly unreliable classifications were removed. We also manually vetted any outliers from more populous classes that were identified with the help of the visualization website.

\section{Physical Oversampling}
\label{apd:physical-oversampling}

A new oversampling method was employed to generate a fainter population of sources compared to the TD sources before oversampling, by applying additional (optical) extinction and (X-ray) absorption to randomly selected TD sources, excluding AGNs, which are the most populous class. 
We first fit the  distributions of $E(B-V)$ values for the TD sources (regardless of their classes) obtained from the 3D extinction maps \citep{2016ApJ...818..130B} with the help of Gaia distances using uniform, Gaussian, Poisson, and gamma distributions. 
A gamma distribution provides the best fit with a probability density of $p(x) = x^{k-1}\frac{e^{-x/\theta}}{\theta^k \Gamma (k)}$, where $k=0.5$ and $\theta=1.5$.  
The corresponding  hydrogen column density values are then calculated using the relation $N_{\rm H}$ (cm$^{-2})= (2.21\pm 0.09) \times 10^{21} A_{\rm V}$ (mag) from \cite{2009MNRAS.400.2050G}, where $A_{\rm V}=3.1 \times E(B-V)$. The photon indices, which are used for calculating absorption corrections assuming an absorbed PL model for the X-ray spectra, were sampled from the distribution\footnote{We followed Equations (1a) and (1b)  from \cite{2022ApJ...941..104Y} for calculating the distributions.} of photon indices for individual sources, if they are available from the CSCv2.0.  
Otherwise, the mean values of photon indices for each class from the CSCv2.0 TD are used, i.e., $\Gamma_{\rm AGN}=1.94$, $\Gamma_{\rm YSO}=2.95$, $\Gamma_{\rm LM-STAR}=5.14$, $\Gamma_{\rm HM-STAR}=3.09$, $\Gamma_{\rm NS}=1.94$, $\Gamma_{\rm CV}=1.61$, $\Gamma_{\rm HMXB}=1.28$, $\Gamma_{\rm LMXB}=1.97$. 
Once the reddening, $E(B-V)$,    hydrogen absorption column, $N_{\rm H}$, and $\Gamma$  are determined, they are applied to modify the fluxes of the randomly selected (from the TD) sources using the method described in Section 3.3 of \cite{2022ApJ...941..104Y}, which was employed to account for the absorption/extinction bias for the AGNs in the TD.

\begin{figure*}
\begin{center}
\includegraphics[scale=0.49,trim=0 0 0 0]{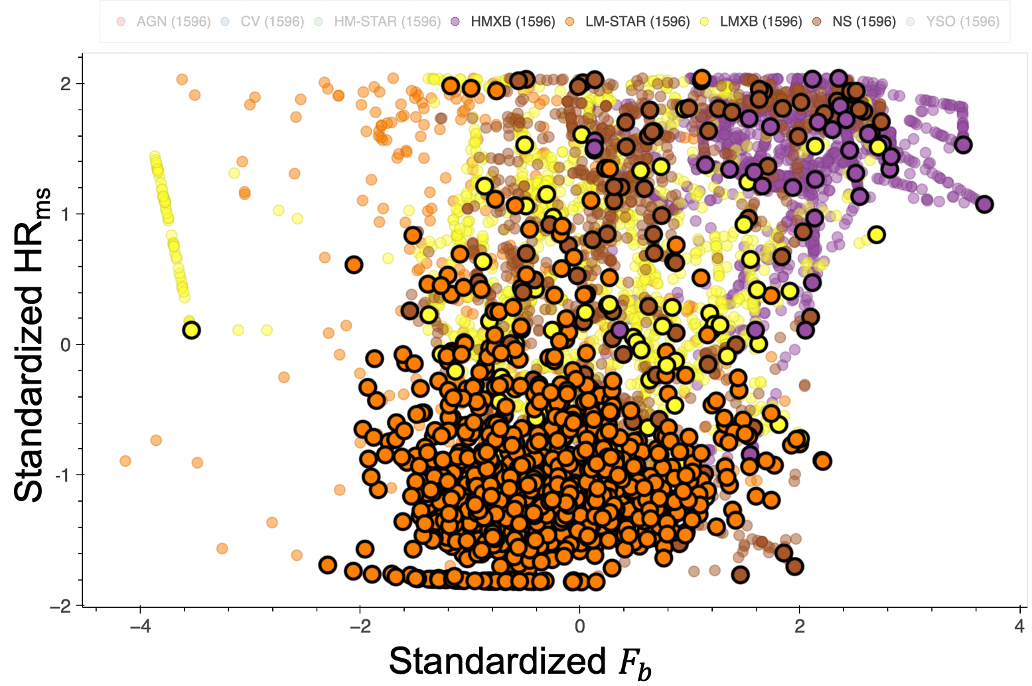}
\includegraphics[scale=0.49,trim=0 0 0 0]{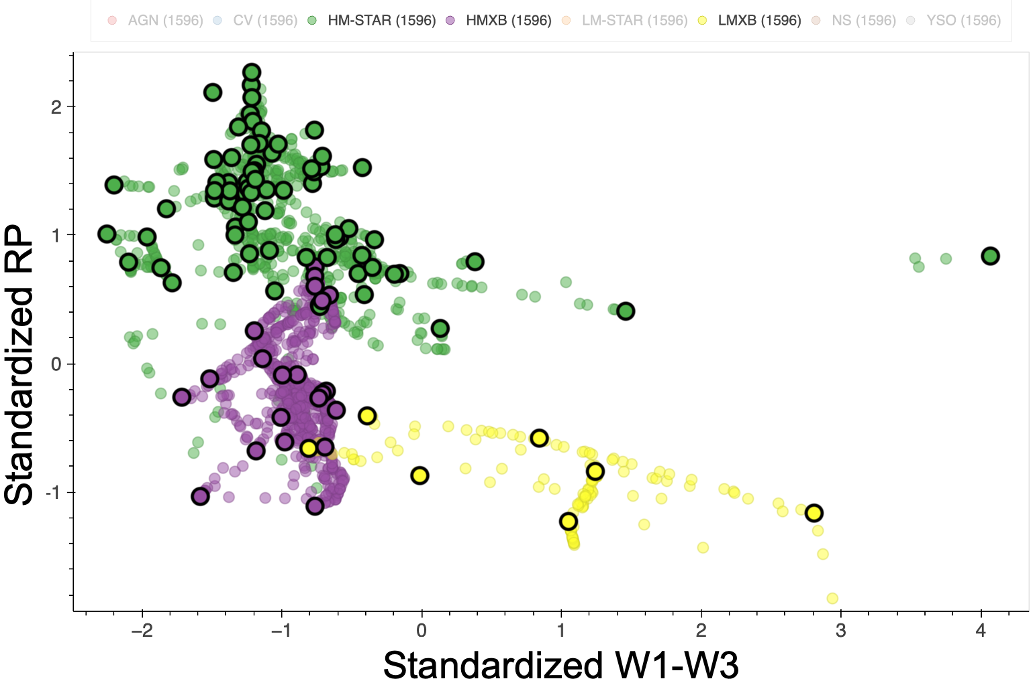}
\caption{Examples of 2D slices of the multidimensional feature space showing the TD content after applying the physical oversampling method to generate synthetic sources and balance the numbers of sources across all classes. The original (real) TD sources have black borders while the synthetic ones do not have them. The features have been preprocessed and standardized, as described in
Section 3.4 in \cite{2022ApJ...941..104Y}. For comparison purposes, the same features as those presented in Figure 3 of \cite{2022ApJ...941..104Y} are used. However, we note that the broadband flux feature $F_{\rm b}$ was excluded from our TD since it is distance dependent.} 
\label{fig:physical-oversample}
\end{center}
\end{figure*}

To address the issue of different artificial structures introduced by the physical oversampling and SMOTE methods, a hybrid oversampling approach is adopted by combining both physical oversampling  (described above)  and the SMOTE method used in \cite{2022ApJ...941..104Y}. Firstly, the less-populated classes of sources from the TD are oversampled using the physical oversampling method until their number reaches $\min (\frac{N_{\rm class} \times 1596}{(1596 - f \times (1596 - N_{\rm class}))}, 1596)$, where 1596 is the number of sources for the most populated class (AGN), $N_{\rm class}$ is the number of sources for the corresponding class before oversampling, and $f$ is the percentage contribution of the oversampled sources from the physical oversampling relative to the SMOTE method, which is set to 0.5. 
In the next step, the sources are further oversampled using the SMOTE method to ensure that each class has the same number of sources as the most populated class, following the method detailed in Section 3.5 of \cite{2022ApJ...941..104Y}. 
Figure \ref{fig:physical-oversample} illustrates examples of employing the new oversampling method described above to the TD sources, using the same 2D feature spaces as those  shown in Figure 3 of \cite{2022ApJ...941..104Y} for comparison.

Since the implemented physical oversampling method does not preserve the correlations between distances and reddening, we removed the broadband flux feature from the TD. 
Since all other fluxes are normalized by the source's broadband flux, they are distance independent. 
For instance, an oversampled source may exhibit significant absorption, resulting in the lack  of optical and NIR fluxes, but it may still be bright in the MIR and hard X-rays since these bands are minimally affected by the applied large absorption. However, in reality, highly absorbed sources are likely to be distant, which causes them to be faint in all of these bands.

\section{MUWCLASS Performance Evaluation}
\label{apd:evaluation}

\begin{figure*}
\begin{center}
\includegraphics[width=400pt,trim=0 0 0 0]{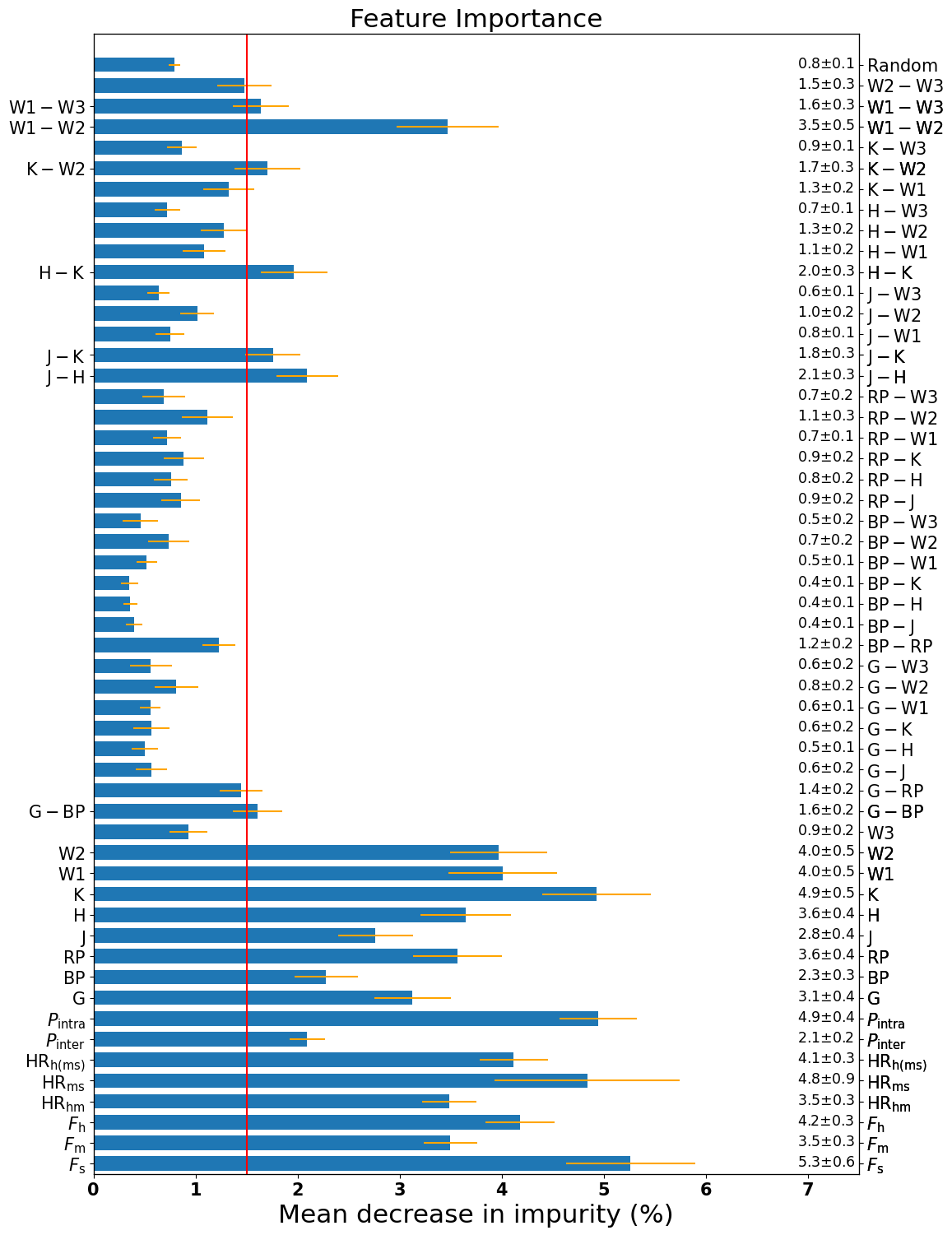}
\caption{
Feature importance (blue bars) and their 1$\sigma$ uncertainties (orange error bars) for all considered features with their names specified next to the vertical axis on the right.  The values and their uncertainties are also shown next to the feature names. The subset of features  used in the ML classification have an importance exceeding that shown by the red line (1.5\%) with their names shown next to the vertical axis on the left. We also added a dummy (random) feature and evaluated its importance (the top feature). 
}
\label{fig:feature-selection}
\end{center}
\end{figure*}

We rerun the feature selection with the updated pipeline and TD. Figure \ref{fig:feature-selection} shows the importance values of all features. A conservative threshold of 1.5\% (two times larger than the random feature sampled from a uniform distribution) was applied to omit the less important features. This resulted in a selection of 23 features,  as listed on the left side of Figure \ref{fig:feature-selection}. 

\begin{figure*}
\begin{center}
\includegraphics[scale=0.4,trim=0 0 0 0]{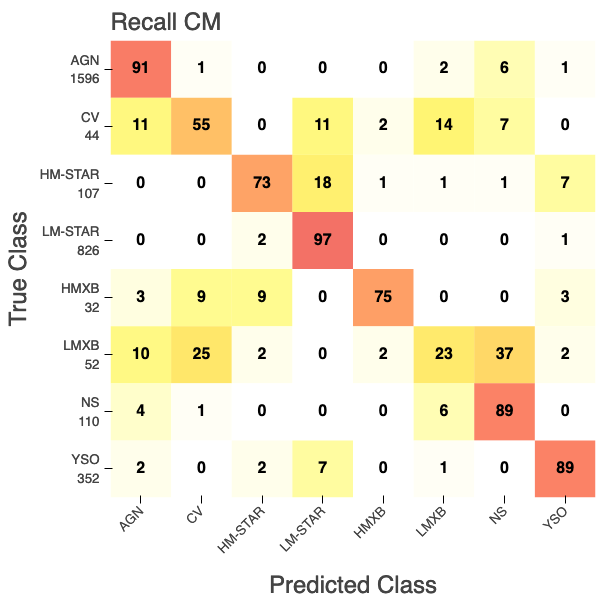}
\includegraphics[scale=0.4,trim=0 0 0 0]{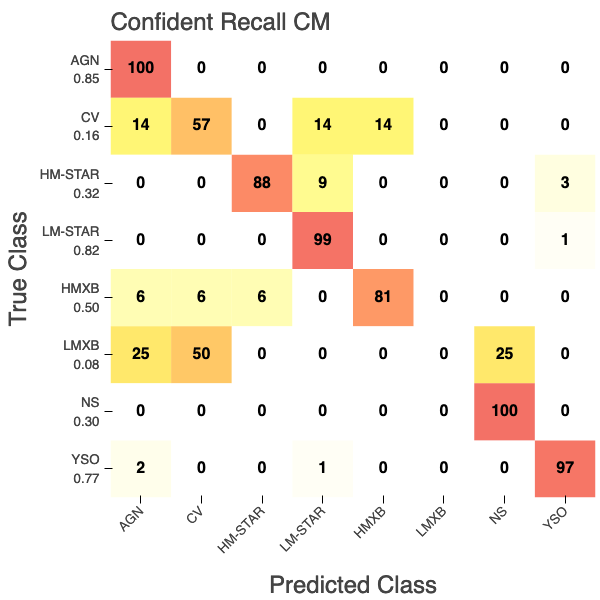}
\includegraphics[scale=0.4,trim=0 0 0 0]{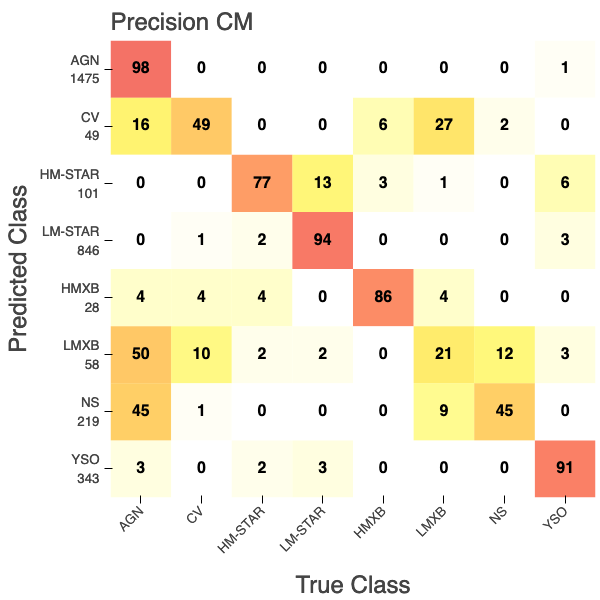}
\includegraphics[scale=0.4,trim=0 0 0 0]{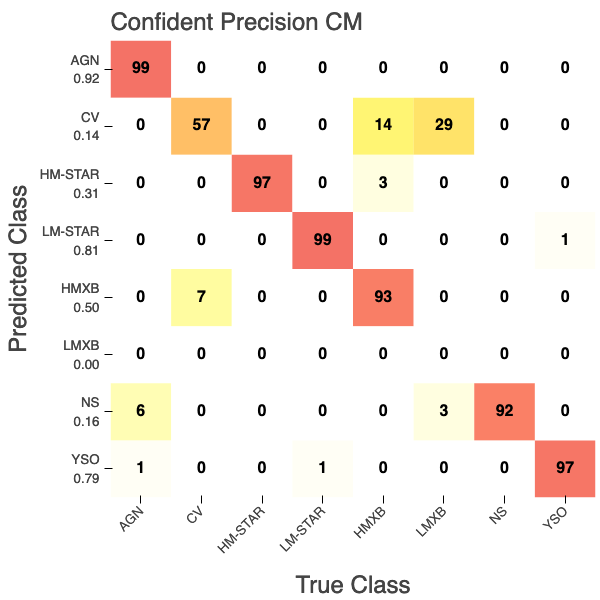}
\caption{The left panels show the normalized confusion matrices (CMs) of all classifications while the right panels show the normalized CMs for the confident classifications (${\rm CT}\ge2$). The upper panels show the normalized recall CMs, and the lower panels show the normalized precision CMs. 
The value within each element of the CM is the percentage of sources in a predicted (true) class, shown on the horizontal axis, that are from the true (predicted) class, shown on the vertical axis, for the normalized recall (precision) CMs.
The values under the class labels along the vertical axis in the left panels are the total numbers of the sources in the corresponding classes while in the right panels these values are the fractions of the sources surviving the confidence cut (CT$\ge2$) for each class. 
The darker the color, the higher the percentage is.} 
\label{fig:LOOCV}
\end{center}
\end{figure*}

The pipeline's performance was assessed through the leave-one-out cross-validation, as explained in Section 4.1 in \cite{2022ApJ...941..104Y}, with 1000 Monte-Carlo samplings conducted for each TD source to account for feature uncertainties. 
In Figure \ref{fig:LOOCV}, the normalized recall (upper row) and precision (lower row) CMs are presented for all classifications (left panel) and for only confident classifications with ${\rm CT}\ge2$ (right panel). The numbers of  sources for each class are indicated below the class name on the vertical axis in the left panel, while the fractions of the sources surviving the confidence cut ${\rm CT}\ge2$ for each class are shown on the vertical axis in the right panel. 
The normalized recall (precision) CMs are obtained by calculating the fraction of true (predicted) sources of the row class that are predicted (true) sources of the column class, as defined in Appendix A in \cite{2022ApJ...941..104Y}.

The overall accuracy for all classifications is 89.7\%, increasing to 98.7\% for confident classifications (CT$>$2). Notably, the LMXB class exhibits the poorest performance among all eight classes, despite not being the smallest in size within the TD. This disparity may stem from the heterogeneous nature of the LMXB class, encompassing both XRBs fueled by accretion and nonaccreting XRBs like spider binaries. Additionally, LMXBs in the TD are usually discovered during outbursts, so that LMXBs tend to be at farther distances, and their MW counterparts are thus too faint to be detected by the MW surveys we are currently using. Consequently, they are prone to be confused with the NS class.

\end{document}